\documentclass[a4paper,11pt]{article}
\pdfoutput=1 % if your are submitting a pdflatex (i.e. if you have
\usepackage{jheppub} % for details on the use of the package, please
\usepackage[T1]{fontenc} % if needed
\usepackage{soul}%to highlight somte text, use \hl{text to be highlighted}
\usepackage{color}% for mutiple soul color

\usepackage{empheq}
\usepackage{mathrsfs} % curly lagrangian
\usepackage{url}
\usepackage[utf8]{inputenc}
\usepackage[toc,page]{appendix}
\usepackage[autostyle]{csquotes}
\usepackage{slashed}
\usepackage{amssymb}
\usepackage{graphicx}
\usepackage{dirtytalk}% quote and double quote \say{text}
\usepackage{longtable}
\usepackage{array}%This Package is for table
\usepackage{bm} %This Package is for math bold
\usepackage{amsmath}
\usepackage{nccmath}
\usepackage{cancel}
\usepackage{mathtools}
\usepackage{amsfonts}
\usepackage{amssymb}
\usepackage[labelfont=bf]{caption}
\usepackage{paralist} %for inline enumerate
\usepackage{hyperref}
\usepackage{cleveref}
\title{\boldmath Logarithmic corrections to black hole entropy in matter coupled $\mathcal{N} \geq 1$ Einstein-Maxwell supergravity}

\author{Sudip Karan}
\author{and Binata Panda}
\affiliation{Department of Physics,\\ Indian Institute of Technology (Indian School of Mines),\\ Dhanbad, Jharkhand-826004, India}

\emailAdd{sudip.karan@ap.ism.ac.in}
\emailAdd{binata@iitism.ac.in}

\abstract{We calculate the first three Seeley-DeWitt coefficients for fluctuation of the massless fields of a $\mathcal{N}=2$ Einstein-Maxwell supergravity theory (EMSGT) distributed into different multiplets in $d=4$ space-time dimensions. By utilizing the Seeley-DeWitt data in the quantum entropy function formalism, we then obtain the logarithmic correction contribution of individual multiplets to the entropy of extremal Kerr-Newman family of black holes. Our results allow us to find the logarithmic entropy corrections for the extremal black holes in a fully matter coupled $\mathcal{N}=2,d=4$ EMSGT, in a particular class of $\mathcal{N}=1,d=4$ EMSGT as consistent decomposition of $\mathcal{N}=2$ multiplets ($\mathcal{N}=2 \to \mathcal{N}=1$) and in $\mathcal{N} \geq 3,d=4$ EMSGTs by decomposing them into $\mathcal{N}=2$ multiplets ($\mathcal{N} \geq 3 \to \mathcal{N}=2$). For completeness, we also obtain logarithmic entropy correction results for the non-extremal Kerr-Newman black holes in the matter coupled $\mathcal{N} \geq 1,d=4$ EMSGTs by employing the same Seeley-DeWitt data into a different Euclidean gravity approach developed in \cite{Sen:2013ns}.}

\begin{document} 
\maketitle
\flushbottom

\section{Introduction}\label{A}
In gravity theories, black holes are awarded an intrinsic entropy given by the eminent semi-classical Bekenstein-Hawking entropy formula \cite{Bekenstein:1973jb,Hawking:1975sh}. To test the visibility of any putative quantum theory of gravity, such as string theory, we need to investigate black holes, especially their entropies from both macroscopic and microscopic sides. On the macroscopic side (or gravity side), black holes are described as classical solutions to some low-energy effective field theory of gravity where the Bekenstein-Hawking entropy formula receives classical higher derivative corrections ($\alpha^\prime$) and quantum or string loop corrections ($g_s$). The higher derivative corrections are encoded by the familiar Wald formula \cite{Wald:1993rw}. The most general quantum corrected form of the Bekenstein-Hawking entropy can be obtained as a series expansion in the inverse power of a large horizon area $\mathcal{A}_H$ (the large-charge limit),\footnote{The fundamental constants like the Boltzmann constant ($k_{B}$), reduced Planck's constant ($\hbar$) and speed of light ($c$) are considered as a unity.}
\begin{equation}\label{qqf}
	S_{\text{BH}}(\mathcal{A}_H) = \frac{\mathcal{A}_H}{4G_N}+ \mathcal{C}\thinspace \text{ln}\thinspace \left(\frac{\mathcal{A}_{H}}{G_N}\right) +\sum_{n=1}^\infty C_n {\mathcal{A}_H}^{-n+1},
\end{equation} 
where $G_N$ is the gravitational constant and $\mathcal{C}, C_n$ are numerical constants that generally depend on different black hole parameters. In the expansion \eqref{qqf}, the leading part is the Bekenstein-Hawking entropy $\frac{\mathcal{A}_H}{4G_N}$, while the subleading parts are quantum corrections to it. The second term with the universal form proportional to $\text{ln}\thinspace \frac{\mathcal{A}_{H}}{G_N}$ is the principal quantum correction, called the logarithmic correction. This particular one-loop correction\footnote{The n-loop quantum corrections to black hole entropy are of the order $g_s\sim\mathcal{A}_H^{-n+1}$, where the one-loop correction is proportional to $\mathcal{A}_H^0, \text{ln}\thinspace \mathcal{A}_H$, etc.} can be computed using just the massless fluctuations (IR data) running in the one-loop \cite{Banerjee:2011oo,Banerjee:2011pp,Sen:2012qq,Sen:2012rr,Bhattacharyya:2012ss,Chowdhury:2014np,Gupta:2014ns,Jeon:2017ij,Karan:2019sk,Sen:2013ns,Keeler:2014nn,Charles:2015nn,Larsen:2015nx,Castro:2018tg,Ferrara:2012bp,Banerjee:2020wbr} and is independent of the UV completion of the theory. Naturally, the logarithmic correction can serve as a \say{laboratory} for testing any proposed quantum completion of the classical gravity theory.

In the extremal (zero-temperature) limit, black holes cease to evaporate and become a stable platform for investigating the logarithmic entropy correction program. Often but not always, extremal black holes resemble BPS states in supergravity (e.g., extremal Reissner-Nordstr\"om black holes in four-dimensional supergravities \cite{Mohaupt:2000mj}) and add more stability for being the lightest charged black holes. Remarkably, the full quantum entropy of an extremal black hole can be obtained from just the near-horizon analysis without knowing details of the entire black hole geometry. The near-horizon geometry of every extremal black hole accommodates an $AdS_2$ part. Hence, the rules of $AdS_2/CFT_1$ conjecture prescribe that the entropy of extremal black holes is equal to the entropy calculated from the full string partition function characterizing the near-horizon geometry. This prescription is recognized as the famous quantum entropy function formalism \cite{Sen:2008wa,Sen:2009wb,Sen:2009wc}, which turns out to be a smart trick to compute quantum corrections to the extremal black hole entropy. So far, quantum entropy function formalism gained immense success by providing logarithmic entropy corrections for many supersymmetric and non-supersymmetric extremal black holes \cite{Banerjee:2011oo,Banerjee:2011pp,Sen:2012rr,Sen:2012qq,Chowdhury:2014np,Gupta:2014ns,Bhattacharyya:2012ss,Karan:2019sk,Jeon:2017ij,Keeler:2014nn,Ferrara:2012bp,Banerjee:2020wbr,Larsen:2015nx}.
Readers are also referred to \cite{Solodukhin:1995na,Solodukhin:1995nb,Fursaev:1995df,Mavromatos:1996kc,Mann:1996bi,Mann:1998hm,Kaul:2000rk,Carlip:2000nv,Govindarajan:2001ee,Gupta:2002bg,Medved:2004eh,Page:2005xp,Banerjee:2008cf,Banerjee:2009fz,Majhi:2009gi,Cai:2010ua,Aros:2010jb,Solodukhin:2010pk} for some earlier techniques to obtain the logarithmic entropy corrections.

At present, it is impossible to compute the full quantum partition function of any gravity theory. One can try to calculate it using supersymmetric localization \cite{Dabholkar:2011ad,Dabholkar:2013ad,Gupta:2013rkg,Dabholkar:2015ad,Murthy:2015sm,Gupta:2015rkg,Murthy:2016sm} as well as perturbatively using saddle point or loop expansion \cite{Banerjee:2011oo,Banerjee:2011pp,Sen:2012rr,Sen:2012qq,Chowdhury:2014np,Gupta:2014ns,Bhattacharyya:2012ss,Karan:2019sk,Jeon:2017ij,Banerjee:2020wbr}. In this paper, we opt for the latter approach where the leading saddle point is nothing but the classical black hole solution and the subleading loop-terms around the saddle point serve as quantum corrections. Entropy computed on the saddle point is the Bekenstein-Hawking entropy or Wald's entropy if the higher derivative corrections are included. At one-loop (next to saddle point), filtering out only massless modes leads to the desired logarithmic entropy correction to the classical black hole entropy. Necessarily, evaluation of the logarithmic entropy corrections requires the analysis of one-loop quantum effective action characterizing only massless fluctuations. The heat kernel method can guide us from here by describing the one-loop effective action in terms of heat kernel of the kinetic operator characterizing the fluctuations. The heat kernel can be further expanded perturbatively in Seeley-DeWitt coefficients \cite{Seeley:1966tt,Seeley:1969uu,DeWitt:1965ff,DeWitt:1967gg,DeWitt:1967hh,DeWitt:1967ii}, which we can compute in terms of background field and geometric invariants following a standard approach \cite{Vassilevich:2003ll}. Most of the other recognized methods\footnote{The current paper only discusses the heat kernel methods that are more relevant and successful in the logarithmic entropy correction calculations. However, in general, the heat kernel can be evaluated by various other available perturbative expansion methods, quasinormal-mode approach \cite{Denef:2010nc}, group-theoretic method \cite{David:2010xn,Gopakumar:2011xi,Lovrekovic:2016ni}, etc.} that compute the heat kernel for black hole entropy corrections are either restricted to fixed backgrounds (constant fields, presence of rotational symmetry, homogeneous space-time, etc.) or involve some complex computations. The eigenfunction expansion method employed in \cite{Banerjee:2011oo,Banerjee:2011pp,Sen:2012qq,Gupta:2014ns} earns excellent success in this direction. However, that is constrained to only the Bertotti-Robinson extremal near-horizon geometry $AdS_2\times S^2$. This eigenfunction expansion method is further customized in from of the on-shell \cite{Keeler:2014nn} and off-shell \cite{Larsen:2015nx} methods. The hybrid method advocated in \cite{Charles:2015nn,Castro:2018tg} is analogous to the standard Seeley-DeWitt computation method \cite{Vassilevich:2003ll} but modified with field redefinition. One needs specialized skills to redefine fields in different gravity theories where it is also impossible to keep track of individual field contributions. This, in turn, destroys the generality within the hybrid method \cite{Charles:2015nn,Castro:2018tg}.

Supergravity theories are particular class of vacuums that can be realized as the low-energy truncation of string theories compactified down to four space-time dimensions. Logarithmic entropy corrections have been extensively investigated in supergravity theories, which already have a well-established microscopic counterpart within string theory \cite{Mandal:2010im,Sen:2014nvt}. But, most of the popularly known examples \cite{Banerjee:2011oo,Banerjee:2011pp,Sen:2012qq,Gupta:2014ns,Larsen:2015nx,Keeler:2014nn,Ferrara:2012bp} of logarithmic entropy corrections in supergravities are for extremal black holes with non-rotating geometry. So far, not many macroscopic results are available for the extremal rotating black holes in $\mathcal{N} \geq 1$ supergravities. This is because the technical approaches used to analyze the one-loop contribution to partition function are mostly best suited for extremal non-rotating backgrounds with the near-horizon geometry $AdS_2\times S^2$. However, there are some reports \cite{Sen:2012rr,Bhattacharyya:2012ss} of logarithmic corrections for extremal rotating black holes in four-dimensional non-supersymmetric Einstein-Maxwell theories. Especially in \cite{Bhattacharyya:2012ss}, the authors have utilized the standard Seeley-DeWitt computation technique \cite{Vassilevich:2003ll} into the framework of quantum entropy function formalism that surprisingly provides a global platform to manage the logarithmic correction results for extremal Kerr-Newman, Kerr and Reissner-Nordstr\"om black holes. Motivated by this success, the current paper aims to extend the logarithmic correction program for all types of extremal black holes (rotating or non-rotating, charged or uncharged) in fully matter-coupled $\mathcal{N} \geq 1$ supergravity theories. We also receive a strong hint from the fact that both the quantum entropy function formalism \cite{Sen:2008wa,Sen:2009wb,Sen:2009wc} and the calculation technique \cite{Vassilevich:2003ll} do not at all rely on supersymmetry.  %We have already gained some initial success in this direction by testing the global framework for extremal black holes in `minimal' (or pure) $\mathcal{N}=2$ \cite{Karan:2019sk} and `non-minimal' $\mathcal{N}=1$ \cite{Banerjee:2020wbr} supergravity theories.
As a way forward, we have already gained some initial success by testing the global framework for extremal black holes in `minimal' (or pure) $\mathcal{N}=2$ \cite{Karan:2019sk} and `non-minimal' $\mathcal{N}=1$ \cite{Banerjee:2020wbr} supergravity theories. But, both of them are not fully matter-coupled, rather special cases.\footnote{The `minimal' $\mathcal{N}=2$ contains only supergravity multiplet, while in the `non-minimal' $\mathcal{N}=1$ a vector multiplet is non-minimally coupled to the supergravity multiplet.} Note that similar logarithmic entropy corrections for non-extremal black holes are also reported in \cite{Charles:2015nn,Castro:2018tg} via the conventional Euclidean gravity approach \cite{Sen:2013ns}.

%detailed explanation is in section 3
The $\mathcal{N}=2,d=4$ Einstein-Maxwell supergravity theory (EMSGT) is significant and arises as an IR limit of type-II string theory compactified on a Calabi–Yau three-fold (e.g., see \cite{Grana:2006mg,Freedman:2012xp}). In such a theory, the backgrounds of a four-dimensional Einstein-Maxwell theory are embedded into a $\mathcal{N}=2$ supergravity so that one can interpret the Kerr-Newman family of black holes\footnote{The Kerr-Newman black holes are general solutions to the Einstein-Maxwell theory of gravity \cite{Adamo:2014lk}.} as solutions of the resultant $\mathcal{N}=2,d=4$ EMSGT \cite{Charles:2015nn,Karan:2019sk,Sen:2012qq}. A general $\mathcal{N}=2,d=4$ EMSGT contains four massless multiplets (supergravity, gravitino, vector and hyper), and the logarithmic correction contributions of these multiplets to the entropy of extremal black holes can be extracted by analyzing quadratic field fluctuations. As discussed, we particularly interested to do this by calculating the Seeley-DeWitt coefficients. The corresponding calculations are unwieldy but manageable if one proceeds systematically. Any motive to extend the logarithmic correction program for the higher-extended EMSGTs will gradually increase the technical complexities, requiring tremendous skills in managing the Seeley-DeWitt trace calculations.  
%The technical complexities will gradually increase for higher-extended EMSGTs, and any motive to extend the logarithmic correction program for them demands tremendous skills to manage the Seeley-DeWitt calculations. 
But, the contents of supergravities are such that they are always kinematically reducible to lower supergravities. Based on this feature, the $\mathcal{N}=2,d=4$ EMSGT results are found to be crucial for carrying out the logarithmic correction program of the extremal Kerr-Newman family of black holes in all the matter coupled $\mathcal{N} \geq 1,d=4$ EMSGTs. This is the central objective of this paper.

The prime technical results of this paper are recorded in two parts. In the first part (see \cref{mul}), we calculate the Seeley-DeWitt coefficients of the supergravity, gravitino, vector, and hyper multiplet of the $\mathcal{N}=2,d=4$ EMSGT by analyzing quadratic fluctuations of their content. The supergravity multiplet results are borrowed from our previous work \cite{Karan:2019sk}, while other multiplet results are freshly calculated by employing the standard manual \cite{Vassilevich:2003ll}. The calculated Seeley-DeWitt coefficients agree with the results of both the eigenfunction expansion method \cite{Sen:2012qq, Keeler:2014nn} and the hybrid field-redefinition approach \cite{Charles:2015nn}. We then make use of the Seeley-DeWitt data in the quantum entropy function framework and derive the logarithmic correction contributions of individual multiplets to the entropy of extremal Kerr-Newman, Kerr and Reissner-Nordstr\"om black holes (see \cref{S5,G16,V12,H12}). In the second part (refer \cref{diss}), we utilize the $\mathcal{N}=2$ multiplet data to obtain logarithmic correction results for the extremal Kerr-Newman family of black holes in three different cases --
\begin{inparaenum}[(i)]
	\item a matter coupled $\mathcal{N}=2,d=4$ EMSGT by coupling the supergravity multiplet with any arbitrary number of vector and hyper multiplets, 
	\item a particular class of matter coupled $\mathcal{N}=1,d=4$ EMSGT obtained as consistent decomposition of $\mathcal{N}=2$ multiplets, and 
	\item the matter coupled $\mathcal{N} \geq 3,d=4$ EMSGTs by decomposing them into $\mathcal{N}=2$ multiplets.
\end{inparaenum}
The results for all three cases are recorded in \cref{mc2,t6,d3}, respectively. The entropy corrections for Reissner-Nordstr\"om black holes match with the results available in \cite{Sen:2012qq,Banerjee:2011pp,Keeler:2014nn,Ferrara:2012bp}, while the Kerr-Newman and Kerr corrections are new reports to the literature. We derive a combined and compact logarithmic correction formula \eqref{d4} for the extremal Kerr-Newman black holes in the matter coupled $\mathcal{N} \geq 2,d=4$ EMSGTs. In comparison, we also extract the logarithmic entropy corrections for non-extremal Kerr-Newman family of black holes in the matter coupled $\mathcal{N} \geq 1,d=4$ EMSGTs using the same $\mathcal{N}= 2$ Seeley-DeWitt data in the Euclidean gravity approach \cite{Sen:2013ns}. The non-extremal results are recorded in \cref{ne9,ne10,ne11}, where the $\mathcal{N} \geq 2$ part is a reproduction of \cite{Charles:2015nn} and the $\mathcal{N} = 1$ part is new results.

The rest of this paper is planned in the following way. \Cref{B} reviews a general framework of evaluating logarithmic correction to the entropy of extremal Kerr-Newman family of black holes using Seeley-DeWitt coefficients in the quantum entropy function formalism. In \cref{mul}, we analyze quadratic fluctuations in the multiplets of $\mathcal{N}=2,d=4$ EMSGT and find their logarithmic correction contributions to the entropy of extremal black holes by calculating corresponding Seeley-DeWitt coefficients. In \cref{diss}, we obtain the logarithmic entropy correction results for the extremal Kerr-Newman family of black holes in the matter coupled $\mathcal{N} \geq 1,d=4$ EMSGTs. We also revisit the same for the non-extremal Kerr-Newman family of black holes via the Euclidean gravity approach \cite{Sen:2013ns}. We end up discussing some highlights and concluding remarks about the work presented in this paper. \Cref{AA} records the equations of motion and identities used in this paper. In \cref{BB}, we present a brief flow chart of comparatively lengthy trace calculations in the non-minimally coupled sectors of $\mathcal{N}=2$ gravitino, vector and hyper multiplets.

%====================================================================================

\section{The general framework}\label{B}
In this section, we start by briefly reviewing the basic algorithm for 
computing logarithmic correction to the entropy of extremal black holes following the quantum entropy function formalism \cite{Sen:2008wa,Sen:2009wb,Sen:2009wc}.
% provide a general structure of computing logarithmic correction to the entropy of extremal black holes following the quantum entropy function formalism \cite{Sen:2008wa,Sen:2009wb,Sen:2009wc}. After briefly reviewing the basic algorithm of the quantum entropy function formalism,
We then analyze the relation between logarithmic entropy correction and the one-loop effective action for massless fields propagating in the near-horizon geometry of extremal black holes. We also shed light on the evaluation of the one-loop effective action from the knowledge of Seeley-DeWitt coefficients for these massless fluctuations. We outline a standard and efficient approach for calculating the Seeley-DeWitt coefficients following the manual \cite{Vassilevich:2003ll}, where the coefficients are demonstrated in terms of different background fields and geometric invariants. Finally, we set up for the near-horizon extremal limits of the Kerr-Newman, Kerr and Reissner-Nordstr\"om black holes after clearly highlighting the key working formula of the logarithmic entropy corrections.

%the one-loop effective can be obtained by evaluating the Seeley-DeWitt coefficients for these massless fluctuations. 

\subsection{Heat kernel  treatment of one-loop effective action and Seeley-DeWitt coefficients}\label{hk}
We consider a 4D theory of gravity characterized by the Euclideanized action $\mathcal{S}$ (with Lagrangian density $\mathscr{L}$),
\begin{equation}\label{hk1}
	\mathcal{S}[g,\xi] = \int \mathrm{d}^4x\sqrt{\text{det}\thinspace g} \thinspace\mathscr{L},
\end{equation}
where $\lbrace\xi\rbrace$ is the matter field configuration propagating through a geometry represented by the metric $g$. 
One can choose ($\bar{g},\bar{\xi}$) as an arbitrary solution to the classical equations of motion of the theory and fluctuate the metric $g$ and fields $\lbrace\xi\rbrace$ around the classical background ($\bar{g},\bar{\xi}$) for small quantum fluctuations $\tilde{\xi}_m = \lbrace \tilde{{g}}, \tilde{\xi} \rbrace $,
\begin{equation}\label{hk2}
	{g} = \bar{g}+ \tilde{{g}},\enspace \xi = \bar\xi + \tilde{\xi}.
\end{equation}
If $\Lambda$ is the differential operator that governs the quadratic fluctuation of the action \eqref{hk1}, then we can express the quadratic fluctuated action form $\delta^2\mathcal{S}$ as
\begin{equation}\label{hk3}
	\delta^2\mathcal{S}[\tilde{{g}},\tilde{\xi}] = \int \mathrm{d}^4x \sqrt{\text{det}\thinspace \bar{g}}\thinspace \tilde{\xi}_m\Lambda \tilde{\xi}_n,
\end{equation}
which necessarily evaluates the quantum corrected Euclidean one-loop effective action $\mathcal{W}$ as \cite{Gibbons:1977ta,Hawking:1978td,Hawking:1977te,Avramidi:1994th,Denardo:1982tb,Peixoto:2001wx}
\begin{equation}\label{hk4}
	e^{-\mathcal{W}}=\int \mathcal{D}[\tilde{g},\tilde{\xi}]\exp(-\delta^2\mathcal{S}[\tilde{{g}},\tilde{\xi}])= (\text{det}\thinspace \Lambda)^{-\chi/2},
\end{equation}
where $\chi = +1$ for boson fluctuations and $\chi = -1$ for fermion fluctuations. $\mathcal{D}[\tilde{g},\tilde{\xi}]$ suggests that the integration \eqref{hk4} is acting over all possible fluctuations of the metric and matter fields. Let us now assume that $\lbrace f_i\rbrace$ and $\lbrace \lambda_i \rbrace$ are respectively the complete basis of eigenfunctions and eigenvalues of the operator $\Lambda$ satisfying the eigenvalue equation $\Lambda f_i(x) = \lambda_i f_i(x)$ with the normalization conditions,
\begin{equation}\label{hk5}
	\begin{gathered}
		\sum_i f_i(x)f_i(y) = \delta^4(x,y),\\
		\int \mathrm{d}^4x \sqrt{\text{det}\thinspace \bar{g}}\thinspace  f_i(x)f_j(x) = \delta_{ij},
	\end{gathered}
\end{equation}
where $\delta^4(x,y)$ and $\delta_{ij}$ are appropriate delta functions. The {heat kernel} $K(x,y;s)$ of the operator $\Lambda$ and its trace $D(s)$ (dubbed as the {heat trace}) are respectively defined as \cite{Vassilevich:2003ll}
\begin{equation}\label{hk6}
	\begin{gathered}
		K(x,y;s) = \sum_i e^{-\lambda_is}f_i(x)f_i(y),\\
		D(s) = \int \mathrm{d}^4x \sqrt{\text{det}\thinspace \bar{g}}\thinspace K(x,x;s)= \sum_i e^{-\lambda_i s},
	\end{gathered}
\end{equation}
where $s$ is an auxiliary parameter (dimension of length square), called the heat kernel parameter or \say{proper time}.  
At this point, the Schwinger-DeWitt proper time representation \cite{Schwinger:1951sp,DeWitt:1975ps} allows us to express the one-loop effective action $\mathcal{W}$ and its Lagrangian density $\Delta\mathscr{L}_{\text{eff}}$ in terms of the heat trace and heat kernel respectively as
\begin{equation}\label{hk7}
	\begin{gathered}
		\mathcal{W} = -\frac{1}{2}\int_\epsilon^\infty \frac{\mathrm{d}s}{s}\chi D(s),\\
		\Delta\mathscr{L}_{\text{eff}}= -\frac{1}{2}\int_\epsilon^\infty \frac{\mathrm{d}s}{s}\chi K(x,x;s).
	\end{gathered}
\end{equation}
where $\epsilon$ is a cut-off ($\epsilon\sim {l_p}^2\sim G_N$) introduced to regulate the UV divergence of $\mathcal{W}$.
This is the central idea about the heat kernel method, where one can evaluate the one-loop effective action via heat kernel (precisely the trace of heat kernel) of the kinetic operator $\Lambda$ without worrying about its eigenfunctions. Furthermore, the heat kernel $K(x,y;s)$ is associated with a power-law asymptotic expansion for small $s$, called the Seeley-DeWitt expansion,
\begin{equation}\label{hk10}
	D(s)\cong \int \mathrm{d}^4x \sqrt{\text{det}\thinspace \bar{g}}\thinspace \sum_{n=0}^\infty s^{n-2}a_{2n}(x),
\end{equation}
where $a_{2n}(x)$ are known as \textit{Seeley-DeWitt coefficients}\cite{Seeley:1966tt,Seeley:1969uu,DeWitt:1965ff,DeWitt:1967gg,DeWitt:1967hh,DeWitt:1967ii}. A standard computation approach of these coefficients is outlined in \cref{P}.

It is also essential to define the zero modes of the operator $\Lambda$ in the present heat kernel treatment. Among the eigenfunctions $f_i(x)$ of $\Lambda$, there may be a particular group of modes for which the eigenvalues $\lambda_i$ vanish, i.e., $\lambda_i=0$. These modes are recognized as zero modes of the operator $\Lambda$ and denoted by the special notation $f^0_{i}(x)$ with the eigenvalue equation,
\begin{equation}\label{hk8}
	\Lambda f^0_{i}(x) = 0.
\end{equation}
For all possible fluctuations of the operator $\Lambda$, the total number of zero modes $n_{\text{zm}}$ can be defined as
\begin{equation}\label{hk9}
	\begin{gathered}
		D(s) = \sum_i e^{-s\lambda_i} = \sideset{}{'}\sum_{i\atop (\lambda_i \neq 0)} e^{-s\lambda_i} + n_{\text{zm}},\\
		n_{\text{zm}} = \sum_i\int \mathrm{d}^4x \sqrt{\text{det}\thinspace \bar{g}}\thinspace f^0_{i}(x)f^0_{i}(x),
	\end{gathered}
\end{equation}
where $\sideset{}{'}\sum$ represents summation over the non-zero modes only. 

%------------------------------------------------------------------------------------
\subsection{A standard computation strategy for the Seeley-DeWitt coefficients}\label{P}
In this paper, our motive is to express the Seeley-DeWitt coefficients in terms of different invariants induced from the background metric, fields and their covariant derivatives present in a particular 4D gravity theory. To fulfill this demand, we cast a standard approach following the manual \cite{Vassilevich:2003ll}.\footnote{For a detailed review, please refer to \cite{Karan:2019sk}.} The present approach assumes that the quadratic order fluctuated action \eqref{hk3} must be customized up to a total derivative for providing the kinetic differential operator $\Lambda$ a minimal and Laplace-type form,
\begin{align}\label{p1}
	\tilde{\xi}_m \Lambda^{mn}\tilde{\xi}_n = \pm\tilde{\xi}_m\big(D^\rho D_\rho I^{mn} +(N^\rho D_\rho)^{mn}+P^{mn}\big)\tilde{\xi}_n,
\end{align}
where $m$ is a mixed index that includes both the tensor indices and the labeling for any arbitrary fluctuation $\xi_m$; $D_\rho$ is the ordinary covariant derivative with Christoffel and spin connections; $I$ is the identity operator in fields space; $N^\rho, P$ are arbitrary matrices constructed from background fields.
In any generic theory of gravity, the fields are interacting and hence we need to redefine the covariant derivative $D_\rho$ by embedding a gauge connection $\omega_\rho$,
\begin{equation}\label{p2}
	\mathcal{D}_\rho \tilde{\xi}_m = D_\rho \tilde{\xi}_m + {(\omega_\rho)_m}^n \tilde{\xi}_n \quad\forall m\neq n.
\end{equation}
Naturally, the connection $\omega_\rho$ vanishes for free or minimally-coupled\footnote{A free or minimal-coupled fluctuation is only coupled to background gravity via the $\sqrt{\text{det}\thinspace \bar{g}}$ term in the fluctuated action form \eqref{hk3}.} fluctuations. The curvature corresponding to the new covariant derivative $\mathcal{D}_\rho$ is
\begin{equation}\label{p3}
	(\Omega_{\rho\sigma})^{mn}\equiv [\mathcal{D}_\rho,\mathcal{D}_\sigma]^{mn} = [D_\rho,D_\sigma]^{mn}+{D_{[\rho}\omega_{\sigma]}} ^{mn}+[\omega_\rho,\omega_\sigma]^{mn}.
\end{equation}
The schematic \eqref{p1} therefore reduces to a more generalized and standard form,
\begin{equation}\label{p4}
	\tilde{\xi}_m \Lambda^{mn}\tilde{\xi}_n = \pm\tilde{\xi}_m\big(\mathcal{D}^\rho\mathcal{D}_\rho I^{mn} +E^{mn}\big)\tilde{\xi}_n,
\end{equation}
where 
\begin{equation}\label{p6}
	(\omega_\rho)^{mn} = \frac{1}{2}(N_\rho)^{mn},\thinspace E^{mn} = P^{mn}-(D^\rho\omega_\rho)^{mn}-(\omega^\rho)^{mp}{(\omega_\rho)_p}^n.
\end{equation}
With this structure, the first three Seeley-DeWitt coefficients for the quadratic fluctuations of any arbitrary 4D theory are obtained by computing $\text{tr}(I)$, $\text{tr}(E)$, $\text{tr}(E^2)$ and $\text{tr}(\Omega_{\rho\sigma}\Omega^{\rho\sigma})$ in the following formulae \cite{Vassilevich:2003ll}
\begin{equation}\label{p7}
	\begin{split}
		(4\pi)^2a_0(x) &= \chi\,\mathrm{tr}(I),\\
		(4\pi)^2a_2(x) &=\frac{\chi}{6}\, \mathrm{tr}(RI+6E),\\
		(4\pi)^2a_4(x) &= \frac{\chi}{360} \mathrm{tr}\Big( 60RE+ 180E^2 + 30\Omega_{\rho\sigma}\Omega^{\rho\sigma}\\
		&\quad +(5 R^2+ 2R_{\mu\nu\rho\sigma}R^{\mu\nu\rho\sigma}-2R_{\mu\nu}R^{\mu\nu})I\Big).
	\end{split}
\end{equation}
where $R$, $R_{\mu\nu}$ and $R_{\mu\nu\rho\sigma}$ are respectively the Ricci scalar, Ricci tensor and Riemann tensor corresponding to the background metric $\bar{g}_{\mu\nu}$.
It is essential to mention that the quadratic fluctuated actions $\eqref{hk3}$ of fermionic fluctuations are generally controlled by first-order operators (e.g., see \cite{Karan:2018ac}). But the special technique of \cite{Sen:2012qq,Peixoto:2001wx} allows us to revise the one-loop determinant of the first-order fermionic operators, denoted by $\slashed{D}$, via the following way
\begin{equation}\label{p8}
	\text{ln det}\thinspace \slashed{D}=\text{ln det}\thinspace \slashed{D}^\dagger = \frac{1}{2}\thinspace \text{ln det}\thinspace\slashed{D}^\dagger\slashed{D}, 
\end{equation}
where $\slashed{D}^\dagger\slashed{D}$ can be structured further to achieve the typical Laplace-type form \eqref{p1}. The treatment \eqref{p8} will double up the integration variables for a complex Dirac spinor operator, which in return  compensates the additional 1/2 factor. Consequently, we can state that the heat kernel treatment primarily considers any fermion as Dirac.  But for a real Majorana spinor operator, one needs to include the extra 1/2 factor in the Seeley-DeWitt formulae \eqref{p7}. Weyl spinors can not fit into this prescription due to the presence of both left and right chiral states, providing $\text{det}(\slashed{D}^\dagger\slashed{D}) \neq (\text{det}\thinspace \slashed{D})^2 \thinspace \text{or}\thinspace (\text{det}\thinspace \slashed{D}^\dagger)^2$. %Reference IR window, page 12, footnote
Therefore, the exclusive values of $\chi$ in the formulae \eqref{p7} for different fields are
\begin{equation}\label{stff3}
	\chi =
	\begin{cases}
		-1 & \text{for Dirac spinors},\\
		-\frac{1}{2} & \text{for Majorana spinors},\\
		+1 & \text{for bosons and scalars}.
	\end{cases}
\end{equation}
Again, the ghost fields that appear in gauge-fixing of the theory follow the reverse of above.

In this paper, we will mainly deal with Einstein-Maxwell supergravity theories that characterize a $R=0$ manifold (see \cref{S2}). For scalar ($\phi$), vector ($a_\mu$), spin-1/2 ($\lambda$) and spin-3/2 ($\psi_{\mu}$) field fluctuations, the covariant derivative commutation relations have the following standard forms
\begin{subequations}
	\begin{align}
		[D_\rho,D_\sigma]\phi &= 0, \label{scalar}\\
		[D_\rho,D_\sigma]a_\mu &= -{R^\nu}_{\mu\rho\sigma}a_\nu,\label{vector}\\
		[D_\rho,D_\sigma]\lambda &= \frac{1}{8}[\gamma^\alpha,\gamma^\beta]R_{\rho\sigma\alpha\beta}\lambda,\label{spinor}\\
		[D_\rho,D_\sigma]\psi_\mu &=-{R^\nu}_{\mu\rho\sigma}\psi_\nu +\frac{1}{8}[\gamma^\alpha,\gamma^\beta]R_{\rho\sigma\alpha\beta}\psi_\mu.\label{RS}
	\end{align}
\end{subequations}
With all these data and following the standard manual provided in this section, the Seeley-DeWitt contributions for the free or minimally-coupled field fluctuations are relatively less complicated and straightforward compute.\footnote{Readers are referred to \cite{Karan:2018ac} for revising the free spin-1/2 and spin-3/2 calculations.} The results for the first three Seeley-DeWitt coefficients can be gathered and combined into a single frame as
\begin{equation}\label{cf}
	\begin{split}
		(4\pi)^2 {a_0}^{\text{min-coupled}}(x) &= (n_0 + 2n_1 -2n_{1/2}-2n_{3/2} ),\\
		(4\pi)^2 {a_2}^{\text{min-coupled}}(x) &= 0,\\
		(4\pi)^2 {a_4}^{\text{min-coupled}}(x) &= \frac{1}{360}\Big\lbrace  \big(2n_0 -26n_1 +\frac{7}{2}n_{1/2}-\frac{233}{2}n_{3/2} \big)R_{\mu\nu\rho\sigma}R^{\mu\nu\rho\sigma}\\
		&\qquad+\big(-2n_0+176n_1+4n_{1/2}+4n_{3/2} \big)R_{\mu\nu}R^{\mu\nu}\Big\rbrace,
	\end{split}
\end{equation}
where $n_0$, $n_1$, $n_{1/2}$ and $n_{3/2}$ are respectively the multiplicities of the real massless free scalar, vector, spin-1/2 Majorana and spin-3/2 Majorana field fluctuations. The results \eqref{cf} play crucial side roles in the calculation of \cref{mul}. 

%------------------------------------------------------------------------------------
\subsection{Seeley-DeWitt coefficient ``recipe'' for logarithmic correction to the entropy of extremal Kerr-Newman black holes}\label{lc}
For extremal black holes, the prescription of quantum entropy function formalism \cite{Sen:2008wa,Sen:2009wb,Sen:2009wc} is a comfortable and efficient method to determine the quantum corrections to the black hole entropy. With the help of this formalism, our aim is to provide a clear structure for computing logarithmic correction to the entropy of extremal Kerr-Newman family of black holes. The Seeley-DeWitt coefficients play a vital role in these logarithmic correction computations. 
%------------------------------------------------------------------------------------------
\subsubsection{General set-up via the quantum entropy function formalism}
%We find it useful to briefly outline the basic algorithm of the quantum entropy function formalism --
The basic algorithm of calculating logarithmic entropy corrections for extremal black holes via the quantum entropy function formalism is presented as follows --
\begin{enumerate}
	\item The entropy of extremal black holes can be computed from their near-horizon analysis and we do not need to know the details of full space-time. The near-horizon geometry of every extremal black hole can be reshaped into the form $AdS_2 \times \mathcal{K}$,
	\begin{equation}\label{near}
		{\mathrm{d}s^2}_{\text{horizon}} = \bar{g}_{\mu\nu}\mathrm{d}x^\mu \mathrm{d}x^\nu = \mathrm{d}s^2_{AdS_2}+ \mathrm{d}s^2_{\mathcal{K}},
	\end{equation}
	where $\mathcal{K}$ is a space described by the angular coordinates or compact coordinates and $\bar{g}_{\mu\nu}$ is the metric describing the extremal near-horizon geometry with coordinates $x^\mu$.
	
	\item The Euclidean string partition function $\mathcal{Z}_{AdS_2 \times \mathcal{K}}$ describing the extremal near-horizon geometry is defined as
	\begin{equation}\label{lc16}
		\mathcal{Z}_{AdS_2 \times \mathcal{K}}= \int \mathcal{D}[\bar{g},\xi]\exp\left(-\mathcal{S}[\bar{g},\xi] \right),
	\end{equation}	
	where $\xi$ collectively denotes all the string fields and $\mathcal{S}$ is the action with Euclideanized time ($t \to -i\theta$). The string fields $\xi$ asymptotically approach to the classical near-horizon geometry \eqref{near}, which serves as a boundary condition for this formalism. The boundary condition modifies the partition function \eqref{lc16} into a standard quantum entropy function form by accommodating a Gibbons-Hawking type boundary term $\text{exp}\left[-iq_k \oint  d\theta \mathcal{A}_\theta^{(k)}\right]$, where $\mathcal{A}_\theta^{(k)}$ are gauge fields on $AdS_2$ and $q_k$ are the corresponding charges \cite{Sen:2008wa,Sen:2009wb,Sen:2009wc,Sen:2012qq,Sen:2012rr}.	 
	\item $AdS_2/CFT_1$ conjecture suggests that the string partition function in $AdS_2$ is equal to the partition function of the boundary $CFT_1$, i.e., $\mathcal{Z}_{AdS_2 \times \mathcal{K}}= \mathcal{Z}_{CFT_1}$. The dual $CFT_1$ lives at the one-dimensional boundary of $AdS_2$ and can be treated as the IR limit of the quantum mechanics associated with the black hole microstates. Therefore, the entropy of the extremal black hole is equal to the entropy calculated from the full near-horizon partition function $\mathcal{Z}_{AdS_2 \times \mathcal{K}}$.
	
	\item $\mathcal{Z}_{AdS_2 \times \mathcal{K}}$ is divergent due to the infinite volume of $AdS_2$. But this divergence can be regulated by inserting an IR cut-off, which limits the boundary of $AdS_2$ by a finite length $\ell$. On the $CFT_1$ side (i.e., microscopic side) if we take $\ell\to \infty$ limit, only the ground state will contribute:
	\begin{equation}
		\lim_{\ell\to \infty}\mathcal{Z}_{CFT_1} = d_{\text{hor}} e^{-\ell E_0},
	\end{equation} 
	where $d_{\text{hor}}$ and $E_0$ are respectively the degeneracy and energy of the states of $CFT_1$. On the $AdS_2$ side (i.e., macroscopic or gravity side), the trick is to choose appropriate coordinates so that the partition function $\mathcal{Z}_{AdS_2 \times \mathcal{K}}$ can be structured in the form of $\mathcal{Z}_{CFT_1}$: 
	\begin{equation}
		\lim_{\ell\to \infty} \mathcal{Z}_{AdS_2 \times \mathcal{K}} =\mathcal{Z}_{AdS_2 \times \mathcal{K}}^{\text{finite}} \cdot e^{-{C}\ell+\mathcal{O}(\ell^{-1})} ,
	\end{equation}
	where ${C}$ is some arbitrary constant and $\mathcal{Z}_{AdS_2 \times \mathcal{K}}^{\text{finite}}$ is the cutoff-insensitive (i.e., $\ell$ independent) finite part. Then using $AdS_2/CFT_1$ conjecture one identifies $\mathcal{Z}_{AdS_2 \times \mathcal{K}}^{\text{finite}}$ with $d_{\text{hor}}$, i.e., $\mathcal{Z}_{AdS_2 \times \mathcal{K}}^{\text{finite}}$ can be recognized as the macroscopic horizon degeneracy of the extremal black hole. This kind of prescription of horizon degeneracy is termed as the \textit{quantum entropy function} \cite{Sen:2008wa,Sen:2009wb,Sen:2009wc}. $\mathcal{Z}_{AdS_2 \times \mathcal{K}}^{\text{finite}}$ reduces to the Wald entropy equality in the classical limit, which checks the fact that the quantum corrected entropy ($S_{\text{BH}}$) of the extremal black hole can be identified as
	\begin{equation}
		S_{\text{BH}} = \ln \mathcal{Z}_{AdS_2 \times \mathcal{K}}^{\text{finite}}.
	\end{equation} 
	
	\item  Computation of the full path integral of any gravity theory is almost an impossible task.
	But one can carry out saddle point expansion \cite{Banerjee:2011oo,Banerjee:2011pp,Sen:2012rr,Sen:2012qq,Chowdhury:2014np,Gupta:2014ns,Bhattacharyya:2012ss,Karan:2019sk,Jeon:2017ij,Keeler:2014nn,Larsen:2015nx,Banerjee:2020wbr} of the string partition function \eqref{lc16} for small fluctuations around the classical near-horizon background \eqref{near}. At one loop, the correction to $\mathcal{Z}_{AdS_2 \times \mathcal{K}}$ can be obtained as 
	\begin{equation}\label{lc17}
		\mathcal{Z}_{AdS_2 \times \mathcal{K}}^{\text{1-loop}}=e^{-\mathcal{W}},
	\end{equation}
	where the one-loop effective action $\mathcal{W}$ is expressed via the kinetic operator for quadratic fluctuations of string fields around the classical near-horizon background.
	
	\item One-loop corrections generally have contributions from all the massive stringy states running into the loop, but the logarithmic corrections are typical one-loop quantum corrections that only depend on the massless states. Hence, the logarithmic corrections to the extremal black holes, denoted as $\Delta S_{\text{BH}}$, can be computed by using the massless fluctuations in the finite part of the one-loop effective action $\mathcal{W}$ \cite{Banerjee:2011oo,Banerjee:2011pp,Sen:2012rr,Sen:2012qq,Chowdhury:2014np,Gupta:2014ns,Bhattacharyya:2012ss,Karan:2019sk,Jeon:2017ij,Keeler:2014nn,Larsen:2015nx,Banerjee:2020wbr}. 
	
	%But the logarithmic corrections to the extremal black holes, denoted as $\Delta S_{\text{BH}}$, are typical one-loop quantum corrections that could be computed by using just the massless fluctuations in the finite part of the one-loop effective action $\mathcal{W}$ \cite{Banerjee:2011oo,Banerjee:2011pp,Sen:2012rr,Sen:2012qq,Chowdhury:2014np,Gupta:2014ns,Bhattacharyya:2012ss,Karan:2019sk,Jeon:2017ij}.	
\end{enumerate}
The heat kernel treatment as discussed in \cref{hk} is an effective path to evaluate the one-loop effective action $\mathcal{W}$. However, the zero mode contribution to $\mathcal{W}$ can not be assessed correctly by the heat kernel treatment because $\mathcal{W}$ cannot preserve its Gaussian structure \eqref{hk4} for zero modes. As a remedy, one can remove zero modes from the heat kernel and substitute them back via a volume factor of different asymptotic symmetry groups responsible for these zero modes \cite{Banerjee:2011oo,Banerjee:2011pp,Sen:2012rr,Sen:2012qq,Gupta:2014ns,Bhattacharyya:2012ss,Chowdhury:2014np,Jeon:2017ij,Sen:2013ns}. This process splits the Euclidean one-loop partition function \eqref{hk4} into two parts,
\begin{equation}\label{lc2}
	e^{-\mathcal{W}} = (\text{det}^\prime\thinspace\Lambda)^{-\chi/2}\cdot \mathcal{Z}_{\mathrm{zero}}(L),
\end{equation}
where the primed part describes determinant over non-zero modes of the operator $\Lambda$ and $\mathcal{Z}_{\mathrm{zero}}(L)$ is the zero-mode integral with an overall length scale $L$. In this setup, $L$ is defined in a way that the non-zero eigenvalues of the Laplace-type operator $\Lambda$ scale as $1/L^{2}$ and consequently, the heat kernel parameter is reset as $\bar{s}=s/L^2$. The non-zero mode part of the one-loop effective action is then extracted as
\begin{equation}\label{lc3}
	\begin{split}
		\mathcal{W}^\prime =\frac{\chi}{2}\thinspace \text{ln}\thinspace \text{det}^\prime\Lambda &= -\frac{\chi}{2}\int_{\epsilon/L^2}^\infty \frac{\mathrm{d}\bar{s}}{\bar{s}}\Big(D(\bar{s})- n_{\text{zm}}\Big)\\
		&= -\frac{1}{2}\bigg(\int \mathrm{d}^4x \sqrt{\text{det}\thinspace \bar{g}}\thinspace a_4(x) - \chi n_{\text{zm}} \bigg)\ln \left(\frac{L^2}{G_N}\right) +\cdots,
	\end{split}
\end{equation}
where the logarithmic term obtained after integrating out only $\bar{s}$ independent part in the perturbative expansion \eqref{hk10} of $D(\bar{s})$ and \say{$\cdots$} indicates the non-logarithmic terms. On the other hand, the zero-mode part $\mathcal{Z}_{\mathrm{zero}}(L)$ can be analyzed by evaluating the Jacobian for the change of integration variables from fields to the parameters labeling zero-mode asymptotic symmetries \cite{Banerjee:2011oo,Banerjee:2011pp,Sen:2012rr,Sen:2012qq,Gupta:2014ns,Bhattacharyya:2012ss,Chowdhury:2014np,Jeon:2017ij,Sen:2013ns}. This analysis assigns a factor of $L^{\beta_{\tilde{\xi}}}$ corresponding to each zero mode, providing
\begin{equation}\label{lc4}
	\mathcal{Z}_{\mathrm{zero}}(L) = L^{ \sum_{\tilde{\xi}}\chi\beta_{\tilde{\xi}} n^0_{\tilde{\xi}}}\cdot\mathcal{Z}_0,
\end{equation}
where $\beta_{\tilde{\xi}}$ are constant numbers computed individually for different fluctuations; $n^0_{\tilde{\xi}}$ are the number of zero modes of any arbitrary fluctuation $\tilde{\xi}$, giving the total zero-mode $n_{\text{zm}}=\sum_{\tilde{\xi}}n^0_{\tilde{\xi}}$;  $\mathcal{Z}_0$ is independent of  $L$. By putting the contributions \eqref{lc3} and \eqref{lc4} together, we present the revised form of one-loop effective action as % Special Refs: 1402.2441, 1707.04208, 1210.6057
\begin{equation}\label{lc5}
	\mathcal{W} = -\frac{1}{2}\bigg( \int \mathrm{d}^4x \sqrt{\text{det}\thinspace \bar{g}}\thinspace a_4(x)+ \sum_{\tilde{\xi}}\chi(\beta_{\tilde{\xi}}- 1)n^0_{\tilde{\xi}}\bigg) \ln \left(\frac{L^2}{G_N}\right) + \cdots,
\end{equation}
which yields the following central formula for finding the logarithmic correction to the entropy of large\footnote{Logarithmic corrections are quantum corrections in the large-charge limit. In this limit, the horizon area of the black holes are scaled as $\mathcal{A}_{H} \sim L^2$, where $L$ is large \cite{Bhattacharyya:2012ss}.} extremal black holes
\begin{equation}\label{lc6}
	\begin{split}
		\Delta S_{\text{BH}} &= \frac{1}{2}(\mathcal{C}_{\text{local}}+\mathcal{C}_{\text{zm}})\ln\left(\frac{\mathcal{A}_{H}}{G_N}\right),\\
		\mathcal{C}_{\text{local}} &= \int_{\text{horizon}} \mathrm{d}^4x \sqrt{\text{det}\thinspace \bar{g}}\thinspace a_4(x),\\
		\mathcal{C}_{\text{zm}}&= \sum_{\tilde{\xi}}\chi (\beta_{\tilde{\xi}}-1)n^0_{\tilde{\xi}}.
	\end{split}
\end{equation}
Therefore, the local contribution in logarithmic correction formula, denoted as $\mathcal{C}_{\text{local}}$, is obtained by calculating the $a_4(x)$ coefficient for the massless fluctuations and then integrating it over the near-horizon geometry of extremal black holes with the aid of appropriate limits (e.g., see \cref{lc13,lc14,lc15} for the extremal Kerr-Newman family of black holes). On the other hand, the correction due to zero-modes\footnote{One should remember that only the $\beta_{\tilde{\xi}}$ terms are related to zero modes in the structure of $\mathcal{C}_{\text{zm}}$, while the (-1) terms arise for subtracting zero modes from the heat kernel.} $\mathcal{C}_{\text{zm}}$ has been enumerated in a variety of works \cite{Banerjee:2011pp,Sen:2012rr,Sen:2012qq,Sen:2013ns} that allow us to use the following compact formula \cite{Charles:2015nn}
\begin{equation}\label{lc7}
	\mathcal{C}_{\text{zm}} = -(3+\mathbb{K}) + 2N_{\text{SUSY}}+ 3\delta,
\end{equation}
where $\mathbb{K}$ is 1 for rotating black holes (e.g., Kerr-Newman, Kerr), 3 otherwise (e.g., Reissner-Nordstr\"om, Schwarzschild); $N_{\text{SUSY}}$ is 4 for BPS black holes, 0 otherwise; $\delta$ is 1 for non-extremal black holes, 0 otherwise.

%------------------------------------------------------------------------------------------------------
\subsubsection{Extremal near-horizon structures of Kerr-Newman family of black holes}
Our ultimate motive is to investigate the extremal near-horizon structures of Kerr-Newman family (Kerr-Newman, Kerr and Reissner-Nordstr\"om) of black holes and extract the integration limits required for their logarithmic entropy corrections. Therefore, we start with the metric describing a general Kerr-Newman solution \cite{Adamo:2014lk} to the 4D Einstein-Maxwell theories, 
{
	\allowdisplaybreaks
	\begin{equation}\label{lc8}
		\begin{split}
			\mathrm{d}s^2 &= \bar{g}_{\mu\nu}\mathrm{d}x^\mu \mathrm{d}x^\nu\\
			&= -\frac{r^2+b^2\thinspace\text{cos}^2\thinspace\psi-2Mr+Q^2}{r^2+b^2\thinspace\text{cos}^2\thinspace\psi}\,\mathrm{d}t^2+\frac{r^2+b^2\thinspace\text{cos}^2\thinspace\psi}{r^2+b^2-2Mr+Q^2}\,\mathrm{d}r^2 \\
			&\quad +\frac{(r^2+b^2\thinspace\text{cos}^2\thinspace\psi)(r^2+b^2)+(2Mr-Q^2)b^2\thinspace\text{sin}^2\thinspace\psi}{r^2+b^2\thinspace\text{cos}^2\thinspace\psi}\thinspace\text{sin}^2\thinspace\psi\, \mathrm{d}\phi^2\\
			&\quad +(r^2+b^2\thinspace\text{cos}^2\thinspace\psi)\,\mathrm{d}\psi^2+\frac{2(Q^2-2Mr)b}{r^2+b^2\thinspace\text{cos}^2\thinspace\psi}\thinspace\text{sin}^2\thinspace\psi \,\mathrm{d}t \, \mathrm{d}\phi,
		\end{split}
\end{equation}}
where $Q$, $M$, $J$ and $b=J/M$ are respectively the charge, mass, angular momentum and Kerr parameter. The above Kerr-Newman metric $\bar{g}_{\mu\nu}$ is described in terms of standard spherical coordinates ($t,r,\psi,\phi$) and serves the following Riemannian and Ricci invariant forms \cite{Henry:2000wd,Cherubini:2002we}
{
	\allowdisplaybreaks
	\begin{align}\label{lc9}
		\begin{split}
			R_{\mu\nu\rho\sigma}R^{\mu\nu\rho\sigma} &= \frac{8}{(r^2+b^2\thinspace\text{cos}^2\thinspace\psi)^6}\Big( 6M^2(r^6-15b^2r^4\thinspace\text{cos}^2\thinspace\psi+ 15b^4r^2\thinspace\text{cos}^4\thinspace\psi\\
			&\quad -b^6\thinspace\text{cos}^6\thinspace\psi)-12MQ^2r(r^4-10r^2b^2\thinspace\text{cos}^2\thinspace\psi+ 5b^4\thinspace\text{cos}^4\thinspace\psi)\\
			&\quad  +Q^4(7r^4-34r^2b^2\thinspace\text{cos}^2\thinspace\psi+7b^4\thinspace\text{cos}^4\thinspace\psi) \Big),\\
			R_{\mu\nu}R^{\mu\nu}&= \frac{4Q^4}{(r^2+b^2\thinspace\text{cos}^2\thinspace\psi)^4}.
		\end{split}
\end{align}}
The apparent singularities for the metric \eqref{lc8} are described as
\begin{equation}\label{lc20}
	r^2+b^2-2Mr+Q^2 = 0,
\end{equation}
which provides the locations of event horizons (outer and inner, respectively) at
\begin{equation}\label{lc21}
	r_{\pm}  = M\pm \sqrt{M^2-Q^2-b^2}.
\end{equation} With our choice,\footnote{$\hbar = c = k_{B} = 1, G_N = 1/16\pi$. Also, it is impossible to access the interior of a black hole and we consider only the outer horizon $r_+$ of area $\mathcal{A}_{H} = 4\pi({r_+}^2+b^2)$ for the Kerr-Newman metric \cite{Bekenstein:2008bk}.} the Bekenstein-Hawking entropy ($S_{\text{BH}}$) and Hawking temperature ($T_{\text{BH}}$) for the Kerr-Newman black hole take the forms,
\begin{equation}\label{lc22}
	\begin{split}
		&S_{\text{BH}}= \frac{\mathcal{A}_{H}}{4G_N}= 16\pi^2\left(2M^2-Q^2+2M\sqrt{M^2-Q^2-b^2} \right),\\
		&T_{\text{BH}}= \frac{1}{\frac{\partial S_{\text{BH}}}{\partial M}}= \frac{1}{32\pi^2}\frac{\sqrt{M^2-Q^2-b^2}}{\left(2M^2-Q^2+2M\sqrt{M^2-Q^2-b^2} \right)}.
	\end{split}
\end{equation}
In the extremal limit, a black hole ceases to evaporate and the Hawking temperature vanishes ($T_{\text{BH}}=0$). This leads us to the extremality condition for the Kerr-Newman black hole \eqref{lc8} as
\begin{equation}\label{lc23}
	M^2 = Q^2+ b^2,
\end{equation}
which yields $r_+=r_-=M$, i.e., the two horizons merge into a single horizon of radius $r_H=M$. The extremality condition \eqref{lc23} also explains the fact that for a vanishing temperature ($T_{\text{BH}}=0$), the black hole still possesses a finite entropy which can be expressed in terms of its parameters as
\begin{equation}\label{lc24}
	S_{\text{BH}} = 16\pi^2 \left(Q^2+2b^2\right).
\end{equation}  
To structure the extremal near-horizon geometry of the Kerr-Newman black hole, we now review the standard trick used in \cite{Bhattacharyya:2012ss}. We consider a new set coordinates ($\tau,\eta,\psi,\zeta$) and set up the following coordinate transformations in terms of a new parameter $\sigma$
\begin{equation}\label{lc11}
	\begin{gathered}
		M^2 = Q^2+ b^2+ \sigma^2, \thinspace r= M+ \sigma\thinspace \text{cosh}\thinspace\eta, \thinspace t = \frac{(2b^2+Q^2)\tau}{\sigma},\\
		\phi = \zeta + \frac{b}{Q^2+2b^2}\left(1-\frac{2M\sigma}{Q^2+2b^2}\right)t.
	\end{gathered}
\end{equation}%source: https://en.wikipedia.org/wiki/Near-horizon_metric
The parameter $\sigma$ serves the desired extremal near-horizon condition as $\sigma \to 0$ (for fixed $\eta$, $\tau$) and therefore reduces the metric \eqref{lc8} to the Euclidean near-horizon geometry of the extremal Kerr-Newman black hole,
\begin{equation}\label{lc12}
	\begin{split}
		{\mathrm{d}s^2}_{\text{horizon}} &=\bar{g}_{\mu\nu}\,\mathrm{d}x^\mu\, \mathrm{d}x^\nu\\
		&= \Big(Q^2 + b^2(1 + \text{cos}^2\thinspace\psi)\Big)(\mathrm{d}\eta^2+\text{sinh}^2\thinspace\eta\, \mathrm{d}\theta^2+\mathrm{d}\psi^2)\\
		&\quad + \frac{(Q^2+2b^2)^2}{\Big(Q^2 + b^2(1 + \text{cos}^2\thinspace\psi)\Big)}\left(\mathrm{d}\zeta-i\frac{2Mb}{(Q^2+2b^2)}(\text{cosh}\thinspace\eta-1)\mathrm{d}\theta\right)^2\text{sin}^2\thinspace\psi,
	\end{split}
\end{equation}
where $\tau = -i\theta$ is set for Euclidean continuation with $\theta$ as the Euclideanized time of period $2\pi$. As expected, the above extremal near-horizon geometry is now structured in the form $AdS_2 \times \mathcal{K}$, where the $AdS_2$ part is parametrized by ($\eta,\theta$) for $0\leq \eta < \infty; \theta \equiv \theta + 2\pi$. For the extremal near-horizon Kerr-Newman metric \eqref{lc12}, one obtains the following finite\footnote{The \say{finite} word depicts the fact that the results \eqref{lc13} dropped all the terms proportional to the IR regulated $\eta$ coordinate as per the requirement in the quantum entropy function formalism \cite{Bhattacharyya:2012ss,Banerjee:2020wbr}.} limits \cite{Bhattacharyya:2012ss}
{
	\allowdisplaybreaks
	\begin{equation}\label{lc13}
		\begin{split}
			\int_{\text{horizon}}  \mathrm{d}^4x \sqrt{\text{det}\thinspace \bar{g}}\thinspace R_{\mu\nu\rho\sigma}R^{\mu\nu\rho\sigma} &= -16\pi^2\Big( 3\mathcal{B}-(8{b^\prime}^6+20{b^\prime}^4+8{b^\prime}^2-1)\mathcal{B}^\prime\Big),\\
			\int_{\text{horizon}}  \mathrm{d}^4x \sqrt{\text{det}\thinspace \bar{g}}R_{\mu\nu}R^{\mu\nu}&= -4\pi^2\Big( 3\mathcal{B} +(8{b^\prime}^2+5)\mathcal{B}^\prime\Big),
		\end{split}
	\end{equation}
}
where
\begin{equation}\nonumber
	{b^\prime}= J/MQ, \mathcal{B} = \frac{2{b^\prime}^2+1}{{b^\prime}({b^\prime}^2+1)^{5/2}}\text{tan}^{-1}\left(\frac{{b^\prime}}{\sqrt{{b^\prime}^2+1}} \right),\thinspace \text{and}\thinspace \mathcal{B}^\prime=\frac{1}{({b^\prime}^2+1)^2(2{b^\prime}^2+1)}.
\end{equation} 
We can also achieve similar limits for the extremal near-horizon Kerr and Reissner-Nordstr\"om black hole backgrounds as special cases of the Kerr-Newman geometry \eqref{lc12}:
\begin{itemize}
	\item[\scalebox{0.6}{$\blacksquare$}] For the extremal Kerr black holes, we need to set $J\neq 0, Q=0$ (i.e., ${b^\prime}\to\infty$) in \cref{lc13}, which provides
	\begin{equation}\label{lc14}
		\begin{split}
			\int_{\text{horizon}} \mathrm{d}^4x \sqrt{\text{det}\thinspace \bar{g}}\thinspace R_{\mu\nu\rho\sigma}R^{\mu\nu\rho\sigma} &= 64\pi^2, \\ \int_{\text{horizon}} \mathrm{d}^4x \sqrt{\text{det}\thinspace \bar{g}}R_{\mu\nu}R^{\mu\nu} &= 0.
		\end{split}
	\end{equation}
	\item[\scalebox{0.6}{$\blacksquare$}] For the extremal Reissner-Nordstr\"om black holes, we need to set $J= 0, Q\neq0$ (i.e., ${b^\prime}\to 0$) in \cref{lc13}, which gives
	\begin{equation}\label{lc15}
		\begin{split}
			\int_{\text{horizon}} \mathrm{d}^4x \sqrt{\text{det}\thinspace \bar{g}}\thinspace R_{\mu\nu\rho\sigma}R^{\mu\nu\rho\sigma} &= -64\pi^2, \\ \int_{\text{horizon}} \mathrm{d}^4x \sqrt{\text{det}\thinspace \bar{g}}R_{\mu\nu}R^{\mu\nu} &= -32\pi^2.
		\end{split}
	\end{equation}
\end{itemize}
The above near-horizon extremal limits (\cref{lc13,lc14,lc15}) of the Riemannian and Ricci invariants are crucial ingredients for computing logarithmic correction (especially the $\mathcal{C}_{\text{local}}$ contribution \eqref{lc6}) to the entropy of Kerr-Newman family of black holes in different 4D Einstein-Maxwell theories. %Note that the discussed extremal near-horizon set-up restricts the Schwarzschild black holes ($J= 0, Q=0$) for possessing an undetermined ${b^\prime}$.

%=======================================================================
%\section{$\mathcal{N}=2,d=4$ Einstein-Maxwell supergravity multiplets and their logarithmic correction contributions to the entropy of extremal black holes}\label{mul}

\section{Seeley-DeWitt coefficients and logarithmic correction contributions of the multiplets of $\mathcal{N}=2,d=4$ EMSGT}\label{mul}

%\section{Seeley-DeWitt coefficients of $\mathcal{N}=2,d=4$ EMSGT multiplets and their logarithmic correction contributions}\label{mul}
{
	\renewcommand{\arraystretch}{1.4}
	\begin{table}[t]% h (here), t (top), b (bottom) and p (page of floats), ht (here top)
		\centering
		\hspace{-0.2in}
		\begin{tabular}{|>{\centering}p{1.3in}|>{\centering}p{.4in}|>{\centering}p{.5in}>{\centering}p{.5in}>{\centering}p{.5in}>{\centering}p{.5in}>{\centering}p{.5in}|}
			\hline
			\textbf{Multiplets}& \textbf{$s_{\text{max}}$} & {$\bm{s=2}$} & {$\bm{s=\frac{3}{2}}$} & {$\bm{s=1}$} & {$\bm{s=\frac{1}{2}}$} & {$\bm{s=0}$} \tabularnewline \hline \hline
			supergravity ($\mathcal{G}_2$)  & 2 &1 & 2&1 & & \tabularnewline 
			gravitino ($\mathcal{G}_{3/2}$)& $\frac{3}{2}$ & & 1&2 &1 & \tabularnewline 
			vector ($\mathcal{G}_1$)& 1 & & &1 &2 & 1+1 \tabularnewline 
			hyper ($\mathcal{G}_{1/2}$)& $\frac{1}{2}$ & & & &2 & 2+2 \tabularnewline \hline
		\end{tabular}
		\caption{Different multiplets and spin content of a $\mathcal{N}=2,d=4$ EMSGT.}\label{tab1}
	\end{table}
}  
Supergravity theories are field theories that emerge as the IR (low energy) limit of superstring theories where supersymmetric matters are coupled to gravity. The field content of these theories is determined by the space-time dimension ($d$) and the amount of supersymmetry ($\mathcal{N}$) they embed. %Extended supergravity theories are embedded with more than one supersymmetry.
%Supergravity theories embedded with more than one supersymmetry are known as extended supergravities.
$\mathcal{N}=2,d=4$ EMSGT is the most simple and significant extended supergravity embedded Einstein-Maxwell theory, which arises as an IR limit of type-II string theory compactified on a Calabi–Yau three-fold \cite{Grana:2006mg,Freedman:2012xp}. One can structure such theory by embedding the general background solutions (Kerr-Newman family of black holes \cite{Adamo:2014lk}) of a 4D Einstein-Maxwell theory into a $\mathcal{N}=2$ supergravity with the condition that the supergravity fields fluctuate around the background of the Einstein-Maxwell theory. This ensures that both the Einstein-Maxwell theory and the $\mathcal{N}=2$ supergravity share common equations of motion and solutions, which in turn justifies the embedding \cite{Charles:2015nn}. Such embedding induces both minimal and non-minimal couplings between the background and matter fluctuations distributed in different multiplets of the resultant theory. 

For a general $\mathcal{N}=2,d=4$ EMSGT, the massless fields are distributed in four kinds of multiplets -- supergravity multiplet ($\mathcal{G}_2$), gravitino multiplet ($\mathcal{G}_{3/2}$), vector multiplet ($\mathcal{G}_1$) and hyper multiplet ($\mathcal{G}_{1/2}$). The spin contents of these multiplets are summarized  in table \ref{tab1}, where each multiplet has equal bosonic and fermionic on-shell degrees of freedom. Interestingly, the quadratic field fluctuations of any particular multiplet do not mix with fluctuations of other multiplets. This permits us to study the quadratic fluctuations of each $\mathcal{N}=2$ multiplet (divided into bosonic and fermionic sectors) separately and calculate the necessary Seeley-DeWitt coefficients. We then employ the Seeley-DeWitt data to obtain the logarithmic correction contributions of individual multiplets to the entropy of extremal Kerr-Newman, Kerr and Reissner-Nordstr\"om black holes. Note that the supergravity multiplet results are collected from \cite{Karan:2019sk}, while calculations for other multiplets are thoroughly executed by pursuing the standard procedure mentioned in \cref{B}. 
\subsection{$\mathcal{N}=2,d=4$ EMSGT: supergravity multiplet}\label{supergravity}
The field contents of the supergravity multiplet ($\mathcal{G}_2$) of a $\mathcal{N}=2,d=4$ EMSGT can be divided into two sectors --
\begin{itemize}
	\item[] \textbf{Bosonic sector ($\mathcal{G}^{b}_2$):} one vector field $A_\mu$, dubbed as {graviphoton}, is minimally coupled to one graviton $g_{\mu\nu}$. This part is identical to the pure 4D Einstein-Maxwell theory;
	\item[] \textbf{Fermionic sector ($\mathcal{G}^{f}_2$):} two massless Majorana gravitini $\lbrace \psi_\mu$, $\phi_\mu\rbrace$ are coupled non-minimally with their Hermitian conjugates $\lbrace\bar{\psi}_\mu$, $\bar{\phi}_\mu\rbrace$ through the background graviphoton field strength $\bar{F}_{\mu\nu}$ and its dual $\bar{H}_{\mu\nu}=-\varepsilon_{\mu\nu\rho\sigma}\bar{F}^{\rho\sigma}$.
\end{itemize}
The action of a pure 4D Einstein-Maxwell theory describing the bosonic-sector of the supergravity multiplet is
\begin{equation}\label{EMT}
	\mathcal{S}_{\text{EM}}= \int \mathrm{d}^4x \sqrt{\text{det}\thinspace {g}} \thinspace \left(\mathcal{R}-F_{\mu\nu}F^{\mu\nu} \right),
\end{equation}
where ${F}_{\mu\nu}\equiv \partial_\mu {A}_\nu-\partial_\nu {A}_\mu$ is the graviphoton field strength tensor and $\mathcal{R}$ is Ricci scalar corresponding to the metric $g_{\mu\nu}$. Any arbitrary classical solution $(\bar{g}_{\mu\nu}, \bar{A}_\mu)$ to the theory \eqref{EMT} satisfies the following equations of motion
\begin{equation}\label{S2}
	R_{\mu\nu}= 2\bar{F}_{\mu\rho}{\bar{F_\nu}}^\rho-\frac{1}{2}\bar{g}_{\mu\nu}\bar{F}_{\rho\sigma}\bar{F}^{\rho\sigma}, R=0,
\end{equation}
where $\bar{F}_{\mu\nu}$ is the strength of the background graviphoton field $\bar{A}_\mu$; $R_{\mu\nu}$ and $R$ are respectively Ricci tensor and Ricci scalar corresponding to the background metric $\bar{g}_{\mu\nu}$. All the equations of motion, Maxwell equations, Bianchi identities satisfied by $\bar{F}_{\mu\nu}$ and it's $\bar{H}_{\mu\nu}$ are recorded in \cref{AA}. As mentioned, the $\mathcal{N}=2,d=4$ EMSGT is structured in such a manner that all the matter fields should fluctuate around the background $(\bar{g}_{\mu\nu}, \bar{A}_\mu)$. For that, the $\mathcal{N}=2,d=4$ EMSGT shares the same equations of motion, solutions, identities as the 4D Einstein-Maxwell theory (or the bosonic-sector ($\mathcal{G}^{b}_2$)) \eqref{EMT}. Also, there will be  various non-minimal couplings in the fluctuation of $\mathcal{N}=2$ matter fields via $\bar{F}_{\mu\nu}$ and $\bar{H}_{\mu\nu}$. For the fluctuations,
\begin{equation}\label{S3}
	g_{\mu\nu} = \bar{g}_{\mu\nu} + \sqrt{2}h_{\mu\nu}, \thinspace A_\mu = \bar{A}_\mu + \frac{1}{2}a_\mu,
\end{equation}
the Seeley-DeWitt coefficients and the logarithmic correction results of the quadratic fluctuated $\mathcal{N}=2$ supergravity multiplet\footnote{For the quadratic fluctuated actions of the bosonic ($\mathcal{G}^{b}_2$) and fermionic ($\mathcal{G}^{f}_2$) sectors of the $\mathcal{N}=2$ supergravity multiplet, readers are referred to eqs. (5.4) and (5.21) of the work \cite{Karan:2019sk}.} ($\mathcal{G}_2$) have already been achieved in our earlier work \cite{Karan:2019sk}. The first three Seeley-DeWitt coefficients are \cite{Karan:2019sk},
%\footnote{For the quadratic order fluctuated forms of the bosonic ($\mathcal{G}^{b}_2$) and fermionic ($\mathcal{G}^{f}_2$) supergravity sectors, readers are referred to eqs. (5.4) and (5.21) of the work \cite{Karan:2019sk}.}
\begin{align}\label{S4}
	\begin{split}
		&(4\pi)^2 a^{\mathcal{G}_2}_0(x) = 0,\\
		&(4\pi)^2 a^{\mathcal{G}_2}_2(x) = 14 \bar{F}_{\mu\nu}\bar{F}^{\mu\nu},\\
		&(4\pi)^2 a^{\mathcal{G}_2}_4(x) = \frac{11}{24}\left( R_{\mu\nu\rho\sigma}R^{\mu\nu\rho\sigma}-4 R_{\mu\nu}R^{\mu\nu} \right),
	\end{split}
\end{align}
and using them in the formulae \eqref{lc6}, the logarithmic correction contributions of the supergravity multiplet ($\mathcal{G}_2$) to the entropy of extremal Kerr-Newman family of black holes are obtained as
\begin{equation}\label{S5}
	\Delta S^{\mathcal{G}_2}_{\text{BH}} =
	\begin{cases}
		\left(\frac{11}{12}\frac{(2{b^\prime}^6+5{b^\prime}^4+4{b^\prime}^2+1)}{({b^\prime}^2+1)^2(2{b^\prime}^2+1)}-2\right) \text{ln}\thinspace \mathcal{A}_{H} & \text{for Kerr-Newman},\\
		-\frac{13}{12}\thinspace \text{ln}\thinspace \mathcal{A}_{H} & \text{for Kerr},\\
		\frac{23}{12}\thinspace \text{ln}\thinspace \mathcal{A}_{H} & \text{for Reissner-Nordstr\"om}.
	\end{cases}
\end{equation}
The above results contain $\mathcal{C}_{\text{zm}}=-4, -4$ and 2 respectively for the extremal Kerr-Newman, Kerr and Reissner-Nordstr\"om black holes, according to the formula \eqref{lc7}. Here extremal Reissner-Nordstr\"om black holes are the only example of a BPS solution in the  $\mathcal{N}=2,d=4$ EMSGT.
%-------------------------------------------------------------------------------------------------------------------- 
\subsection{$\mathcal{N}=2,d=4$ EMSGT: gravitino multiplet}\label{gravitino}
The field contents of the gravitino multiplet ($\mathcal{G}_{3/2}$) of a $\mathcal{N}=2,d=4$ EMSGT can be split into the following sectors --
\begin{itemize}
	\item[] \textbf{Bosonic sector ($\mathcal{G}^{b}_{3/2}$):} two vector fields are minimally coupled to the background gravity,
	\item[] \textbf{Fermionic sector ($\mathcal{G}^{f}_{3/2}$):} one Majorana gravitino $\psi_\mu$, one Majorana gaugino $\lambda$ and their Hermitian conjugates ($\bar{\psi}_\mu$, $\bar{\lambda}$) are non-minimally coupled via the background graviphoton field strength $\bar{F}_{\mu\nu}$.
\end{itemize}
The quadratic order fluctuated action describing the fermionic-sector ($\mathcal{G}^{f}_{3/2}$) is 
\begin{equation}\label{G1}
	\begin{split}
		\delta^2\mathcal{S}_{\mathcal{G}^{f}_{3/2}} &= \frac{2}{\kappa^2}\int \mathrm{d}^4x \sqrt{\text{det}\thinspace \bar{g}}\thinspace \mathscr{L}_{\mathcal{G}^{f}_{3/2}},\\
		\mathscr{L}_{\mathcal{G}^{f}_{3/2}} &= -\bar{\psi}_\mu\gamma^{\mu\rho\nu} D_\rho\psi_{\nu}- \bar{\lambda} \gamma^\rho D_\rho \lambda-\frac{1}{2\sqrt{2}} \bar{\psi}_\mu \hat{\mathcal{F}}\gamma^\mu \lambda-\frac{1}{2\sqrt{2}}\bar{\lambda}\gamma^\nu \hat{\mathcal{F}}\psi_\nu,
	\end{split}
\end{equation}
where $\kappa =1/\sqrt{2}$ and $\hat{\mathcal{F}} = \bar{F}_{\alpha\beta}\gamma^\alpha\gamma^\beta$.\footnote{The useful definitions and identities related to the 4D gamma matrices $\gamma^\mu$ are listed in \cref{AA}.}
We adopted the action from\footnote{See eq. (2.41) of \cite{Charles:2015nn}.} \cite{Charles:2015nn} and rescaled the gravitino field as $\psi_\mu \to \sqrt{2}\psi_\mu$ so that both the gravitino and gaugino fields have the same normalization state in their kinetic parts of the operator form \eqref{G6}. We then cast the harmonic gauge $\gamma^\mu \psi_\mu =0$ and gauge fix the action \eqref{G1} by inserting the gauge-fixing term,
\begin{align}\label{G2}
	\frac{1}{\kappa^2}\int \mathrm{d}^4x \sqrt{\text{det}\thinspace \bar{g}} \thinspace(\bar{\psi}_\mu \gamma^\mu)\gamma^\rho D_\rho(\gamma^\nu \psi_\nu),
\end{align}
followed by the ghost action \cite{Banerjee:2011oo},
\begin{align}\label{Gx}
	\mathcal{S}_{\mathcal{G}^{f}_{3/2},\text{ghost}} = \frac{1}{\kappa^2}\int \mathrm{d}^4x \sqrt{\text{det}\thinspace \bar{g}} \thinspace (\bar{\tilde{b}}\gamma^\mu D_\mu \tilde{c} + \bar{\tilde{e}}\gamma^\mu D_\mu \tilde{e}),
\end{align}
where $\tilde{b}$, $\tilde{c}$, and $\tilde{e}$ are three bosonic ghosts introduced to the system due to the choice of the particular gauge fixing \eqref{G2}.
These ghost fields have no interaction with any other fields of the fermionic-sector ($\mathcal{G}^{f}_{3/2}$), and hence we will evaluate their contributions separately. The standard prescription of \cref{P} assists us from here. Essentially the gauge fixed action (without accounting for the ghost action \eqref{Gx}) provides a first-order Dirac-type operator $\slashed{D}$,\footnote{In \cref{G3}, the fermionic operator is set as Dirac to proceed with the standard set-up described in \cref{P}. But in the end, we extract back the Seeley-DeWitt coefficients (see the results \eqref{G12}) for the actual Majorana spinors described in the action \eqref{G1} by setting $\chi=-1/2$ according to the relation \eqref{stff3}. The same treatment is followed throughout this paper for other Majorana fermions present in the $\mathcal{N}=2, d=4$ EMSGT.}
\begin{align}\label{G3}
	\begin{gathered}
		\delta^2\mathcal{S}_{\mathcal{G}^{f}_{3/2}} = \frac{2}{\kappa^2}\int \mathrm{d}^4x \sqrt{\text{det}\thinspace \bar{g}} \thinspace \tilde{\xi}_m\slashed{D}^{mn} \tilde{\xi}_n,\\
		\tilde{\xi}_m\slashed{D}^{mn} \tilde{\xi}_n = \frac{i}{2}\bar{\psi}_\mu\gamma^\nu\gamma^\rho\gamma^\mu D_\rho\psi_{\nu}- i\bar{\lambda} \gamma^\rho D_\rho \lambda-\frac{i}{2\sqrt{2}} \bar{\psi}_\mu \hat{\mathcal{F}}\gamma^\mu \lambda-\frac{i}{2\sqrt{2}}\bar{\lambda}\gamma^\nu\hat{\mathcal{F}}\psi_\nu,
	\end{gathered}
\end{align}
from which one needs to structure the necessary operator $\Lambda$ in the following manner:
{
	\allowdisplaybreaks
	%\begin{equation}
	\begin{align}\label{G4}
		\tilde{\xi}_m \Lambda^{mn}\tilde{\xi}_n &= \tilde{\xi}_m (\slashed{D}^{mp})^\dagger{\slashed{D}_p}^n\tilde{\xi}_n \nonumber\\
		&= -\bar{\psi}_{\mu}\bigg\lbrace\frac{1}{4}\gamma^\alpha \gamma^\rho\gamma^\mu\gamma^\nu\gamma^\sigma\gamma_\alpha D_\rho D_\sigma- \frac{1}{8} \hat{\mathcal{F}}\gamma^\mu\gamma^\nu \hat{\mathcal{F}}\bigg\rbrace\psi_{\nu} \nonumber\\
		&\qquad -\bar{\lambda}\bigg\lbrace \gamma^\rho\gamma^\sigma D_\rho D_\sigma- \frac{1}{8} \gamma^\alpha\hat{\mathcal{F}}\hat{\mathcal{F}}\gamma_\alpha\bigg\rbrace\lambda \nonumber\\
		&\qquad+\bar{\psi}_{\mu}\bigg\lbrace \frac{1}{4\sqrt{2}}\gamma^\alpha\gamma^\rho\gamma^\mu(D_\rho \hat{\mathcal{F}}+
		\hat{\mathcal{F}}D_\rho)\gamma_\alpha + \frac{1}{2\sqrt{2}}\hat{\mathcal{F}}\gamma^\mu\gamma^\rho D_\rho\bigg\rbrace\lambda \nonumber\\
		&\qquad-\bar{\lambda}\bigg\lbrace \frac{1}{4\sqrt{2}}\gamma^\alpha \hat{\mathcal{F}} \gamma^\nu \gamma^\rho\gamma_\alpha D_\rho + \frac{1}{2\sqrt{2}} \gamma^\rho\gamma^\nu (D_\rho \hat{\mathcal{F}} + \hat{\mathcal{F}} D_\rho)\bigg\rbrace \psi_{\nu}. 
	\end{align}
	%\end{equation}
}
Make sure to consider the fact that in our choice of 4D Euclidean space-time, $\gamma^\mu$ matrices are Hermitian, $(\gamma^\rho D_\rho)^\dagger = -\gamma^\rho D_\rho$, $(\gamma^\nu\gamma^\rho\gamma^\mu D_\rho)^\dagger = -\gamma^\nu\gamma^\rho\gamma^\mu D_\rho$, $({\hat{\mathcal{F}}})^\dagger = \hat{\mathcal{F}}$. The operator form \eqref{G4} is not of Laplace-type as prescribed in the schematic \eqref{p1} because the principal kinetic terms are not appropriately contracted to provide the Laplacian $D^\rho D_\rho$. To rectify this issue, we set up the following relations for the gravitino $\psi_{\mu}$ and gaugino $\lambda$ fluctuations
\begin{subequations}
	\begin{align}
		\gamma^\alpha \gamma^\rho\gamma^\mu\gamma^\nu\gamma^\sigma\gamma_\alpha D_\rho D_\sigma\psi_\mu &= (4\mathbb{I}_4\bar{g}^{\mu\nu}D^\rho D_\rho +2\gamma^\alpha\gamma^\beta {R^{\mu\nu}}_{\alpha\beta}\nonumber\\
		& \qquad -2\gamma^\mu\gamma^\alpha {R^\nu}_\alpha+2\gamma^\nu\gamma^\alpha {R^\mu}_\alpha-4\mathbb{I}_4R^{\mu\nu})\psi_\mu, \label{idd1}\\
		\gamma^\rho \gamma^\sigma D_\rho D_\sigma\lambda &= \mathbb{I}_4 D^\rho D_\rho\lambda,\label{idd2}
	\end{align}
\end{subequations}
and restructure $\Lambda$ into the required Laplace-type form,
{
	\allowdisplaybreaks
	%\begin{equation}\label{G6}
	\begin{align}\label{G6}
		\tilde{\xi}_m \Lambda^{mn}\tilde{\xi}_n &= -\mathbb{I}_4\bar{g}^{\mu\nu}\bar{\psi}_{\mu}D^\rho D_\rho\psi_{\nu}- \mathbb{I}_4\bar{\lambda}  D^\rho D_\rho \lambda \nonumber\\
		& \quad -\bar{\psi}_{\mu}\bigg\lbrace \frac{1}{2}\gamma^\alpha\gamma^\beta {R^{\mu\nu}}_{\alpha\beta}-\frac{1}{2}\gamma^\mu\gamma^\alpha {R^\nu}_\alpha+\frac{1}{2}\gamma^\nu\gamma^\alpha {R^\mu}_\alpha -\mathbb{I}_4 R^{\mu\nu} \nonumber\\
		& \quad - \frac{1}{8} \gamma^\alpha\gamma^\beta\gamma^\mu\gamma^\nu \gamma^\theta\gamma^\phi \bar{F}_{\alpha\beta}\bar{F}_{\theta\phi}\bigg\rbrace\psi_{\nu}+\frac{1}{8} \bar{\lambda}\gamma^\sigma\gamma^\alpha\gamma^\beta\gamma^\theta \gamma^\phi\gamma_\sigma \bar{F}_{\alpha\beta}\bar{F}_{\theta\phi}  \lambda \nonumber\\
		&\quad  +\frac{1}{4\sqrt{2}}\bar{\psi}_{\mu}\gamma^\sigma\gamma^\rho\gamma^\mu \gamma^\alpha\gamma^\beta \gamma_\sigma (D_\rho \bar{F}_{\alpha\beta}) \lambda- \frac{1}{2\sqrt{2}} \bar{\lambda}\gamma^\rho\gamma^\nu\gamma^\alpha\gamma^\beta (D_\rho \bar{F}_{\alpha\beta}) \psi_\nu \nonumber\\
		&\quad+\bar{\psi}_{\mu}\bigg\lbrace \frac{1}{4\sqrt{2}}\gamma^\sigma\gamma^\rho\gamma^\mu
		\gamma^\alpha\gamma^\beta\gamma_\sigma \bar{F}_{\alpha\beta} + \frac{1}{2\sqrt{2}}\gamma^\alpha\gamma^\beta\gamma^\mu\gamma^\rho \bar{F}_{\alpha\beta}\bigg\rbrace (D_\rho\lambda) \nonumber\\
		&\quad-\bar{\lambda}\bigg\lbrace \frac{1}{4\sqrt{2}}\gamma^\sigma \gamma^\alpha\gamma^\beta \gamma^\nu \gamma^\rho\gamma_\sigma \bar{F}_{\alpha\beta} + \frac{1}{2\sqrt{2}} \gamma^\rho\gamma^\nu \gamma^\alpha\gamma^\beta \bar{F}_{\alpha\beta}\bigg\rbrace (D_\rho\psi_{\nu}). 
\end{align}}
%\end{equation}
The derivations \eqref{idd1} and \eqref{idd2} are executed by employing the commutation relations \eqref{RS} and \eqref{spinor} and also considering up to total derivative terms. Since the operator form \eqref{G6} is now comparable with the Laplace-type schematic \eqref{p1}, we extract out the matrices $I$, $P$ and $\omega^\rho (=\frac{1}{2}N^\rho)$ as
\begin{equation}\label{G7}
	\begin{split}
		\tilde{\xi}_m I^{mn}\tilde{\xi}_n &= \bar{\psi}_{\mu}\mathbb{I}_4 \bar{g}^{\mu\nu}\psi_{\nu}+\bar{\lambda}\mathbb{I}_4\lambda,\\
		\tilde{\xi}_m P^{mn}\tilde{\xi}_n &= \bar{\psi}_{\mu}\bigg\lbrace \frac{1}{2}\gamma^\alpha\gamma^\beta {R^{\mu\nu}}_{\alpha\beta}-\frac{1}{2}\gamma^\mu\gamma^\alpha {R^\nu}_\alpha+\frac{1}{2}\gamma^\nu\gamma^\alpha {R^\mu}_\alpha -\mathbb{I}_4 R^{\mu\nu}\\
		& \quad - \frac{1}{8} \gamma^\alpha\gamma^\beta\gamma^\mu\gamma^\nu \gamma^\theta\gamma^\phi \bar{F}_{\alpha\beta}\bar{F}_{\theta\phi}\bigg\rbrace\psi_{\nu}-\frac{1}{8} \bar{\lambda}\gamma^\sigma\gamma^\alpha\gamma^\beta\gamma^\theta \gamma^\phi\gamma_\sigma \bar{F}_{\alpha\beta}\bar{F}_{\theta\phi}  \lambda\\
		&\quad  -\frac{1}{4\sqrt{2}}\bar{\psi}_{\mu}\gamma^\sigma\gamma^\rho\gamma^\mu \gamma^\alpha\gamma^\beta \gamma_\sigma (D_\rho \bar{F}_{\alpha\beta}) \lambda+\frac{1}{2\sqrt{2}} \bar{\lambda}\gamma^\rho\gamma^\nu\gamma^\alpha\gamma^\beta (D_\rho \bar{F}_{\alpha\beta}) \psi_\nu,\\
		\tilde{\xi}_m {(\omega^\rho)}^{mn}\tilde{\xi}_n &= -\bar{\psi}_{\mu}\bigg\lbrace  \frac{1}{4\sqrt{2}}\gamma^\alpha\gamma^\beta\gamma^\mu\gamma^\rho \bar{F}_{\alpha\beta}+\frac{1}{8\sqrt{2}}\gamma^\sigma\gamma^\rho\gamma^\mu
		\gamma^\alpha\gamma^\beta\gamma_\sigma \bar{F}_{\alpha\beta} \bigg\rbrace \lambda\\
		&\quad+\bar{\lambda}\bigg\lbrace \frac{1}{4\sqrt{2}} \gamma^\rho\gamma^\nu \gamma^\alpha\gamma^\beta \bar{F}_{\alpha\beta}+\frac{1}{8\sqrt{2}}\gamma^\sigma \gamma^\alpha\gamma^\beta \gamma^\nu \gamma^\rho\gamma_\sigma \bar{F}_{\alpha\beta} \bigg\rbrace \psi_{\nu}.
	\end{split}
\end{equation} 
%Also, the field connection $\omega^\rho$ is recognized as
%\begin{align}\label{G8}
%\tilde{\xi}_m {(\omega^\rho)}^{mn}\tilde{\xi}_n &=  -\frac{1}{4\sqrt{2}}\bar{\psi}_{\mu}\bigg\lbrace \gamma^\alpha\gamma^\beta\gamma^\mu\gamma^\rho \bar{F}_{\alpha\beta} + \frac{1}{2}\gamma^\tau\gamma^\rho\gamma^\mu
%\gamma^\alpha\gamma^\beta\gamma_\tau \bar{F}_{\alpha\beta}\bigg\rbrace \lambda \nonumber\\
%&\quad+\frac{1}{4\sqrt{2}}\bar{\lambda}\bigg\lbrace  \gamma^\rho\gamma^\nu \gamma^\alpha\gamma^\beta %\bar{F}_{\alpha\beta}+\frac{1}{2}\gamma^\tau \gamma^\alpha\gamma^\beta \gamma^\nu \gamma^\rho\gamma_\tau %\bar{F}_{\alpha\beta} \bigg\rbrace \psi_{\nu}.
%\end{align}
All the above data are sufficient to express $\Lambda$ in the standard form \eqref{p4}, where the matrices $E$ and $\Omega_{\rho\sigma}$ hold the following expressions (according to the formulae \eqref{p6} and \eqref{p3})
{
	\allowdisplaybreaks
	\begin{align}\label{G9}
		\tilde{\xi}_m E^{mn}\tilde{\xi}_n &= \bar{\psi}_{\mu}\Bigg\lbrace \frac{1}{2}\gamma^\alpha\gamma^\beta {R^{\mu\nu}}_{\alpha\beta}-\frac{1}{2}\gamma^\mu\gamma^\alpha {R^\nu}_\alpha+\frac{1}{2}\gamma^\nu\gamma^\alpha {R^\mu}_\alpha -\mathbb{I}_4 R^{\mu\nu} \nonumber\\
		& \qquad - \frac{1}{8} \gamma^\alpha\gamma^\beta\gamma^\mu\gamma^\nu \gamma^\theta\gamma^\phi \bar{F}_{\alpha\beta}\bar{F}_{\theta\phi}+\frac{1}{32}\Big(\gamma^\alpha\gamma^\beta\gamma^\mu\gamma^\rho+\frac{1}{2}\gamma^\sigma\gamma^\rho\gamma^\mu\gamma^\alpha\gamma^\beta\gamma_\sigma \Big)\nonumber\\
		&\qquad \quad\times \Big(\gamma_\rho\gamma^\nu\gamma^\theta\gamma^\phi+ \frac{1}{2}\gamma^\tau\gamma^\theta\gamma^\phi\gamma^\nu\gamma_\rho\gamma_\tau\Big)\bar{F}_{\alpha\beta}\bar{F}_{\theta\phi}\Bigg\rbrace\psi_{\nu}\nonumber\\
		&\quad +\bar{\lambda}\Bigg\lbrace -\frac{1}{8}\gamma^\sigma\gamma^\alpha\gamma^\beta\gamma^\theta \gamma^\phi\gamma_\sigma \bar{F}_{\alpha\beta}\bar{F}_{\theta\phi}+ \frac{1}{32}\Big(\gamma^\rho\gamma^\mu\gamma^\alpha\gamma^\beta+ \frac{1}{2}\gamma^\sigma\gamma^\alpha\gamma^\beta\gamma^\mu\gamma^\rho\gamma_\sigma \Big)\nonumber\\
		&\qquad\quad \times\Big(\gamma^\theta\gamma^\phi\gamma_\mu\gamma_\rho + \frac{1}{2}\gamma^\tau\gamma_\rho\gamma_\mu\gamma^\theta\gamma^\phi\gamma_\tau\Big)\bar{F}_{\alpha\beta}\bar{F}_{\theta\phi}\Bigg\rbrace \lambda\nonumber\\
		& \quad +\bar{\psi}_\mu \Bigg\lbrace \frac{1}{4\sqrt{2}}\gamma^\alpha\gamma^\beta\gamma^\mu\gamma^\rho (D_\rho \bar{F}_{\alpha\beta})-\frac{1}{8\sqrt{2}}\gamma^\sigma\gamma^\rho\gamma^\mu\gamma^\alpha\gamma^\beta\gamma_\sigma (D_\rho \bar{F}_{\alpha\beta})\Bigg\rbrace \lambda \nonumber\\
		& \quad +\bar{\lambda} \Bigg\lbrace \frac{1}{4\sqrt{2}}\gamma^\rho\gamma^\nu\gamma^\alpha\gamma^\beta (D_\rho \bar{F}_{\alpha\beta})- \frac{1}{8\sqrt{2}} \gamma^\sigma\gamma^\alpha\gamma^\beta\gamma^\nu\gamma^\rho\gamma_\sigma (D_\rho \bar{F}_{\alpha\beta})\Bigg\rbrace \psi_\nu,
	\end{align}
}
{
	\allowdisplaybreaks
	\begin{align}\label{G10}
		\tilde{\xi}_m (\Omega_{\rho\sigma})^{mn}\tilde{\xi}_n &= \bar{\psi}_{\mu}\Bigg\lbrace \mathbb{I}_4{R^{\mu\nu}}_{\rho\sigma}+\frac{1}{4}\bar{g}^{\mu\nu}\gamma^\alpha\gamma^\beta R_{\rho\sigma\alpha\beta}-\frac{1}{32}\left(\Big(\gamma^\alpha\gamma^\beta\gamma^\mu\gamma_\rho+\frac{1}{2}\gamma^\tau\gamma_\rho\gamma^\mu\gamma^\alpha\gamma^\beta\gamma_\tau \Big)\right. \nonumber\\
		&\qquad \quad \left.\times \Big(\gamma_\sigma\gamma^\nu\gamma^\theta\gamma^\phi+ \frac{1}{2}\gamma^\lambda\gamma^\theta\gamma^\phi\gamma^\nu\gamma_\sigma\gamma_\lambda\Big)\bar{F}_{\alpha\beta}\bar{F}_{\theta\phi}-(\rho\leftrightarrow\sigma)\right)\Bigg\rbrace\psi_{\nu} \nonumber\\
		&\quad +\bar{\lambda}\Bigg\lbrace \frac{1}{4}\gamma^\alpha\gamma^\beta R_{\rho\sigma\alpha\beta}-\frac{1}{32}\left(\Big(\gamma_\rho\gamma^\mu\gamma^\alpha\gamma^\beta+\frac{1}{2}\gamma^\tau\gamma^\alpha\gamma^\beta\gamma^\mu\gamma_\rho\gamma_\tau\Big)\right.\nonumber\\
		&\qquad\quad \left.\times \Big(\gamma^\theta\gamma^\phi\gamma_\mu\gamma_\sigma+\frac{1}{2}\gamma^\lambda\gamma_\sigma\gamma_\mu\gamma^\theta\gamma^\phi\gamma_\lambda\Big)\bar{F}_{\alpha\beta}\bar{F}_{\theta\phi}-(\rho\leftrightarrow\sigma)\right)\Bigg\rbrace \lambda \nonumber\\
		& \quad -\bar{\psi}_{\mu}\Bigg\lbrace \frac{1}{4\sqrt{2}}\Big(\gamma^\alpha\gamma^\beta\gamma^\mu\gamma_\sigma  + \frac{1}{2}\gamma^\tau\gamma_\sigma\gamma^\mu
		\gamma^\alpha\gamma^\beta\gamma_\tau \Big)(D_\rho\bar{F}_{\alpha\beta})-(\rho\leftrightarrow\sigma)\Bigg\rbrace \lambda \nonumber\\
		&\quad+\bar{\lambda}\Bigg\lbrace \frac{1}{4\sqrt{2}}\Big(\gamma_\sigma\gamma^\nu\gamma^\alpha\gamma^\beta  + \frac{1}{2}\gamma^\tau\gamma^\alpha\gamma^\beta
		\gamma^\nu\gamma_\sigma\gamma_\tau \Big)(D_\rho\bar{F}_{\alpha\beta})-(\rho\leftrightarrow\sigma)\Bigg\rbrace \psi_{\nu}.
\end{align}}
At this point, our major task is to calculate $\text{tr}(E)$, $\text{tr}(E)$, $\text{tr}(E^2)$ and $\text{tr}(\Omega_{\rho\sigma}\Omega^{\rho\sigma})$. These trace calculations are quite lengthy but straightforward; hence one needs to proceed systematically. The required trace results are (see \cref{BB} for calculation details),
\begin{equation}\label{G11}
	\begin{split}
		\text{tr}(I) &=4\times 4+4= 20,\\
		\text{tr}(E) &= -8\bar{F}_{\mu\nu}\bar{F}^{\mu\nu},\\
		\text{tr}(E^2) &= 2R_{\mu\nu\rho\sigma}R^{\mu\nu\rho\sigma}-5 R_{\mu\nu}R^{\mu\nu}-2R_{\mu\nu\rho\sigma}\bar{F}^{\mu\nu}\bar{F}^{\rho\sigma}+10(\bar{F}_{\mu\nu}\bar{F}^{\mu\nu})^2,\\
		\text{tr}(\Omega_{\rho\sigma}\Omega^{\rho\sigma})  &= -\frac{13}{2} R_{\mu\nu\rho\sigma}R^{\mu\nu\rho\sigma}+10 R_{\mu\nu}R^{\mu\nu}+12 R_{\mu\nu\rho\sigma}\bar{F}^{\mu\nu}\bar{F}^{\rho\sigma} -60(\bar{F}_{\mu\nu}\bar{F}^{\mu\nu})^2.
	\end{split}
\end{equation}
Putting all these trace data in the formulae \eqref{p7} and also setting $\chi=-1/2$ to account for the Majorana gravitino and gaugino, we obtain the first three Seeley-DeWitt coefficients for the fermionic-sector (without the ghost contribution),
\begin{equation}\label{G12}
	\begin{split}
		(4\pi)^2 {a_0}^{\mathcal{G}^{f}_{3/2},\text{no-ghost}}(x) &= -10,\\
		(4\pi)^2 {a_2}^{\mathcal{G}^{f}_{3/2},\text{no-ghost}}(x) &= 4 \bar{F}_{\mu\nu}\bar{F}^{\mu\nu},\\
		(4\pi)^2 {a_4}^{\mathcal{G}^{f}_{3/2},\text{no-ghost}}(x) &=-\frac{1}{144}\left(41 R_{\mu\nu\rho\sigma}R^{\mu\nu\rho\sigma}-128 R_{\mu\nu}R^{\mu\nu} \right).
	\end{split}
\end{equation}
Additionally, one needs to include the contribution of the three minimally-coupled spin-1/2 Majorana ghosts in the action \eqref{Gx}. The combined Seeley-DeWitt coefficients for the ghosts fields are $a_{2n}^{\mathcal{G}^{f}_{3/2},\text{ghost}}=-3a_{2n}^{1/2}$, where the single spinor contribution $a_{2n}^{1/2}$ can be accessed from the formulae \eqref{cf} (set $n_{1/2}=1,n_0=n_1=n_{3/2} = 0$).
%We then aim to evaluate the Seeley-DeWitt contribution of the three minimally-coupled spin-1/2 Majorana ghosts in the action \eqref{Gx}. Their combined Seeley-DeWitt contribution to the fermionic-sector is $a_{2n}^{\mathcal{G}^{f}_{3/2},\text{ghost}}=-3a_{2n}^{1/2}$, where the single spinor contribution $a_{2n}^{1/2}$ can be obtained from the formulae \eqref{cf} (set $n_{1/2}=1,n_0=n_1=n_{3/2} = 0$). 
Here the minus sign reflects that the ghost fields counter the conventional spin-statistics of real fields. Thus, the net contribution of the fermionic-sector ($\mathcal{G}^{f}_{3/2}$) is %$a_{2n}^{\mathcal{G}^{f}_{3/2}}=a_{2n}^{\mathcal{G}^{f}_{3/2},\text{no-ghost}}+ a_{2n}^{\mathcal{G}^{f}_{3/2},\text{ghost}}$. 
\begin{equation}
	a_{2n}^{\mathcal{G}^{f}_{3/2}}=a_{2n}^{\mathcal{G}^{f}_{3/2},\text{no-ghost}}+ a_{2n}^{\mathcal{G}^{f}_{3/2},\text{ghost}}.
\end{equation}
On the other hand, the Seeley-DeWitt coefficients of the bosonic-sector ($\mathcal{G}^{b}_{3/2}$) with two minimally-coupled vector fields are obtained as
\begin{equation}\label{G14}
	\begin{split}
		&(4\pi)^2 {a_0}^{\mathcal{G}^{b}_{3/2}}(x) = 4,\\
		&(4\pi)^2 {a_2}^{\mathcal{G}^{b}_{3/2}}(x) = 0,\\
		&(4\pi)^2 {a_4}^{\mathcal{G}^{b}_{3/2}}(x) =-\frac{1}{90}\left(13 R_{\mu\nu\rho\sigma}R^{\mu\nu\rho\sigma}-88 R_{\mu\nu}R^{\mu\nu} \right).
	\end{split}
\end{equation}
Adding up the contributions $a_{2n}^{\mathcal{G}^{f}_{3/2}}$ and $a_{2n}^{\mathcal{G}^{b}_{3/2}}$ necessarily lead to the first three Seeley-DeWitt coefficients of the whole gravitino multiplet ($\mathcal{G}_{3/2}$),
\begin{equation}\label{G15}
	\begin{split}
		&(4\pi)^2 a^{\mathcal{G}_{3/2}}_0(x) = 0,\\
		&(4\pi)^2 a^{\mathcal{G}_{3/2}}_2(x) = 4 \bar{F}_{\mu\nu}\bar{F}^{\mu\nu},\\
		&(4\pi)^2 a^{\mathcal{G}_{3/2}}_4(x) =-\frac{11}{24}\left( R_{\mu\nu\rho\sigma}R^{\mu\nu\rho\sigma}-4 R_{\mu\nu}R^{\mu\nu} \right).
	\end{split}
\end{equation}
Finally, we employ the $a^{\mathcal{G}_{3/2}}_4(x)$ data into the formula \eqref{lc6} and find the logarithmic correction contribution of a single $\mathcal{N}=2$ gravitino multiplet to the entropy of extremal Kerr-Newman, Kerr and Reissner-Nordstr\"om black holes (for the limits \eqref{lc13}, \eqref{lc14} and \eqref{lc15}, respectively). The results are
\begin{equation}\label{G16}
	\Delta S^{\mathcal{G}_{3/2}}_{\text{BH}} =
	\begin{cases}
		-\frac{11}{12}\frac{(2{b^\prime}^6+5{b^\prime}^4+4{b^\prime}^2+1)}{({b^\prime}^2+1)^2(2{b^\prime}^2+1)}\thinspace \text{ln}\thinspace \mathcal{A}_{H} & \text{for Kerr-Newman},\\
		-\frac{11}{12}\thinspace \text{ln}\thinspace \mathcal{A}_{H} & \text{for Kerr},\\
		-\frac{11}{12}\thinspace \text{ln}\thinspace \mathcal{A}_{H} & \text{for Reissner-Nordstr\"om}.
	\end{cases}
\end{equation}

%-----------------------------------------------------------------------------------------------------------------
\subsection{$\mathcal{N}=2,d=4$ EMSGT: vector multiplet}\label{Vector}
The field contents of the vector multiplet ($\mathcal{G}_{1}$) of a $\mathcal{N}=2,d=4$ EMSGT can be distributed into two sectors --
\begin{itemize}
	\item[] \textbf{Bosonic sector ($\mathcal{G}^{b}_{1}$):} one complex scalar field $\phi_c$ is non-minimally coupled to a vector field $a_\mu$ through the background graviphoton field strength $\bar{F}^{\mu\nu}$,
	\item[] \textbf{Fermionic sector ($\mathcal{G}^{f}_{1}$):} two minimally-coupled Majorana gauginos.
\end{itemize}
To begin with the bosonic-sector, our immediate task is to open up the complex scalar field $\phi_c$ into its real and imaginary parts ($\phi_R$ and $\phi_I$, respectively) \cite{Sen:2012qq},
\begin{equation}\label{V1}
	\phi_c = \frac{1}{2}(\phi_R-i\phi_I).
\end{equation}
The quadratic order fluctuated bosonic-sector action for the fluctuations $\tilde{\xi}_m = \lbrace  a_\mu,\phi_R, \phi_I \rbrace$ is expressed as 
\begin{equation}\label{V2}
	\begin{split}
		\delta^2\mathcal{S}_{\mathcal{G}^{b}_{1}} &= \int \mathrm{d}^4x \sqrt{\text{det}\thinspace \bar{g}}\thinspace \mathscr{L}_{\mathcal{G}^{b}_{1}},\\
		\mathscr{L}_{\mathcal{G}^{b}_{1}} &= -\frac{1}{32}f_{\mu\nu}f^{\mu\nu}-\frac{1}{16}D_\rho\phi_R D^\rho\phi_R-\frac{1}{16}D_\rho\phi_I D^\rho\phi_I+\frac{1}{64}{\phi_R}^2\bar{H}_{\mu\nu}\bar{H}^{\mu\nu}\\
		& \quad-\frac{1}{16}{\phi_I}^2\bar{F}_{\mu\nu}\bar{F}^{\mu\nu}-\frac{i}{16}\phi_R \phi_I\bar{F}_{\mu\nu}\bar{H}^{\mu\nu}+\frac{i}{16}\phi_Rf_{\mu\nu}\bar{H}^{\mu\nu}+\frac{1}{8}\phi_If_{\mu\nu}\bar{F}^{\mu\nu},
	\end{split}
\end{equation}
where $f_{\mu\nu}=\partial_\mu a_\nu-\partial_\nu a_\mu$ is the strength tensor of the vector fluctuation $a_\mu$. To proceed with the strategy of \cref{P}, we customize the quadratic action \eqref{V2} further up to total derivatives. This involves the use of commutation relations \eqref{scalar} and \eqref{vector} as well as the inclusion of a gauge-fixing term to gauge fix the vector field fluctuated part $f_{\mu\nu}f^{\mu\nu}$,
\begin{align}\label{V4}
	-\frac{1}{16}\int \mathrm{d}^4x \sqrt{\text{det}\thinspace \bar{g}}\thinspace (D_\mu a^\mu) (D_\nu a^\nu).
\end{align}
The relevant Laplace-type kinetic operator $\Lambda$ holds the form,
%We then present the following Laplace-type form for the kinetic operator $\Lambda$, 
\begin{equation}\label{V3}
	\begin{split}
		\delta^2\mathcal{S}_{\mathcal{G}^{b}_{1}} &= \frac{1}{16}\int \mathrm{d}^4x \sqrt{\text{det}\thinspace \bar{g}} \thinspace \tilde{\xi}_m\Lambda^{mn} \tilde{\xi}_n,\\
		\tilde{\xi}_m\Lambda^{mn} \tilde{\xi}_n &= \bar{g}^{\mu\nu}a_\mu D^\rho D_\rho a_\nu + \phi_R D^\rho D_\rho\phi_R + \phi_I D^\rho D_\rho\phi_I-a_\mu R^{\mu\nu}a_\nu\\
		& \quad +\frac{1}{4}\phi_R \bar{H}_{\mu\nu}\bar{H}^{\mu\nu}\phi_R-\phi_I \bar{F}_{\mu\nu}\bar{F}^{\mu\nu}\phi_I-\frac{i}{2}\phi_R \bar{F}_{\mu\nu}\bar{H}^{\mu\nu}\phi_I-\frac{i}{2}\phi_I \bar{F}_{\mu\nu}\bar{H}^{\mu\nu}\phi_R\\
		&\quad + i a_\mu\bar{H}^{\mu\rho} (D_\rho \phi_R) +i \phi_R \bar{H}^{\rho\nu}(D_\rho a_\nu)+ 2 a_\mu\bar{F}^{\mu\rho} (D_\rho \phi_I) +2 \phi_I \bar{F}^{\rho\nu}(D_\rho a_\nu),
	\end{split}
\end{equation}
and provides the results for the matrices $I$, $P$ and $\omega^\rho\thinspace (= \frac{1}{2}N^\rho)$, 
\begin{equation}\label{V5}
	\begin{split}
		\tilde{\xi}_m I^{mn}\tilde{\xi}_n &=  a_\mu \bar{g}^{\mu\nu}a_\nu+\phi_R \phi_R + \phi_I\phi_I,\\
		\tilde{\xi}_m P^{mn}\tilde{\xi}_n &= -a_\mu R^{\mu\nu}a_\nu +\frac{1}{4}\phi_R \bar{H}_{\mu\nu}\bar{H}^{\mu\nu}\phi_R-\phi_I \bar{F}_{\mu\nu}\bar{F}^{\mu\nu}\phi_I\\
		&\qquad -\frac{i}{2}\phi_R \bar{F}_{\mu\nu}\bar{H}^{\mu\nu}\phi_I-\frac{i}{2}\phi_I \bar{F}_{\mu\nu}\bar{H}^{\mu\nu}\phi_R,\\
		\tilde{\xi}_m {(\omega^\rho)}^{mn}\tilde{\xi}_n &= \frac{i}{2} a_\mu\bar{H}^{\mu\rho} \phi_R +\frac{i}{2} \phi_R \bar{H}^{\rho\nu} a_\nu+  a_\mu\bar{F}^{\mu\rho} \phi_I + \phi_I \bar{F}^{\rho\nu} a_\nu. 
	\end{split}
\end{equation}
Consequently, we can now utilize the above data into the formulae \eqref{p6} and \eqref{p3} for deriving the expressions of $E$ and $\Omega_{\rho\sigma}$. The results are
\begin{equation}\label{V6}
	\begin{split}
		&\tilde{\xi}_m E^{mn}\tilde{\xi}_n =  a_\mu\Big\lbrace -R^{\mu\nu}+\bar{F}^{\mu\rho}\bar{F^\nu}_\rho-\frac{1}{4}\bar{H}^{\mu\rho}\bar{H^\nu}_\rho\Big\rbrace a_\nu,\\
		&\tilde{\xi}_m (\Omega_{\rho\sigma})^{mn}\tilde{\xi}_n = a_\mu\Big\lbrace {R^{\mu\nu}}_{\rho\sigma}+ \left({\bar{F^\mu}}_{\rho}{\bar{F_{\sigma}}}^\nu-{\bar{F^\mu}}_{\sigma}{\bar{F_{\rho}}}^\nu\right)-\frac{1}{4}\left({\bar{H^\mu}}_{\rho}{\bar{H_{\sigma}}}^\nu-{\bar{H^\mu}}_{\sigma}{\bar{H_{\rho}}}^\nu\right)\Big\rbrace a_\nu \\
		&\qquad\qquad\enspace +\frac{i}{2}a_\mu (D^\mu \bar{H}_{\rho\sigma})\phi_R-\frac{i}{2}\phi_R (D^\nu \bar{H}_{\rho\sigma})a_\nu+a_\mu (D^\mu \bar{F}_{\rho\sigma})\phi_I-\phi_I (D^\nu \bar{F}_{\rho\sigma})a_\nu,
	\end{split}
\end{equation}
where simplifications were done with the help of Maxwell-Bianchi equations.\footnote{See \cref{AA}.} We, therefore, compute the following needful trace values (see \cref{BB} for details)
\begin{equation}\label{V7}
	\begin{split}
		&\text{tr}(I) = 4+1+1=6, \\
		&\text{tr}(E) = 0, \\
		&\text{tr}(E^2) = 0,\\
		&\text{tr}(\Omega_{\rho\sigma}\Omega^{\rho\sigma}) = -R_{\mu\nu\rho\sigma}R^{\mu\nu\rho\sigma}+ 2R_{\mu\nu}R^{\mu\nu},
	\end{split}
\end{equation}
which allow us to find the first three Seeley-DeWitt coefficients for the bosonic-sector (without the ghost contribution) with the help of the formulae \eqref{p7},
\begin{equation}\label{V8}
	\begin{split}
		(4\pi)^2 {a_0}^{\mathcal{G}^{b}_{1},\text{no-ghost}}(x) &= 6,\\
		(4\pi)^2 {a_2}^{\mathcal{G}^{b}_{1},\text{no-ghost}}(x) &= 0,\\
		(4\pi)^2 {a_4}^{\mathcal{G}^{b}_{1},\text{no-ghost}}(x) &=-\frac{1}{60}\left(3 R_{\mu\nu\rho\sigma}R^{\mu\nu\rho\sigma}-8 R_{\mu\nu}R^{\mu\nu} \right).
	\end{split}
\end{equation}
Here $\chi=1$ is chosen to account for the vector and scalar fluctuations. Also, the gauge-fixing term \eqref{V4} produces two scalar ghost fields $b$ and $c$ via the Lagrangian \cite{Banerjee:2011oo},
\begin{align}\label{V9}
	\mathscr{L}_{\mathcal{G}^{b}_{1},\thinspace\text{ghost}} = b D^\rho D_\rho c,
\end{align}
The above scalar ghosts are minimally-coupled and their Seeley-DeWitt contribution is thus $a_{2n}^{\mathcal{G}^{b}_{1},\text{ghost}}=-2a_{2n}^s$, where the single scalar field results $a_{2n}^s$ can be accessed from the formulae \eqref{cf} (set $n_0=1,n_1=n_{1/2}=n_{3/2}=0$). 
%providing a Seeley-DeWitt contribution $a_{2n}^{\mathcal{G}^{b}_{1},\text{ghost}}=-2a_{2n}^s$, where the single scalar field contribution $a_{2n}^s$ can be accessed from the formulae \eqref{cf} (set $n_0=1,n_1=n_{1/2}=n_{3/2}=0$). 
Hence, the net Seeley-DeWitt contribution of the bosonic-sector ($\mathcal{G}^{b}_{1}$) becomes
\begin{equation}
	a_{2n}^{\mathcal{G}^{b}_{1}}=a_{2n}^{\mathcal{G}^{b}_{1},\text{no-ghost}} + a_{2n}^{\mathcal{G}^{b}_{1},\text{ghost}}.
\end{equation}
On the other hand, the fermionic-sector ($\mathcal{G}^{f}_{1}$) casts two minimally-coupled spin-1/2 gaugino fields and provides,
\begin{equation}\label{V10}
	\begin{split}
		(4\pi)^2 {a_0}^{\mathcal{G}^{f}_{1}}(x) &= -4,\\
		(4\pi)^2 {a_2}^{\mathcal{G}^{f}_{1}}(x) &= 0,\\
		(4\pi)^2 {a_4}^{\mathcal{G}^{f}_{1}}(x) &=\frac{1}{360}\left(7 R_{\mu\nu\rho\sigma}R^{\mu\nu\rho\sigma}+8 R_{\mu\nu}R^{\mu\nu} \right).
	\end{split}
\end{equation}
By combining the contributions of bosonic and fermionic sectors, the Seeley-DeWitt coefficients of the full vector multiplet ($\mathcal{G}_{1}$) are obtained as
\begin{equation}\label{V11}
	\begin{split}
		&(4\pi)^2 a^{\mathcal{G}_{1}}_0(x) = 0,\\
		&(4\pi)^2 a^{\mathcal{G}_{1}}_2(x) = 0,\\
		&(4\pi)^2 a^{\mathcal{G}_{1}}_4(x) =-\frac{1}{24}\left(R_{\mu\nu\rho\sigma}R^{\mu\nu\rho\sigma}-4 R_{\mu\nu}R^{\mu\nu} \right).
	\end{split}
\end{equation}
Finally, the $a^{\mathcal{G}_{1}}_4(x)$ result needs to be utilized in the set-up of \cref{lc} for calculating the logarithmic entropy correction contributions of a single $\mathcal{N}=2$ vector multiplet. The results are
%We end up with the contribution of a single vector multiplet in logarithmic correction to the entropy of extremal Kerr-Newman family of black holes, by embedding the Seeley-DeWitt data \eqref{V11} into the prescription of \cref{lc}, are
\begin{equation}\label{V12}
	\Delta S^{\mathcal{G}_{1}}_{\text{BH}} =
	\begin{cases}
		-\frac{1}{12}\frac{(2{b^\prime}^6+5{b^\prime}^4+4{b^\prime}^2+1)}{({b^\prime}^2+1)^2(2{b^\prime}^2+1)}\thinspace \text{ln}\thinspace \mathcal{A}_{H} & \text{for Kerr-Newman},\\
		-\frac{1}{12}\thinspace \text{ln}\thinspace \mathcal{A}_{H} & \text{for Kerr},\\
		-\frac{1}{12}\thinspace \text{ln}\thinspace \mathcal{A}_{H} & \text{for Reissner-Nordstr\"om}.
	\end{cases}
\end{equation}
%-------------------------------------------------------------------------------------------------  
\subsection{$\mathcal{N}=2,d=4$ EMSGT: hyper multiplet}\label{hyper}
The field contents of the hyper multiplet ($\mathcal{G}_{1/2}$) of a $\mathcal{N}=2,d=4$ EMSGT is also distributed into bosonic and fermionic sectors --
\begin{itemize}
	\item[] \textbf{Bosonic sector ($\mathcal{G}^{b}_{1/2}$):} four minimally-coupled real scalars,
	\item[] \textbf{Fermionic sector ($\mathcal{G}^{f}_{1/2}$):} two spin-1/2 Weyl spinors $\lbrace\zeta_1,\zeta_2\rbrace$ are coupled non-minimally to their opposite chiralities $\lbrace\zeta^1,\zeta^2\rbrace$ via the background graviphoton strength $\bar{F}^{\mu\nu}$.
\end{itemize}
We begin with the fermionic-sector and define a Majorana spinor corresponding to each Weyl fermions by combining with their chiral partner \cite{Sen:2012qq}, 
\begin{equation}\label{H1}
	\lambda_1 = \zeta_1 + \zeta^1, \thinspace \lambda_2 = \zeta_2 + \zeta^2,
\end{equation}
where the newly formed Majorana spinors $\lambda_1$ and $\lambda_2$ are coupled through the following action (up to quadratic order)
\begin{equation}\label{H2}
	\begin{gathered}
		\delta^2\mathcal{S}_{\mathcal{G}^{f}_{1/2}} = -\frac{1}{2}\int \mathrm{d}^4x \sqrt{\text{det}\thinspace \bar{g}} \thinspace  \mathscr{L}_{\mathcal{G}^{f}_{1/2}},\\
		\mathscr{L}_{\mathcal{G}^{f}_{1/2}} = \bar{\lambda}_1\gamma^\rho D_\rho \lambda_1+\bar{\lambda}_2\gamma^\rho D_\rho \lambda_2- \frac{1}{4} \bar{\lambda}_1\bar{F}_{\alpha\beta}\gamma^\alpha\gamma^\beta\lambda_2+\frac{1}{4} \bar{\lambda}_2\gamma^\alpha\gamma^\beta\bar{F}_{\alpha\beta}\lambda_1.
	\end{gathered}
\end{equation}
By expressing $\mathscr{L}_{\mathcal{G}^{f}_{1/2}}=\tilde{\xi}_m\slashed{D}^{mn} \tilde{\xi}_n$, one finds the form of relevant first-order Dirac-type operator $\slashed{D}$ as
\begin{equation}\label{H3}
	\tilde{\xi}_m\slashed{D}^{mn} \tilde{\xi}_n = i\bar{\lambda}_1\gamma^\rho D_\rho \lambda_1+i\bar{\lambda}_2\gamma^\rho D_\rho \lambda_2- \frac{i}{4} \bar{\lambda}_1\hat{\mathcal{F}}\lambda_2+\frac{i}{4} \bar{\lambda}_2\hat{\mathcal{F}}\lambda_1.
\end{equation}
Note that the operator $\slashed{D}$ is Hermitian since $\gamma^\mu$, $\hat{\mathcal{F}}=\bar{F}_{\alpha\beta}\gamma^\alpha\gamma^\beta$ are Hermitian and $(\gamma^\rho D_\rho)^\dagger = -\gamma^\rho D_\rho$. From here, one needs to follow up the typical fermionic strategy of structuring a second-order operator $\Lambda$ by squaring the Dirac-type operator form \eqref{H3}:
\begin{align}\label{H4}
	\tilde{\xi}_m \Lambda^{mn}\tilde{\xi}_n &= \tilde{\xi}_m (\slashed{D}^{mp})^\dagger{\slashed{D}_p}^n\tilde{\xi}_n \nonumber\\
	&= -\bar{\lambda}_1\bigg\lbrace\gamma^\rho\gamma^\sigma D_\rho D_\sigma- \frac{1}{4} \hat{\mathcal{F}} \hat{\mathcal{F}}\bigg\rbrace\lambda_1-\bar{\lambda}_2\bigg\lbrace\gamma^\rho\gamma^\sigma D_\rho D_\sigma- \frac{1}{4} \hat{\mathcal{F}} \hat{\mathcal{F}}\bigg\rbrace\lambda_2 \nonumber\\
	&\qquad+\bar{\lambda}_1\bigg\lbrace \frac{1}{2}\gamma^\rho(D_\rho \hat{\mathcal{F}}+
	\hat{\mathcal{F}}D_\rho) + \frac{1}{2}\hat{\mathcal{F}}\gamma^\rho D_\rho\bigg\rbrace\lambda_2 \nonumber\\
	&\qquad-\bar{\lambda}_2\bigg\lbrace \frac{1}{2} \hat{\mathcal{F}}\gamma^\rho D_\rho + \frac{1}{2} \gamma^\rho (D_\rho \hat{\mathcal{F}} + \hat{\mathcal{F}} D_\rho)\bigg\rbrace \lambda_1.
\end{align}
The above operator form is not of Laplace-type as depicted in the schematic \eqref{p1}. But we can repair it using the operator identity \eqref{idd2} and achieve the desired Laplace-type form of $\Lambda$ as
\begin{align}\label{H5}
	\tilde{\xi}_m \Lambda^{mn}\tilde{\xi}_n &= -\mathbb{I}_4 \bar{\lambda}_1 D^\rho D_\rho \lambda_1-\mathbb{I}_4 \bar{\lambda}_2 D^\rho D_\rho \lambda_2 +\frac{1}{4}\bar{\lambda}_1\hat{\mathcal{F}}\hat{\mathcal{F}}\lambda_1 \nonumber\\
	&\quad +\frac{1}{4}\bar{\lambda}_2\hat{\mathcal{F}}\hat{\mathcal{F}}\lambda_2+\frac{1}{2}\bar{\lambda}_1\gamma^\rho (D_\rho \hat{\mathcal{F}})\lambda_2-\frac{1}{2}\bar{\lambda}_2\gamma^\rho (D_\rho \hat{\mathcal{F}})\lambda_1 \nonumber\\
	&\quad  +\frac{1}{2}\bar{\lambda}_1 (\gamma^\rho \hat{\mathcal{F}}+ \hat{\mathcal{F}}\gamma^\rho)D_\rho\lambda_2-\frac{1}{2}\bar{\lambda}_2 (\gamma^\rho \hat{\mathcal{F}}+ \hat{\mathcal{F}}\gamma^\rho)D_\rho\lambda_1.
\end{align}
It is now quite easy to extract out $I$, $P$ and $\omega^\rho$ ($=\frac{1}{2}N^\rho$),
{
	\allowdisplaybreaks
	%\begin{equation}
	\begin{align}\label{H6}
		\tilde{\xi}_m I^{mn}\tilde{\xi}_n &= \bar{\lambda}_1\mathbb{I}_4 \lambda_1+ \bar{\lambda}_2\mathbb{I}_4 \lambda_2, \nonumber\\
		\tilde{\xi}_m P^{mn}\tilde{\xi}_n &= -\frac{1}{4}\bar{\lambda}_1\gamma^\alpha\gamma^\beta\gamma^\theta\gamma^\phi\bar{F}_{\alpha\beta}\bar{F}_{\theta\phi}\lambda_1-\frac{1}{4}\bar{\lambda}_2\gamma^\alpha\gamma^\beta\gamma^\theta\gamma^\phi\bar{F}_{\alpha\beta}\bar{F}_{\theta\phi}\lambda_2 \nonumber\\
		&\quad -\frac{1}{2}\bar{\lambda}_1\gamma^\rho\gamma^\alpha\gamma^\beta (D_\rho \bar{F}_{\alpha\beta})\lambda_2+\frac{1}{2}\bar{\lambda}_2\gamma^\rho\gamma^\alpha\gamma^\beta (D_\rho \bar{F}_{\alpha\beta})\lambda_1, \\
		\tilde{\xi}_m {(\omega^\rho)}^{mn}\tilde{\xi}_n &= -\bar{\lambda}_1\bigg\lbrace \frac{1}{4}\gamma^\rho\gamma^\alpha\gamma^\beta \bar{F}_{\alpha\beta}+ \frac{1}{4}\gamma^\alpha\gamma^\beta\gamma^\rho\bar{F}_{\alpha\beta}\bigg\rbrace\lambda_2 \nonumber\\
		&\quad +\bar{\lambda}_2\bigg\lbrace \frac{1}{4}\gamma^\rho\gamma^\alpha\gamma^\beta \bar{F}_{\alpha\beta}+ \frac{1}{4}\gamma^\alpha\gamma^\beta\gamma^\rho\bar{F}_{\alpha\beta}\bigg\rbrace\lambda_1. \nonumber
\end{align}}
and then jot down the expressions for the matrices $\Omega_{\rho\sigma}$ and $E$ (using \cref{p3,p6}). The simplified results are
{
	\allowdisplaybreaks
	%\begin{subequations}
	\begin{align}\label{H7}
		\tilde{\xi}_m E^{mn}\tilde{\xi}_n &= \frac{1}{16}\bar{\lambda}_1 \gamma^\rho\gamma^\alpha\gamma^\beta\gamma^\theta\gamma^\phi\gamma_\rho \bar{F}_{\alpha\beta}\bar{F}_{\theta\phi}\lambda_1 + \frac{1}{16}\bar{\lambda}_2 \gamma^\rho\gamma^\alpha\gamma^\beta\gamma^\theta\gamma^\phi\gamma_\rho \bar{F}_{\alpha\beta}\bar{F}_{\theta\phi}\lambda_2,\nonumber\\
		\tilde{\xi}_m {(\Omega_{\rho\sigma})}^{mn}\tilde{\xi}_n &= \bar{\lambda}_1\bigg\lbrace \frac{1}{4}\gamma^\alpha\gamma^\beta R_{\rho\sigma\alpha\beta}-\frac{1}{16}\Big((\gamma_\rho\gamma^\alpha\gamma^\beta+\gamma^\alpha\gamma^\beta\gamma_\rho) \nonumber\\
		&\qquad\quad \times (\gamma_\sigma\gamma^\theta\gamma^\phi+\gamma^\theta\gamma^\phi\gamma_\sigma)\bar{F}_{\alpha\beta}\bar{F}_{\theta\phi}-(\rho\leftrightarrow\sigma)\Big)\bigg\rbrace \lambda_1 \nonumber\\
		& \quad +\bar{\lambda}_2\bigg\lbrace \frac{1}{4}\gamma^\alpha\gamma^\beta R_{\rho\sigma\alpha\beta}-\frac{1}{16}\Big((\gamma_\rho\gamma^\alpha\gamma^\beta+\gamma^\alpha\gamma^\beta\gamma_\rho) \\
		&\qquad\quad \times (\gamma_\sigma\gamma^\theta\gamma^\phi+\gamma^\theta\gamma^\phi\gamma_\sigma)\bar{F}_{\alpha\beta}\bar{F}_{\theta\phi}-(\rho\leftrightarrow\sigma)\Big)\bigg\rbrace \lambda_2 \nonumber\\
		& \quad -\bar{\lambda}_{1}\bigg\lbrace \frac{1}{4}(\gamma_\sigma\gamma^\alpha\gamma^\beta + \gamma^\alpha\gamma^\beta\gamma_\sigma)(D_\rho\bar{F}_{\alpha\beta})-(\rho\leftrightarrow\sigma)\bigg\rbrace \lambda_2 \nonumber\\
		& \quad +\bar{\lambda}_{2}\bigg\lbrace \frac{1}{4}(\gamma_\sigma\gamma^\alpha\gamma^\beta + \gamma^\alpha\gamma^\beta\gamma_\sigma)(D_\rho\bar{F}_{\alpha\beta})-(\rho\leftrightarrow\sigma)\bigg\rbrace \lambda_1.\nonumber
\end{align}}
The absence of any cross-terms involving $\lambda_1$ and $\lambda_2$ in the expression of $E$ is not a direct consequence -- the type of coupling between $\lambda_1$ and $\lambda_2$ makes the cross-terms to vanish while simplifying $E$ with the help of the relations \eqref{H6}. As usual, our ultimate task is to calculate the trace values (refer to \cref{BB}),
\begin{equation}\label{H8}
	\begin{split}
		\text{tr}(I) &=4+4= 8,\\
		\text{tr}(E) &= -4\bar{F}_{\mu\nu}\bar{F}^{\mu\nu},\\
		\text{tr}(E^2) &= 4(\bar{F}_{\mu\nu}\bar{F}^{\mu\nu})^2-2 R_{\mu\nu}R^{\mu\nu},\\
		\text{tr}(\Omega_{\rho\sigma}\Omega^{\rho\sigma})  &= - R_{\mu\nu\rho\sigma}R^{\mu\nu\rho\sigma}+16 R_{\mu\nu}R^{\mu\nu} -24(\bar{F}_{\mu\nu}\bar{F}^{\mu\nu})^2,
	\end{split}
\end{equation}
and insert them into the formulae \eqref{p7} for finding the Seeley-DeWitt coefficients of the fermionic-sector ($\mathcal{G}^{f}_{1/2}$):
\begin{equation}\label{H9}
	\begin{split}
		(4\pi)^2 a^{\mathcal{G}^{f}_{1/2}}_0(x) &= -4,\\
		(4\pi)^2 a^{\mathcal{G}^{f}_{1/2}}_2(x) &= 2 \bar{F}_{\mu\nu}\bar{F}^{\mu\nu},\\
		(4\pi)^2 a^{\mathcal{G}^{f}_{1/2}}_4(x) &=\frac{1}{360}\left(7 R_{\mu\nu\rho\sigma}R^{\mu\nu\rho\sigma}-52 R_{\mu\nu}R^{\mu\nu} \right).
	\end{split}
\end{equation}
The above result considered $\chi=-1/2$ to account for the Majorana fermions $\lambda_{1}$ and $\lambda_{2}$.
On the other hand, the bosonic-sector ($\mathcal{G}^{b}_{1/2}$) has four minimally-coupled scalars and the corresponding Seeley-DeWitt contributions $a_{2n}^{\mathcal{G}^{b}_{1/2}}$ can be accessed by setting $n_0=4,n_1=n_{1/2}=n_{3/2}=0$ in the formulae \eqref{cf}. Adding up the $\mathcal{G}^{f}_{1/2}$ and $\mathcal{G}^{b}_{1/2}$ contributions, the first three Seeley-DeWitt coefficients of the hyper multiplet ($\mathcal{G}_{1/2}$) are presented as
\begin{equation}\label{H11}
	\begin{split}
		&(4\pi)^2 a^{\mathcal{G}_{1/2}}_0(x) = 0,\\
		&(4\pi)^2 a^{\mathcal{G}_{1/2}}_2(x) = 2 \bar{F}_{\mu\nu}\bar{F}^{\mu\nu},\\
		&(4\pi)^2 a^{\mathcal{G}_{1/2}}_4(x) =\frac{1}{24}\left( R_{\mu\nu\rho\sigma}R^{\mu\nu\rho\sigma}-4 R_{\mu\nu}R^{\mu\nu} \right).
	\end{split}
\end{equation}
The above Seeley-DeWitt data, especially the $a^{\mathcal{G}_{1/2}}_4(x)$, are all we need (in the structure of \cref{lc}) for calculating the contribution of a single $\mathcal{N}=2$ hyper multiplet in logarithmic corrections to the entropy of extremal Kerr-Newman, Kerr and Reissner-Nordstr\"om black holes. The results are 
\begin{equation}\label{H12}
	\Delta S^{\mathcal{G}_{1/2}}_{\text{BH}} =
	\begin{cases}
		\frac{1}{12}\frac{(2{b^\prime}^6+5{b^\prime}^4+4{b^\prime}^2+1)}{({b^\prime}^2+1)^2(2{b^\prime}^2+1)}\thinspace \text{ln}\thinspace \mathcal{A}_{H} & \text{for Kerr-Newman},\\
		\frac{1}{12}\thinspace \text{ln}\thinspace \mathcal{A}_{H} & \text{for Kerr},\\
		\frac{1}{12}\thinspace \text{ln}\thinspace \mathcal{A}_{H} & \text{for Reissner-Nordstr\"om}.
	\end{cases}
\end{equation}

All the Seeley-DeWitt results recorded in \cref{S4,G15,V11,H11} are consistent with the eigenfunction expansion technique results \cite{Sen:2012qq,Keeler:2014nn} for $AdS_2\times S^2$ background and also with the results of hybrid field redefinition approach \cite{Charles:2015nn}. Interestingly, the $a_4(x)$ results are noticed to be fully electro-magnetic duality invariant, i.e., they do not include the terms proportional to $R_{\mu\nu\rho\sigma}\bar{F}^{\mu\nu}\bar{F}^{\rho\sigma}$ and $(\bar{F}_{\mu\nu}\bar{F}^{\mu\nu})^2$. 
%\hl{The $a_4(x)$ coefficients are fully expressed in background Riemannian and Ricci invariants and hence provide a global platform to investigate logarithmic correction for all the Kerr-Newman family of black holes.}
It is also essential to mention that the logarithmic corrections \eqref{G16}, \eqref{V12} and \eqref{H12} are obtained by incorporating only the $\mathcal{C}_{\text{local}}$ contribution for the corresponding multiplets. This is because the $\mathcal{C}_{\text{zm}}$ contribution \eqref{lc7} is fixed for a particular $\mathcal{N}=2,d=4$ EMSGT and already added in the supergravity multiplet results \eqref{S5}, while the $\mathcal{C}_{\text{local}}$ contribution is different for individual multiplets. The extremal Reissner-Nordstr\"om logarithmic correction results are in accordance with \cite{Sen:2012qq,Keeler:2014nn}.

%=====================================================================
%----------------------------------------------------------------------
\section{Discussions}\label{diss}
In this section, we aim to provide some crucial applications of the results obtained in \cref{mul} for different $\mathcal{N} = 2$ multiplets. %This includes \hl{finding} logarithmic corrections to the entropy of extremal Kerr-Newman, Kerr and Reissner-Nordstr\"om black holes in a matter coupled $\mathcal{N} = 2,d=4$ EMSGT, in a matter coupled $\mathcal{N}=1,d=4$ \hl{EMSGT} as consistent decomposition of the $\mathcal{N}=2,d=4$ EMSGT multiplets and in matter coupled $\mathcal{N} \geq 3,d=4$ EMSGTs by decomposing them into $\mathcal{N}=2,d=4$ EMSGT multiplets. 
This includes finding logarithmic corrections to the entropy of extremal Kerr-Newman, Kerr and Reissner-Nordstr\"om black holes in the matter coupled $\mathcal{N} = 2$ EMSGT, $\mathcal{N}=1$ EMSGT as truncation of $\mathcal{N}=2$ multiplets and $\mathcal{N} \geq 3$ EMSGTs by decomposing into $\mathcal{N}=2$ multiplets. In this way, we achieve our central goal of carrying out the logarithmic entropy correction program for extremal black holes in all $\mathcal{N} \geq 1$ matter coupled EMSGTs. %In comparison, we also extract out the logarithmic correction results for the Kerr-Newman family of non-extremal black holes in the same matter coupled $\mathcal{N} \geq 1,d=4$ EMSGTs. 
Additionally, we explore the same results for the non-extremal Kerr-Newman family of black holes and compare them with the extremal results. To achieve this goal, we follow the Euclidean gravity approach \cite{Sen:2013ns} where one needs to analyze the coefficients of trace anomalies from the $a_4(x)$ results. %This non-extremal part is not our central motive in this paper, and it does not base on the quantum entropy function formalism used for the extremal black holes. 
Note that the extremal Reissner-Nordstr\"om black holes are $\frac{1}{\mathcal{N}}$-BPS solutions in the $\mathcal{N} \geq 2,d=4$ EMSGTs, while all other black holes are non-BPS in the $\mathcal{N} \geq 1,d=4$ EMSGTs. We end up discussing some concluding remarks about our results.

\subsection{Logarithmic correction to the entropy of extremal black holes in the matter coupled $\mathcal{N} = 2,d=4$ EMSGT}\label{mc}
A pure or `minimal' $\mathcal{N} = 2,d=4$ EMSGT includes only a supergravity multiplet \cite{Karan:2019sk}. At any point, one can generalize it further by coupling additional matter multiplets such as $n_V$ number of vector multiplets and $n_H$ number of hyper multiplets to the supergravity multiplet. This generalized theory is called the matter coupled $\mathcal{N} = 2,d=4$ EMSGT in which several non-minimal couplings are distributed so that each multiplet behaves as a minimal one. The Seeley-DeWitt coefficients and logarithmic correction contributions of the $\mathcal{N}=2$ supergravity, vector and hyper multiplets have been evaluated and recorded in \cref{supergravity,Vector,hyper}, respectively.
Putting these results together, we present the form of $a_4(x)$ coefficient in the matter coupled $\mathcal{N}=2,d=4$ EMSGT as
\begin{equation}\label{mc1}
	(4\pi)^2 a^{\mathcal{N}=2}_4(x) =\frac{1}{24}\big(11-n_V+n_H\big) \big( R_{\mu\nu\rho\sigma}R^{\mu\nu\rho\sigma}-4 R_{\mu\nu}R^{\mu\nu} \big),
\end{equation}
followed by the logarithmic correction to the entropy of extremal Kerr-Newman, Kerr and Reissner-Nordstr\"om black holes,
\begin{equation}\label{mc2}
	\Delta S^{\mathcal{N}=2}_{\text{BH}} =
	\begin{cases}
		\left(\frac{1}{12}\big(11-n_V+n_H\big)\frac{(2{b^\prime}^6+5{b^\prime}^4+4{b^\prime}^2+1)}{({b^\prime}^2+1)^2(2{b^\prime}^2+1)}-2\right) \text{ln}\thinspace \mathcal{A}_{H} &\text{for Kerr-Newman},\\
		\frac{1}{12}\big(-13-n_V+n_H\big)\thinspace \text{ln}\thinspace \mathcal{A}_{H} &\text{for Kerr},\\
		\frac{1}{12}\big(23-n_V+n_H\big)\text{ln}\thinspace \mathcal{A}_{H} \hspace{-0.1in}&\text{for Reissner-Nordstr\"om}.
	\end{cases}
\end{equation}
The extremal Reissner-Nordstr\"om black hole result agrees with \cite{Sen:2012qq}, where calculations were done using the eigenfunction expansion method for $AdS_2\times S^2$ background. %The extremal Kerr-Newman and Kerr black hole results are original reports to the literature. 
The formulae \eqref{mc2} are actually the generalization of the `minimal' $\mathcal{N} = 2,d=4$ EMSGT results provided in \cite{Karan:2019sk}.
%----------------------------------------------------------------------------------------------------------------------

\subsection{Logarithmic correction to the entropy of extremal black holes in the matter coupled $\mathcal{N}=1,d=4$ EMSGT}\label{truncation}
Any general $\mathcal{N}=1$ EMSGT consists of four massless multiplets -- the graviton (${g}_2$), gravitino (${g}_{3/2}$), vector (${g}_1$) and chiral (${g}_{1/2}$) multiplets. The spin contents of these multiplets in $d=4$ space-time dimensions are recorded in \cref{tab2}. Analyzing the spin content of both \cref{tab1} and \cref{tab2}, it is possible to disintegrate the $\mathcal{N}=2,d=4$ EMSGT multiplets into the multiplets of $\mathcal{N}=1,d=4$ EMSGT as following
\begin{align}\label{t1}
	\begin{split}
		\mathcal{G}_{2} &= g_2 + g_{3/2},\\
		\mathcal{G}_{3/2} &= g_{3/2} + g_1,\\
		\mathcal{G}_{1} &= g_1 + g_{1/2},\\
		\mathcal{G}_{1/2} &= 2g_{1/2}. 
	\end{split}
\end{align}
This particular class of $\mathcal{N}=1$ EMSGT is purely obtained as consistent kinematic decomposition of $\mathcal{N}=2$ multiplets, where the interactions in different multiplets are constrained by some assumptions. However, there is no restriction on the multiplicity of any multiplet content. There must be at least one vector ($g_{1}$) and chiral ($g_{1/2}$) multiplets. Readers are referred to \cite{Ferrara:2012bp,Andrianopoli:2002gm} and references therein for all technical details and related assumptions for the reduction $\mathcal{N}=2 \to \mathcal{N}=1$. The $\mathcal{N}=2$ data from \cref{S5,G16,V12,H12} can help us solve the equalities \eqref{t1} and lead to logarithmic entropy correction contributions of the $\mathcal{N}=1,d=4$ multiplets as recorded in \cref{tab4}. In a matter coupled $\mathcal{N}=1$ EMSGT, a single supergravity multiplet is minimally coupled to $n_v$ number of vector multiplets and $n_c$ number of chiral multiplets. Therefore, the logarithmic corrections to the entropy of Kerr-Newman, Kerr and Reissner-Nordstr\"om black holes in the matter coupled $\mathcal{N}=1,d=4$ EMSGT (accordant with the truncation \eqref{t1}) are obtained as
\begin{equation}\label{t6}
	\Delta S^{\mathcal{N}=1}_{\text{BH}} =
	\begin{cases}
		\left(\frac{1}{24}\big(41-3n_v+n_c\big)\frac{(2{b^\prime}^6+5{b^\prime}^4+4{b^\prime}^2+1)}{({b^\prime}^2+1)^2(2{b^\prime}^2+1)}-2\right) \text{ln}\thinspace\mathcal{A}_{H} &\text{for Kerr-Newman},\\
		\frac{1}{24}\big(-7-3n_v+n_c\big)\thinspace \text{ln}\thinspace \mathcal{A}_{H} &\text{for Kerr},\\
		\frac{1}{24}\big(65-3n_v+n_c\big)\thinspace \text{ln}\thinspace \mathcal{A}_{H} &\text{for Reissner-Nordstr\"om}.
	\end{cases}
\end{equation}
The above extremal Reissner-Nordstr\"om result is consistent with \cite{Ferrara:2012bp}.
%The extremal Reissner-Nordstr\"om result is consistent with \cite{Ferrara:2012bp}, while the Kerr-Newman and Kerr results are latest to the literature.
{
	\renewcommand{\arraystretch}{1.4}
	\begin{table}[t]% h (here), t (top), b (bottom) and p (page of floats), ht (here top)
		\centering
		\hspace{-0.2in}
		\begin{tabular}{|>{\centering}p{1.3in}|>{\centering}p{.4in}|>{\centering}p{.5in}>{\centering}p{.5in}>{\centering}p{.5in}>{\centering}p{.5in}>{\centering}p{.5in}|}
			\hline
			\textbf{Multiplets}& \textbf{$s_{\text{max}}$} & {$\bm{s=2}$} & {$\bm{s=\frac{3}{2}}$} & {$\bm{s=1}$} & {$\bm{s=\frac{1}{2}}$} & {$\bm{s=0}$} \tabularnewline \hline\hline
			supergravity (${g}_2$)  & 2 &1 & 1& & & \tabularnewline 
			gravitino (${g}_{3/2}$)& $\frac{3}{2}$ & & 1&1 & & \tabularnewline 
			vector (${g}_1$)& 1 & & &1 &1 &  \tabularnewline 
			chiral (${g}_{1/2}$)& $\frac{1}{2}$ & & & &1 & 1+1 \tabularnewline \hline
		\end{tabular}
		\caption{Different multiplets and spin content of a $\mathcal{N}=1,d=4$ EMSGT.}\label{tab2}
	\end{table}
}
{
	\renewcommand{\arraystretch}{1.8}
	\begin{table}[t]% h (here), t (top), b (bottom) and p (page of floats), ht (here top)
		\centering
		\hspace{-0.2in}
		\begin{tabular}{|>{\centering}p{1.4in}|>{\centering}p{1.6in}>{\centering}p{.8in}>{\centering}p{.9in}|}
			\hline
			\textbf{$\mathcal{N}=1$ multiplet contributions} & \textbf{Kerr-Newman} & \textbf{Kerr} & \textbf{Reissner-Nordstr\"om} \tabularnewline \hline\hline
			$\Delta S^{g_{2}}_{\text{BH}}/\ln \mathcal{A}_{H}$& $\frac{41}{24}\frac{(2{b^\prime}^6+5{b^\prime}^4+4{b^\prime}^2+1)}{({b^\prime}^2+1)^2(2{b^\prime}^2+1)}-2$ & $-\frac{7}{24}$ & $\frac{65}{24}$ \tabularnewline 
			$\Delta S^{g_{3/2}}_{\text{BH}}/\ln \mathcal{A}_{H}$ & $-\frac{19}{24}\frac{(2{b^\prime}^6+5{b^\prime}^4+4{b^\prime}^2+1)}{({b^\prime}^2+1)^2(2{b^\prime}^2+1)}$ & $-\frac{19}{24}$ & $-\frac{19}{24}$ \tabularnewline 
			$\Delta S^{g_{1}}_{\text{BH}}/\ln \mathcal{A}_{H}$ & $-\frac{1}{8}\frac{(2{b^\prime}^6+5{b^\prime}^4+4{b^\prime}^2+1)}{({b^\prime}^2+1)^2(2{b^\prime}^2+1)}$ & $-\frac{1}{8}$ & $-\frac{1}{8}$  \tabularnewline 
			$\Delta S^{g_{1/2}}_{\text{BH}}/\ln \mathcal{A}_{H}$ &$\frac{1}{24}\frac{(2{b^\prime}^6+5{b^\prime}^4+4{b^\prime}^2+1)}{({b^\prime}^2+1)^2(2{b^\prime}^2+1)}$ & $\frac{1}{24}$ & $\frac{1}{24}$ \tabularnewline \hline
		\end{tabular}
		\caption{Logarithmic correction contributions of the multiplets of the decomposed $\mathcal{N}=1,d=4$ EMSGT.}\label{tab4}
	\end{table}
}
%-----------------------------------------------------------------------------------------
\subsection{Logarithmic correction to the entropy of extremal black holes in the matter coupled $\mathcal{N} \geq 3,d=4$ EMSGTs}\label{decomposition}
For any general $\mathcal{N} \geq 3,d=4$ EMSGT, the massless content of supergravity multiplet (${\mathcal{G}_2}^{\mathcal{N}\geq 3}$) includes one graviton, $\mathcal{N}$ number of gravitini and $\frac{1}{2}\mathcal{N}(\mathcal{N}-1)$ number of vector fields \cite{Freedman:2012xp}. At any point, it is possible to decompose them by distributing the graviton, two gravitini and one vector field
into a $\mathcal{N}=2$ supergravity multiplet ($\mathcal{G}_{2}$). Then the remaining content of the $\mathcal{N} \geq 3$ supergravity multiplet (${\mathcal{G}_2}^{\mathcal{N}\geq 3}$), vector multiplet (${\mathcal{G}_1}^{\mathcal{N}\geq 3}$) and hyper multiplet (${\mathcal{G}_{1/2}}^{\mathcal{N}\geq 3}$) can be decomposed into $(\mathcal{N}-2)$ number of $\mathcal{N}=2$ gravitino multiplets ($\mathcal{G}_{3/2}$) and a particular combination of $\mathcal{N}=2$ vector ($\mathcal{G}_{1}$) and $\mathcal{N}=2$ hyper ($\mathcal{G}_{1/2}$) multiplets.
% The supergravity multiplet (${\mathcal{G}_2}^{\mathcal{N}\geq 3}$) content of the $\mathcal{N} \geq 3,d=4$ EMSGTs includes one graviton and $\mathcal{N}$ number of gravitini (e.g., see \cite{Freedman:2012xp}). At any point, it is possible to decompose them by distributing the graviton and two gravitini into a $\mathcal{N}=2,d=4$ EMSGT supergravity multiplet ($\mathcal{G}_{2}$), and the remaining gravitini into $(\mathcal{N}-2)$ number of $\mathcal{N}=2,d=4$ EMSGT gravitino multiplets ($\mathcal{G}_{3/2}$). By carefully analyzing the spin content of the matter coupled $\mathcal{N} \geq 3,d=4$ EMSGTs, one can also decompose the remaining species, vector (${\mathcal{G}_1}^{\mathcal{N}\geq 3}$) and hyper (${\mathcal{G}_{1/2}}^{\mathcal{N}\geq 3}$) multiplets into a particular combination of $\mathcal{N}=2,d=4$ EMSGT vector ($\mathcal{G}_{1}$) and hyper ($\mathcal{G}_{1/2}$) multiplets. 
Collectively, one can report the following multiplet decompositions for the matter coupled $\mathcal{N} \geq 3,d=4$ EMSGTs\footnote{Since $\mathcal{N} \geq 9,d=4$ EMSGTs are deserts of more than one graviton and higher spin ($s \geq $5/2) interactions, the decomposition \eqref{d1} is valid for the $\mathcal{N} < 9,d=4$ EMSGTs only.}
\begin{align}\label{d1}
	\begin{gathered}
		{\mathcal{G}_2}^{\mathcal{N}\geq 3}+ {\mathcal{G}_1}^{\mathcal{N}\geq 3}+ {\mathcal{G}_{1/2}}^{\mathcal{N}\geq 3}=\mathcal{G}_{2} + (\mathcal{N}-2)\mathcal{G}_{3/2}+ n_V \mathcal{G}_{1} + n_H \mathcal{G}_{1/2},
	\end{gathered}
\end{align}
where $n_V$ and $n_H$ are respectively multiplicities of the $\mathcal{N}=2$ vector ($\mathcal{G}_{1}$) and $\mathcal{N}=2$ hyper ($\mathcal{G}_{1/2}$) multiplets for any particular  $\mathcal{N}\geq 3 \to \mathcal{N}=2$ decomposition. For example \cite{Keeler:2014nn,Charles:2015nn,Ferrara:2012bp}, $\mathcal{N}=3 \to \mathcal{N}=2$ has $n_V = n_H$, $\mathcal{N}=4 \to \mathcal{N}=2$ has $n_V = n_H+1$, $\mathcal{N}=5 \to \mathcal{N}=2$ has $n_V = 3, n_H=1$, $\mathcal{N}=6 \to \mathcal{N}=2$ has $n_V = 7, n_H=4$, $\mathcal{N}=8 \to \mathcal{N}=2$ has $n_V = 15, n_H=10$, etc.  
%a decomposed $\mathcal{N}=3,d=4$ EMSGT have $n_V = n_H$; a decomposed $\mathcal{N}=4,d=4$ EMSGT have $n_V = n_H+1$; a decomposed $\mathcal{N}=5,d=4$ EMSGT have $n_V = 3, n_H=1$; a decomposed $\mathcal{N}=6,d=4$ EMSGT have $n_V = 7, n_H=4$; a decomposed $\mathcal{N}=8,d=4$ EMSGT have $n_V = 15, n_H=10$, etc \cite{Keeler:2014nn,Charles:2015nn,Ferrara:2012bp}. 
However, we speculate that the numbers $n_V$ and $n_H$ for the decomposed matter coupled $\mathcal{N}\geq 3,d=4$ EMSGTs are related by a general formula,
\begin{equation}\label{d5}
	n_H-n_V = 3-\mathcal{N}.
\end{equation}
The $\mathcal{N}=2$ logarithmic correction data of \cref{S5,V12,H12}, along with the relation \eqref{d5}, help us to solve the decomposition \eqref{d1}. This essentially provides the logarithmic correction to the entropy of extremal Kerr-Newman, Kerr and Reissner-Nordstr\"om black holes in the matter coupled $\mathcal{N}\geq 3,d=4$ EMSGTs,
\begin{equation}\label{d3}
	\Delta S^{\mathcal{N}\geq 3}_{\text{BH}} =
	\begin{cases}
		\left((3-\mathcal{N})\frac{(2{b^\prime}^6+5{b^\prime}^4+4{b^\prime}^2+1)}{({b^\prime}^2+1)^2(2{b^\prime}^2+1)}-2\right) \text{ln}\thinspace \mathcal{A}_{H} &\text{for Kerr-Newman},\\
		(1-\mathcal{N})\thinspace \text{ln}\thinspace \mathcal{A}_{H} &\text{for Kerr},\\
		(4-\mathcal{N})\thinspace \text{ln}\thinspace \mathcal{A}_{H} &\text{for Reissner-Nordstr\"om}.
	\end{cases}
\end{equation}
The extremal Reissner-Nordstr\"om corrections are in alliance with \cite{Banerjee:2011pp} where calculations were encountered by the eigenfunction expansion method for $\mathcal{N}=4,8$ EMSGTs. Also, the Reissner-Nordstr\"om formula for all $\mathcal{N}\geq 3$ EMSGTs is the same as estimated in \cite{Keeler:2014nn,Ferrara:2012bp}. 

It is also possible to fit both the matter coupled $\mathcal{N}=2,d=4$ EMSGT and $\mathcal{N}\geq 3,d=4$ EMSGTs into a single frame. By merging the results \eqref{mc2} and \eqref{d3}, the combined logarithmic correction formula for extremal Kerr-Newman family of black holes in the matter coupled $\mathcal{N}\geq 2,d=4$ EMSGTs is presented as
\begin{equation}\label{d4}
	\Delta S^{\mathcal{N}\geq 2}_{\text{BH}} =\left(\frac{1}{12}\big(11-11(\mathcal{N}-2)-n_V+n_H\big)\frac{(2{b^\prime}^6+5{b^\prime}^4+4{b^\prime}^2+1)}{({b^\prime}^2+1)^2(2{b^\prime}^2+1)}+\frac{\mathcal{C}_{\text{zm}}}{2}\right) \text{ln}\thinspace \mathcal{A}_{H}.
\end{equation}
The above formula  is general, compact, and one of the central interest of this paper. It provides the logarithmic correction results of extremal Kerr-Newman (${b^\prime}\to$\thinspace finite and non-zero), Kerr (${b^\prime}\to\infty$) and Reissner-Nordstr\"om (${b^\prime}\to 0$) black holes in the matter coupled $\mathcal{N}\geq 2,d=4$ EMSGTs. Here the $\mathcal{C}_{\text{zm}}$ contribution for a particular choice of extremal black hole is obtained by setting $\delta=0$ in the formula \eqref{lc7}. 

\subsection{Logarithmic correction to the entropy of non-extremal black holes in the matter coupled $\mathcal{N} \geq 1,d=4$ EMSGTs}\label{ne}
The Euclidean gravity approach \cite{Sen:2013ns} (also reviewed in \cite{Charles:2015nn}) aids in extending the logarithmic entropy correction program for non-extremal black holes via the following formula 
\begin{equation}\label{ne1}
	\Delta S_{\text{BH}}^{\text{non-ext}} = \frac{1}{2}(\mathcal{C}_{\text{local}}+\mathcal{C}_{\text{zm}})\,\text{ln}\thinspace \left(\frac{\mathcal{A}_H}{G_N}\right),
\end{equation}
where the local contribution can be described as a density of the Seeley-DeWitt coefficient $a_4(x)$ over the whole black hole geometry including the near-horizon,
\begin{equation}\label{ne2}
	\mathcal{C}_{\text{local}} = \int_{\text{full geometry}} \mathrm{d}^4x \sqrt{\text{det}\thinspace \bar{g}}\thinspace a_4(x).
\end{equation} 
It is also well understood that the logarithmic correction formula \eqref{ne1} stores all trace anomaly data of the stress-energy tensor into the Seeley-DeWitt coefficient $a_4(x)$ (see \cite{Keeler:2014nn}, for instance). And one can express \cite{Charles:2015nn},
\begin{align}\label{ne3}
	(4\pi)^2a_4(x) = c W_{\mu\nu\rho\sigma}W^{\mu\nu\rho\sigma}-a E_4, 
\end{align}
where the Weyl tensor square $W_{\mu\nu\rho\sigma}W^{\mu\nu\rho\sigma}$ and the 4D Euler density $E_4$ are particular trace anomalies with coefficients $c$ and $a$, respectively. For any background geometry, $W_{\mu\nu\rho\sigma}W^{\mu\nu\rho\sigma}$ and $E_4$ have the typical forms,
\begin{align}\label{ne4}
	\begin{gathered}
		W_{\mu\nu\rho\sigma}W^{\mu\nu\rho\sigma} = R_{\mu\nu\rho\sigma}R^{\mu\nu\rho\sigma}-2R_{\mu\nu}R^{\mu\nu} + \frac{1}{3}R^2,\\
		E_4 = R_{\mu\nu\rho\sigma}R^{\mu\nu\rho\sigma}-4R_{\mu\nu}R^{\mu\nu} + R^2.
	\end{gathered}	
\end{align}
The coefficients of the trace anomalies ($c, a$) are numerical constants that depend on the type of field content and their coupling to the background. With the help of the data available in \cite{Sen:2013ns}, we present the following necessary limits of the trace anomalies over the whole non-extremal Kerr-Newman geometry \eqref{lc8} (in the same Euclidean continuation $t \to -i\theta$)
\begin{equation}\label{ne5}
	\begin{split}
		\int_{\text{full geometry}} \mathrm{d}^4x \sqrt{\text{det}\thinspace \bar{g}}\thinspace E_4 &= 64\pi^2,\\
		\int_{\text{full geometry}} \mathrm{d}^4x \sqrt{\text{det}\thinspace \bar{g}}\thinspace W_{\mu\nu\rho\sigma}W^{\mu\nu\rho\sigma} &= 64\pi^2 + \frac{\pi \beta Q^4}{b^5 r_H^4(b^2+r_H^2)}\bigg(3 b^5r_H + 2b^3 r_H^3+ 3 b r_H^5 \\
		& \qquad +3(b^2-r_H^2)(b^2+r_H^2)^2\thinspace \text{tan}^{-1}\Big(\frac{b}{r_H}\Big)\bigg),
	\end{split}
\end{equation}
where $b=J/M$. The period $\beta$ of the periodic coordinate $\theta$ and the outer horizon radius $r_H$ hold the forms,
\begin{equation}\label{ne6}
	\begin{split}
		r_H &= r_+ = M + \sqrt{M^2-Q^2-b^2},\\
		\beta &=\frac{1}{T_{BH}}= \frac{32\pi^2}{\sqrt{M^2-Q^2-b^2}}\left(2M^2-Q^2+2M\sqrt{M^2-Q^2-b^2} \right).
	\end{split}
\end{equation}
To accomplish the logarithmic correction program for the non-extremal Kerr-Newman family of black holes, one needs to extract the trace anomaly data $(c,a)$ from the particular $a_4(x)$ form \eqref{ne3} and then insert them in the $\mathcal{C}_{\text{local}}$ formula \eqref{ne2} along with the non-extremal limits \eqref{ne5}. For the $\mathcal{C}_{\text{zm}}$ contributions, we use the same general formula \eqref{lc7}.
%-------------------------------------------------------------------
{
	\renewcommand{\arraystretch}{1.5}
	\begin{table}[t]% h (here), t (top), b (bottom) and p (page of floats), ht (here top)
		\centering
		\hspace{-0.2in}
		\begin{tabular}{|>{\centering}p{1.7in}|>{\centering}p{.5in}>{\centering}p{2.2in}|}
			\hline
			\textbf{EMSGTs} & {$\bm{c}$} & {$\bm{a}$} \tabularnewline \hline \hline
			$\mathcal{N}=1,d=4$ EMSGT & 0 & $-\frac{1}{48}\big(41-3n_v+n_c\big) $ \tabularnewline 
			$\mathcal{N}\geq 2,d=4$ EMSGTs & 0 & $-\frac{1}{24}\big(11-11(\mathcal{N}-2)-n_V+n_H\big)$ \tabularnewline \hline
		\end{tabular}
		\caption{Coefficients of trace anomalies $(c,a)$ data for the matter coupled $\mathcal{N}\geq 1,d=4$ EMSGTs.}\label{tab3}
	\end{table}
} 
%=============================================================================================
The $a_4(x)$ coefficients for the matter coupled $\mathcal{N} = 1$ EMSGT (\cref{truncation}) and $\mathcal{N} \geq 2$ EMSGTs (\cref{decomposition}) are extracted as
\begin{align}
	(4\pi)^2 a^{\mathcal{N}= 1}_4(x) &=\frac{1}{48}\big(41-3n_v+n_c\big) \big( R_{\mu\nu\rho\sigma}R^{\mu\nu\rho\sigma}-4 R_{\mu\nu}R^{\mu\nu} \big),\label{ne7}\\
	(4\pi)^2 a^{\mathcal{N} \geq 2}_4(x) &=\frac{1}{24}\big(11-11(\mathcal{N}-2)-n_V+n_H\big) \big( R_{\mu\nu\rho\sigma}R^{\mu\nu\rho\sigma}-4 R_{\mu\nu}R^{\mu\nu} \big),\label{ne8}
\end{align}
which yields the $c$ and $a$ values as recorded in \cref{tab3}. Therefore, the logarithmic entropy corrections for the non-extremal Kerr-Newman family of black holes in the matter coupled $\mathcal{N} \geq 1,d=4$ EMSGTs are obtained as
\begin{align}
	\Delta S^{\mathcal{N}= 1 \thinspace \text{non-ext}}_{\text{BH}} &=
	\begin{cases}
		\frac{1}{24}(29-3n_v+n_c) \ln \mathcal{A}_{H} &\text{for rotating BHs},\\
		\frac{1}{24}(5-3n_v+n_c) \ln \mathcal{A}_{H} &\text{for non-rotating BHs}.
	\end{cases}\label{ne9}\\
	\Delta S^{\mathcal{N}=2 \thinspace \text{non-ext}}_{\text{BH}} &=
	\begin{cases}
		\frac{1}{12}\big(5-n_V+n_H \big) \ln \mathcal{A}_{H} &\text{for rotating BHs},\\
		\frac{1}{12}\big(-7-n_V+n_H \big) \ln \mathcal{A}_{H} &\text{for non-rotating BHs},
	\end{cases}\label{ne10}\\
	\Delta S^{\mathcal{N}\geq 3 \thinspace \text{non-ext}}_{\text{BH}} &=
	\begin{cases}
		\left(\frac{5}{2}-\mathcal{N}\right) \ln \mathcal{A}_{H} &\text{for rotating BHs},\\
		\left(\frac{3}{2}-\mathcal{N}\right) \ln \mathcal{A}_{H} &\text{for non-rotating BHs}.
	\end{cases}\label{ne11}
\end{align}
Here $\mathcal{C}_{\text{zm}}$ contributions for the particular choice of non-extremal rotating (Kerr-Newman and Kerr) and non-rotating (Reissner-Nordstr\"om and Schwarzschild) black holes are considered by setting $N_{\text{SUSY}}=0$ and $\delta=1$ in the formula \eqref{lc7}. The $\mathcal{N} \geq 2$ part (\cref{ne10,ne11}) accurately reproduces the results of \cite{Charles:2015nn}.
%, while the $\mathcal{N} = 1$ part (\cref{ne9}) is new reports to the literature.
%===============================================================================
\subsection{Conclusions and remarks}
The particular technical set-up employed in the current paper successfully provided logarithmic entropy corrections for all types (rotating, non-rotating, charged and uncharged) of extremal black holes in the matter coupled $\mathcal{N} \geq 1,d=4$ EMSGTs. For this achievement, the key role is played by the third Seeley-DeWitt coefficient $a_4(x)$. The $a_4(x)$ results of different multiplets of the $\mathcal{N}=2,d=4$ EMSGT, calculated via the standard computation approach of \cref{P}, are found to be manageable only in terms of background Riemannian and Ricci invariants. Therefore, the results are applicable for all back hole backgrounds of the theory in proper limits of black hole parameters. Utilization of the background-geometric-invariant form of $a_4(x)$ data into the quantum entropy function formalism (described in \cref{lc}) serves as a global platform to understand the logarithmic entropy corrections for all extremal Kerr-Newman family (Kerr-Newman, Kerr and Reissner-Nordstr\"om) of black holes. Also, the $\mathcal{N} \geq 3 \to \mathcal{N}=2$ decompositions are executed and managed so that we ended up with a generalized and compact logarithmic correction formula (only in terms of $\mathcal{N}$) for each black hole in all matter coupled $\mathcal{N} \geq 3,d=4$ EMSGTs. Quantum corrections to the entropy of extremal rotating black holes in matter-coupled supergravities were previously unknown but successfully investigated in this paper. The results of extremal Kerr-Newman and Kerr black holes are new reports to the literature. All these depict the main highlights and novelty of the work presented in this paper.  

In a very similar fashion, we also extracted logarithmic entropy corrections for the full non-extremal Kerr-Newman family (Kerr-Newman, Kerr, Reissner-Nordstr\"om and Schwarzschild) of black holes in the matter coupled $\mathcal{N} \geq 1,d=4$ EMSGTs. The used $a_4(x)$ data are common, but the Euclidean gravity approach is different (see \cref{ne}). Most of these non-extremal results are previously known; however, the corrections in the $\mathcal{N} = 1,d=4$ EMSGT are novel. Comparison between the extremal (\cref{mc2,t6,d3}) and non-extremal (eqs. \eqref{ne9}, \eqref{ne10} and \eqref{ne11}) results leads us to an interesting remark. When the non-extremal Kerr-Newman family of black holes are interpreted as solutions of matter coupled $\mathcal{N} \geq 1,d=4$ EMSGTs, the logarithmic corrections to their entropy become fully universal. But in the extremal limit, only the Kerr and Reissner-Nordstr\"om results are universal, while entropy corrections for the Kerr-Newman black holes become dependent on black hole parameters (via $b^\prime=\frac{J}{MQ}$). This seems quite surprising and may prove to be vital information for understanding the quantum entropy of black holes inside microscopic theories in the future. For any quantum theory of gravity, it is always challenging to serve microscopic answers to the quantum corrected entropy of black holes. Moreover, the complex nature of the calculated logarithmic correction results will enhance the technical difficulties a lot more in their microscopic reproduction. The calculated macroscopic logarithmic correction to the Bekenstein-Hawking entropy of extremal Reissner-Nordstr\"om black holes in $\mathcal{N}=4$ and $\mathcal{N}=8$ EMSGTs (originally reported in \cite{Banerjee:2011pp}) already have an exact microscopic agreement in string theory (see \cite{Mandal:2010im,Sen:2014nvt} and references therein). No such concrete microscopic answers (particularly in string theory) for any other calculated non-extremal and remaining extremal corrections are reported to date. However, it will always be interesting to overcome the technical barriers and complexities and explore the microscopic results as future progress.

%---------------------------------------------------------------------------
\acknowledgments

The authors are grateful to Rajesh Kumar Gupta, Ashoke Sen, Anthony Charles and Finn Larsen for sharing their expertise during the course of this research. We also want to thank Gourav Banerjee for his fruitful comments on this manuscript. 
%-------------------------------------------------------------------
\appendix
\section{Gamma matrices, EOMs and identities}\label{AA}
%This paper assumes the Euclidean signature of the metric and the following units
%\begin{align}\label{i1}
%	\begin{gathered}
%		\bar{g}_{\mu\nu} = \mathrm{diag}\,(+,+,+,+),\\
%		\hbar = c = k_{B} = 1,G_N=1/{16\pi}.
%	\end{gathered}
%\end{align}
%Throughout, Greek alphabets (e.g., $\mu,\nu,\cdots$) represent the 4D space-time ($d=4$) indices.
In this paper, all the gamma matrices $\gamma^\mu$ are Euclidean, Hermitian, and satisfy the following relations
{
	\allowdisplaybreaks
	\begin{equation}\label{i2}
		\begin{split}
			\left\lbrace \gamma^\mu,\gamma^\nu\right\rbrace &= 2\bar{g}^{\mu\nu}\mathbb{I}_4, \thinspace \left\lbrace \gamma^\mu,\gamma^5\right\rbrace = 0,\\
			\gamma^{\mu\nu\rho\sigma}&=-\gamma^5\varepsilon^{\mu\nu\rho\sigma}, \thinspace \gamma^5 = -\frac{1}{4!}\varepsilon^{\mu\nu\rho\sigma}\gamma_\mu\gamma_\nu\gamma_\rho\gamma_\sigma,\\
			\gamma^{\mu\nu} &= \frac{1}{2}(\gamma^\mu\gamma^\nu-\gamma^\nu\gamma^\mu), \thinspace\gamma^{\mu\rho\nu} = \frac{1}{2}(\gamma^\mu\gamma^\rho\gamma^\nu-\gamma^\nu\gamma^\rho\gamma^\mu).
		\end{split}
\end{equation}}
$\mathbb{I}_4$ is the identity matrix in the Clifford algebra satisfied by the 4D gamma matrices.
%The Levi-Civita contractions for the Euclidean metric are defined as
%\begin{equation}\label{i3}
%	\varepsilon^{\mu_1 \cdots \mu_m\rho_1\cdots\rho_n}\varepsilon_{\nu_1\cdots\nu_m\rho_1\cdots\rho_n} = m!n!\delta^{[\mu_1}_{[\nu_1}\delta^{\mu_2}_{\nu_2}\cdots \delta^{\mu_m]}_{\nu_m]}.
%\end{equation}
The equations of motion and Bianchi identities, satisfied by the background graviphoton field strength tensor $\bar{F}_{\mu\nu}$ and its dual $\bar{H}_{\mu\nu}=-\varepsilon_{\mu\nu\rho\sigma}\bar{F}^{\rho\sigma}$, are expressed as
{
	\allowdisplaybreaks
	\begin{equation}\label{i4}
		\begin{split}
			\bar{F}_{\mu\rho}{\bar{F_\nu}}^\rho &= \frac{1}{2}R_{\mu\nu}+\frac{1}{4}\bar{g}_{\mu\nu}\bar{F}_{\rho\sigma}\bar{F}^{\rho\sigma}, \thinspace R= 0,\\
			\bar{H}_{\mu\rho}{\bar{H_\nu}}^\rho &= -2R_{\mu\nu}+\bar{g}_{\mu\nu}\bar{F}_{\rho\sigma}\bar{F}^{\rho\sigma}, \\ \bar{H}_{\mu\nu}\bar{H}^{\mu\nu} &= 4 \bar{F}_{\mu\nu}\bar{F}^{\mu\nu},\\
			D_\mu \bar{F}^{\mu\nu} &=0, \thinspace D_{[\mu}\bar{F}_{\nu\rho]}=0,\\
			D_\mu \bar{H}^{\mu\nu} &=0, \thinspace D_{[\mu}\bar{H}_{\nu\rho]}=0.
		\end{split}
\end{equation}}
Starting with \eqref{i4} and using the gravitational Bianchi $R_{\mu[\nu\rho\sigma]}=0$, we formulate a list of identities,
{
	\allowdisplaybreaks
	\begin{equation}\label{id1}
		\begin{split}
			\bar{F}^{\mu\rho}\bar{F^\nu}_\rho \bar{H}_{\mu\sigma}{\bar{H_\nu}}^\sigma &= \bar{F}^{\mu\rho}{\bar{H^\nu}}_\rho \bar{H}_{\mu\sigma}\bar{F_\nu}^\sigma= (\bar{F}_{\mu\nu}\bar{F}^{\mu\nu})^2-R_{\mu\nu}R^{\mu\nu},\\
			\bar{F}_{\mu\nu}\bar{H}^{\mu\nu}\bar{F}_{\rho\sigma}\bar{H}^{\rho\sigma}&= 4(\bar{F}_{\mu\nu}\bar{F}^{\mu\nu})^2-4R_{\mu\nu}R^{\mu\nu},\\
			R_{\mu\rho\nu\sigma}\bar{F}^{\mu\nu}\bar{F}^{\rho\sigma}&= \frac{1}{2}R_{\mu\nu\rho\sigma}\bar{F}^{\mu\nu}\bar{F}^{\rho\sigma},\\
			R_{\mu\rho\nu\sigma}\bar{H}^{\mu\nu}\bar{H}^{\rho\sigma}&=\frac{1}{2}R_{\mu\nu\rho\sigma}\bar{H}^{\mu\nu}\bar{H}^{\rho\sigma}= 2R_{\mu\nu\rho\sigma}\bar{F}^{\mu\nu}\bar{F}^{\rho\sigma}-4R_{\mu\nu}R^{\mu\nu},
		\end{split}
	\end{equation}
}
followed by the following derivative relations (up to a total derivative)
{
	\allowdisplaybreaks
	\begin{equation}\label{id2}
		\begin{split}
			(D_\rho \bar{F}_{\mu\nu})(D^\rho \bar{F}^{\mu\nu})=2(D_\mu \bar{{F}_\rho}^\nu)(D_\nu \bar{F}^{\rho\mu}) &= R_{\mu\nu\rho\sigma}\bar{F}^{\mu\nu}\bar{F}^{\rho\sigma}-R_{\mu\nu}R^{\mu\nu},\\
			(D_\rho \bar{H}_{\mu\nu})(D^\rho \bar{H}^{\mu\nu})=2(D_\mu {\bar{H_\rho}}^\nu)(D_\nu \bar{H}^{\rho\mu}) &= 4R_{\mu\nu\rho\sigma}\bar{F}^{\mu\nu}\bar{F}^{\rho\sigma}-4R_{\mu\nu}R^{\mu\nu}.
		\end{split}
	\end{equation}
}
%======================================================================================
\section{Seeley-DeWitt trace calculations for multiplets of $\mathcal{N}=2,d=4$ EMSGT}\label{BB}
In this section, we will present a brief list of comparatively lengthy trace calculations of the non-minimally coupled cases\footnote{For the minimally-coupled cases, we have already provided a combined Seeley-DeWitt formula \eqref{cf}.} of the $\mathcal{N}=2,d=4$ EMSGT. This includes the fermionic-sector ($\mathcal{G}^{f}_{3/2}$) of gravitino multiplet, the bosonic-sector ($\mathcal{G}^{b}_{1}$) of vector multiplet and the fermionic-sector ($\mathcal{G}^{f}_{1/2}$) of hyper multiplet. These trace calculations are undoubtedly lengthy and tedious but not much difficult to manage. One needs to proceed systematically by executing every term step-by-step with the help of gamma identities, Maxwell-Bianchi identities, background equations of motion mentioned in \cref{AA}. The trace calculations also need a distinctive trick to follow: if $I^{mn}$ is the effective metric corresponding to any two arbitrary matrices $A^{mn}$ and $B^{mn}$, then ${A^m}_n = A^{mp}I_{pn}$, $\text{tr}(A) = {A^m}_m$, $\text{tr}(AB) = {A^m}_n {B^n}_m$ and $\text{tr}(AB) = \text{tr}(BA)$. As mentioned, our ultimate goal is to compute and express the traces in terms of the background invariants  like $R_{\mu\nu\rho\sigma}R^{\mu\nu\rho\sigma}$, $R_{\mu\nu}R^{\mu\nu}$, $R_{\mu\nu\rho\sigma}\bar{F}^{\mu\nu}\bar{F}^{\rho\sigma}$, $\bar{F}_{\mu\nu}\bar{F}^{\mu\nu}$, $(\bar{F}_{\mu\nu}\bar{F}^{\mu\nu})^2$, etc.
%----------------------------------------------------------------------------------------------
\subsection*{$\mathcal{N}=2,d=4$ EMSGT: gravitino multiplet}\label{a}
For the fermionic-sector fluctuations $\tilde{\xi}_m = \lbrace \psi_\mu,\lambda \rbrace $ of the $\mathcal{N}=2$ gravitino multiplet (\cref{gravitino}), we define the following traces
\begin{subequations}\label{a11}
	\begin{align}
		&\text{tr}(E) = \text{tr}\Big({E^{\psi_\mu}}_{\psi_\nu}+ {E^{\lambda}}_{\lambda}+ {E^{\psi_\mu}}_{\lambda}+ {E^{\lambda}}_{\psi_\nu}\Big),\\
		&\text{tr}(E^2) = \text{tr}\Big({E^{\psi_\mu}}_{\psi_\nu}{E^{\psi_\nu}}_{\psi_{\mu^\prime}}+ {E^{\lambda}}_{\lambda}{E^{\lambda}}_{\lambda}+ {E^{\psi_\mu}}_{\lambda}{E^{\lambda}}_{\psi_\nu}+ {E^{\lambda}}_{\psi_\nu}{E^{\psi_\nu}}_{\lambda}\Big),\\
		&\text{tr}(\Omega_{\rho\sigma}\Omega^{\rho\sigma}) = \text{tr}\Big({(\Omega_{\rho\sigma})^{\psi_\mu}}_{\psi_\nu}{(\Omega^{\rho\sigma})^{\psi_\nu}}_{\psi_{\mu^\prime}}+ {(\Omega_{\rho\sigma})^{\lambda}}_{\lambda}{(\Omega^{\rho\sigma})^{\lambda}}_{\lambda}\nonumber\\
		&\qquad\qquad\qquad\quad + {(\Omega_{\rho\sigma})^{\psi_\mu}}_{\lambda}{(\Omega^{\rho\sigma})^{\lambda}}_{\psi_\nu}+ {(\Omega_{\rho\sigma})^{\lambda}}_{\psi_\nu}{(\Omega^{\rho\sigma})^{\psi_\nu}}_{\lambda}\Big).
	\end{align}
\end{subequations}
%Note that $\psi$, $\lambda$ associated with $E$ and $\Omega_{\rho\sigma}$ are just field labels; one should not consider them as particular tensor indices.
Simplifying \cref{G9} further, all the components of $E$ are expressed as
{
	\allowdisplaybreaks
	\begin{subequations}\label{a12}
		\begin{align}
			{E^{\psi_\mu}}_{ \psi_\nu} &= \underbrace{\frac{1}{8}\gamma^\alpha\gamma^\beta\gamma^\mu\gamma_\nu\gamma^\theta\gamma^\phi \bar{F}_{\alpha\beta}\bar{F}_{\theta\phi}}_{X_1}\underbrace{+\frac{1}{2}\gamma^\theta\gamma^\phi\gamma^\alpha\gamma^\mu \bar{F}_{\theta\phi}\bar{F}_{\nu\alpha}}_{X_2} \nonumber\\
			& \quad \underbrace{-\frac{1}{8} \bar{g}^\mu_\nu\gamma^\alpha\gamma^\beta \gamma^\theta\gamma^\phi \bar{F}_{\alpha\beta}\bar{F}_{\theta\phi}}_{X_3}\underbrace{+\gamma^\theta\gamma^\phi \bar{F^\mu}_{\nu}\bar{F}_{\theta\phi} +\frac{1}{2} \gamma^\alpha\gamma^\beta {R^{\mu}}_{\nu\alpha\beta}}_{X_4} \nonumber\\
			& \quad \underbrace{-\frac{1}{4}\gamma^\mu\gamma^\alpha R_{\nu\alpha} + \frac{3}{4}\gamma_\nu\gamma^\alpha {R^\mu}_\alpha}_{X_5}\underbrace{-\frac{3}{2}\mathbb{I}_4 {R^\mu}_\nu-\frac{1}{2}\mathbb{I}_4\bar{g^\mu}_\nu\bar{F}_{\alpha\beta}\bar{F}^{\alpha\beta}}_{X_6}, \\
			{E^{\lambda}}_{\lambda} &= \underbrace{\frac{1}{2}\gamma^\alpha\gamma^\beta\gamma^\theta\gamma^\phi \bar{F}_{\alpha\beta}\bar{F}_{\theta\phi}}_{X_7},\\
			{E^{\psi_\mu}}_{ \lambda} &= \underbrace{\frac{1}{4\sqrt{2}}\gamma^\alpha\gamma^\beta\gamma^\mu\gamma^\rho D_\rho \bar{F}_{\alpha\beta}}_{X_8}\underbrace{-\frac{1}{8\sqrt{2}}\gamma^\sigma\gamma^\rho\gamma^\mu\gamma^\alpha\gamma^\beta\gamma_\sigma D_\rho \bar{F}_{\alpha\beta}}_{X_9},\\
			{E^{\lambda}}_{ \psi_\nu} &= \underbrace{\frac{1}{4\sqrt{2}}\gamma^\rho\gamma_\nu\gamma^\alpha\gamma^\beta D_\rho \bar{F}_{\alpha\beta}}_{X_{10}}\underbrace{- \frac{1}{8\sqrt{2}} \gamma^\sigma\gamma^\alpha\gamma^\beta\gamma_\nu\gamma^\rho\gamma_\sigma D_\rho \bar{F}_{\alpha\beta}}_{X_{11}}
		\end{align}
	\end{subequations}
}
Then,
{
	\allowdisplaybreaks
	\begin{equation}\label{a13}
		\begin{split}
			&\text{tr}(X_1) = -4\bar{F}_{\mu\nu}\bar{F}^{\mu\nu},\\ 
			&\text{tr}(X_2) = 4\bar{F}_{\mu\nu}\bar{F}^{\mu\nu},\\ 
			&\text{tr}(X_3) = 4\bar{F}_{\mu\nu}\bar{F}^{\mu\nu},\\ 
			&\text{tr}(X_4) = 0,
		\end{split}
		\hspace{0.5in}
		\begin{split}
			&\text{tr}(X_5) = 0,\\ 
			&\text{tr}(X_6) = -8\bar{F}_{\mu\nu}\bar{F}^{\mu\nu},\\ 
			&\text{tr}(X_7) = -4\bar{F}_{\mu\nu}\bar{F}^{\mu\nu},\\ 
			&\text{tr}(X_8) =  0,
		\end{split}
		\hspace{0.5in}
		\begin{split}
			&\text{tr}(X_9) = 0,\\ 
			&\text{tr}(X_{10}) = 0,\\ 
			&\text{tr}(X_{11}) = 0.
		\end{split}
	\end{equation}
}
Therefore,
{
	\allowdisplaybreaks
	\begin{equation}\label{a14}
		\begin{split}
			&\text{tr}({E^{\psi_\mu}}_{\psi_\nu})=\sum_{i=1}^6 \text{tr}(X_i)= -4\bar{F}_{\mu\nu}\bar{F}^{\mu\nu},\\
			&\text{tr}({E^\lambda}_{\lambda})= \text{tr}(X_7)= -4\bar{F}_{\mu\nu}\bar{F}^{\mu\nu},
		\end{split}
		\hspace{0.5in}
		\begin{split}
			& \text{tr}({E^{\psi_\mu}}_{\lambda})= \sum_{i=8}^9 \text{tr}(X_i) = 0,\\
			&  \text{tr}({E^{\lambda}}_{\psi_\nu})= \sum_{i=10}^{11} \text{tr}(X_i) = 0.
		\end{split}
	\end{equation}
}
Also,
{
	\allowdisplaybreaks
	\begin{equation}\label{a15}
		\begin{split}
			\text{tr}(X_1X_1) &=  8(\bar{F}_{\mu\nu}\bar{F}^{\mu\nu})^2-4R_{\mu\nu}R^{\mu\nu},\\
			\text{tr}(X_2X_2) &=  8(\bar{F}_{\mu\nu}\bar{F}^{\mu\nu})^2-4R_{\mu\nu}R^{\mu\nu},\\
			\text{tr}(X_3X_3) &=  2(\bar{F}_{\mu\nu}\bar{F}^{\mu\nu})^2-R_{\mu\nu}R^{\mu\nu},\\
			\text{tr}(X_4X_4) &=  2R_{\mu\nu\rho\sigma}R^{\mu\nu\rho\sigma}+8(\bar{F}_{\mu\nu}\bar{F}^{\mu\nu})^2\\
			&\qquad +8R_{\mu\nu\rho\sigma}\bar{F}^{\mu\nu}\bar{F}^{\rho\sigma},\\
			\text{tr}(X_5X_5) &=  3R_{\mu\nu}R^{\mu\nu},\\
			\text{tr}(X_6X_6) &=  4(\bar{F}_{\mu\nu}\bar{F}^{\mu\nu})^2+9R_{\mu\nu}R^{\mu\nu},\\
			\text{tr}(X_7X_7) &=  8(\bar{F}_{\mu\nu}\bar{F}^{\mu\nu})^2-4R_{\mu\nu}R^{\mu\nu},\\
			\text{tr}(X_1X_2) &=   2R_{\mu\nu}R^{\mu\nu}-4(\bar{F}_{\mu\nu}\bar{F}^{\mu\nu})^2,\\
			\text{tr}(X_1X_3) &=   R_{\mu\nu}R^{\mu\nu}-2(\bar{F}_{\mu\nu}\bar{F}^{\mu\nu})^2,\\
			\text{tr}(X_1X_4) &=   6R_{\mu\nu}R^{\mu\nu}-4(\bar{F}_{\mu\nu}\bar{F}^{\mu\nu})^2\\
			& \qquad -4R_{\mu\nu\rho\sigma}\bar{F}^{\mu\nu}\bar{F}^{\rho\sigma},\\
			\text{tr}(X_1X_5) &= 0 ,\\
			\text{tr}(X_1X_6) &=  2(\bar{F}_{\mu\nu}\bar{F}^{\mu\nu})^2,
		\end{split}
		\hspace{0.15in}
		\begin{split}
			\text{tr}(X_2X_3) &=  2(\bar{F}_{\mu\nu}\bar{F}^{\mu\nu})^2-R_{\mu\nu}R^{\mu\nu},\\
			\text{tr}(X_2X_4) &=  -2R_{\mu\nu}R^{\mu\nu}-4(\bar{F}_{\mu\nu}\bar{F}^{\mu\nu})^2,\\
			\text{tr}(X_2X_5) &=  5R_{\mu\nu}R^{\mu\nu},\\
			\text{tr}(X_2X_6) &=  -3R_{\mu\nu}R^{\mu\nu}-2(\bar{F}_{\mu\nu}\bar{F}^{\mu\nu})^2,\\
			\text{tr}(X_3X_4) &=  0,\\
			\text{tr}(X_3X_5) &=  0,\\
			\text{tr}(X_3X_6) &=  -2(\bar{F}_{\mu\nu}\bar{F}^{\mu\nu})^2,\\
			\text{tr}(X_4X_5) &=  -8R_{\mu\nu}R^{\mu\nu},\\
			\text{tr}(X_4X_6) &=  0,\\
			\text{tr}(X_5X_6) &=  -3R_{\mu\nu}R^{\mu\nu},\\
			\text{tr}(X_8X_{10}) &= R_{\mu\nu}R^{\mu\nu}-R_{\mu\nu\rho\sigma}\bar{F}^{\mu\nu}\bar{F}^{\rho\sigma},\\
			\text{tr}(X_8X_{11}) &= 0,\\
			\text{tr}(X_9X_{10}) &= 0,\\
			\text{tr}(X_9X_{11}) &= 0.
		\end{split}
	\end{equation}
}
Hence,
{
	\allowdisplaybreaks
	\begin{equation}\label{a16}
		\begin{split}
			&\text{tr}({E^{\psi_\mu}}_{\psi_\nu}{E^{\psi_\nu}}_{\psi_{\mu^\prime}}) = \sum_{i,j=1}^6 \text{tr}(X_i X_j) = 2R_{\mu\nu\rho\sigma}R^{\mu\nu\rho\sigma}-3R_{\mu\nu}R^{\mu\nu}+2(\bar{F}_{\mu\nu}\bar{F}^{\mu\nu})^2,\\
			&\text{tr}({E^{\lambda}}_{\lambda}{E^{\lambda}}_{\lambda}) = \text{tr}(X_7 X_7) = 8(\bar{F}_{\mu\nu}\bar{F}^{\mu\nu})^2-4R_{\mu\nu}R^{\mu\nu},\\
			&\text{tr}({E^{\psi_\mu}}_{\lambda}{E^{\lambda}}_{\psi_\nu})=\text{tr}({E^{\lambda}}_{\psi_\nu}{E^{\psi_\nu}}_{\lambda})= \sum_{i=8}^9\sum_{j=10}^{11} \text{tr}(X_i X_j)= R_{\mu\nu}R^{\mu\nu}-R_{\mu\nu\rho\sigma}\bar{F}^{\mu\nu}\bar{F}^{\rho\sigma}.
		\end{split}
	\end{equation}
}
Similarly, \cref{G10} provides all the components of $\Omega_{\rho\sigma}$ as
{
	\allowdisplaybreaks
	\begin{subequations}\label{a17}
		\begin{align}
			{(\Omega_{\rho\sigma})^{\psi_\mu}}_{\psi_\nu}&=\underbrace{\mathbb{I}_4{R^{\mu}}_{\nu\rho\sigma}}_{Y_1}+\underbrace{\frac{1}{4}\bar{g^\mu}_\nu\gamma^\alpha\gamma^\beta R_{\rho\sigma\alpha\beta}}_{Y_2}\nonumber\\
			&\quad\underbrace{-\frac{1}{32}\Big(\gamma^\alpha\gamma^\beta\gamma^\mu\gamma_\rho+\frac{1}{2}\gamma^\tau\gamma_\rho\gamma^\mu\gamma^\alpha\gamma^\beta\gamma_\tau \Big) \Big(\gamma_\sigma\gamma_\nu\gamma^\theta\gamma^\phi+ \frac{1}{2}\gamma^\lambda\gamma^\theta\gamma^\phi\gamma_\nu\gamma_\sigma\gamma_\lambda\Big)\bar{F}_{\alpha\beta}\bar{F}_{\theta\phi}}_{Y_3}\nonumber\\
			&\quad\underbrace{+\frac{1}{32}\Big(\gamma^\alpha\gamma^\beta\gamma^\mu\gamma_\sigma+\frac{1}{2}\gamma^\tau\gamma_\sigma\gamma^\mu\gamma^\alpha\gamma^\beta\gamma_\tau \Big) \Big(\gamma_\rho\gamma_\nu\gamma^\theta\gamma^\phi+ \frac{1}{2}\gamma^\lambda\gamma^\theta\gamma^\phi\gamma_\nu\gamma_\rho\gamma_\lambda\Big)\bar{F}_{\alpha\beta}\bar{F}_{\theta\phi}}_{Y_4},\\
			{(\Omega_{\rho\sigma})^{\lambda}}_{\lambda}&= \underbrace{\frac{1}{4}\gamma^\alpha\gamma^\beta R_{\rho\sigma\alpha\beta}}_{Y_5} \nonumber\\
			&\quad\underbrace{-\frac{1}{32}\Big(\gamma_\rho\gamma^\mu\gamma^\alpha\gamma^\beta+\frac{1}{2}\gamma^\tau\gamma^\alpha\gamma^\beta\gamma^\mu\gamma_\rho\gamma_\tau\Big)\Big(\gamma^\theta\gamma^\phi\gamma_\mu\gamma_\sigma+\frac{1}{2}\gamma^\lambda\gamma_\sigma\gamma_\mu\gamma^\theta\gamma^\phi\gamma_\lambda\Big)\bar{F}_{\alpha\beta}\bar{F}_{\theta\phi}}_{Y_6} \nonumber\\
			&\quad\underbrace{+\frac{1}{32}\Big(\gamma_\sigma\gamma^\mu\gamma^\alpha\gamma^\beta+\frac{1}{2}\gamma^\tau\gamma^\alpha\gamma^\beta\gamma^\mu\gamma_\sigma\gamma_\tau\Big)\Big(\gamma^\theta\gamma^\phi\gamma_\mu\gamma_\rho+\frac{1}{2}\gamma^\lambda\gamma_\rho\gamma_\mu\gamma^\theta\gamma^\phi\gamma_\lambda\Big)\bar{F}_{\alpha\beta}\bar{F}_{\theta\phi}}_{Y_7}, \\
			{(\Omega_{\rho\sigma})^{\psi_\mu}}_{\lambda}&= \underbrace{-\frac{1}{4\sqrt{2}}\Big(\gamma^\alpha\gamma^\beta\gamma^\mu\gamma_\sigma  + \frac{1}{2}\gamma^\tau\gamma_\sigma\gamma^\mu
				\gamma^\alpha\gamma^\beta\gamma_\tau \Big)(D_\rho\bar{F}_{\alpha\beta})}_{Y_8} \nonumber\\
			&\quad \underbrace{+\frac{1}{4\sqrt{2}}\Big(\gamma^\alpha\gamma^\beta\gamma^\mu\gamma_\rho  + \frac{1}{2}\gamma^\tau\gamma_\rho\gamma^\mu
				\gamma^\alpha\gamma^\beta\gamma_\tau \Big)(D_\sigma\bar{F}_{\alpha\beta})}_{Y_9},\\
			{(\Omega_{\rho\sigma})^{\lambda}}_{\psi_{\nu}}&= \underbrace{\frac{1}{4\sqrt{2}}\Big(\gamma_\sigma\gamma_\nu\gamma^\alpha\gamma^\beta  + \frac{1}{2}\gamma^\tau\gamma^\alpha\gamma^\beta
				\gamma_\nu\gamma_\sigma\gamma_\tau \Big)(D_\rho\bar{F}_{\alpha\beta})}_{Y_{10}} \nonumber\\
			&\quad \underbrace{-\frac{1}{4\sqrt{2}}\Big(\gamma_\rho\gamma_\nu\gamma^\alpha\gamma^\beta  + \frac{1}{2}\gamma^\tau\gamma^\alpha\gamma^\beta
				\gamma_\nu\gamma_\rho\gamma_\tau \Big)(D_\sigma\bar{F}_{\alpha\beta})}_{Y_{11}}.
		\end{align}
	\end{subequations}
}
Then,
{
	\allowdisplaybreaks
	\begin{equation}\label{a18}
		\begin{split}
			\text{tr}(Y_1Y_1) &=  -4R_{\mu\nu\rho\sigma}R^{\mu\nu\rho\sigma},\\
			\text{tr}(Y_2Y_2) &=  -2R_{\mu\nu\rho\sigma}R^{\mu\nu\rho\sigma},\\
			\text{tr}(Y_3Y_3) &=  3R_{\mu\nu}R^{\mu\nu}-2(\bar{F}_{\mu\nu}\bar{F}^{\mu\nu})^2,\\
			\text{tr}(Y_4Y_4) &=  3R_{\mu\nu}R^{\mu\nu}-2(\bar{F}_{\mu\nu}\bar{F}^{\mu\nu})^2,\\
			\text{tr}(Y_5Y_5) &=  -\frac{1}{2} R_{\mu\nu\rho\sigma}R^{\mu\nu\rho\sigma},\\
			\text{tr}(Y_6Y_6) &=  3R_{\mu\nu}R^{\mu\nu}-2(\bar{F}_{\mu\nu}\bar{F}^{\mu\nu})^2,\\
			\text{tr}(Y_7Y_7) &=  3R_{\mu\nu}R^{\mu\nu}-2(\bar{F}_{\mu\nu}\bar{F}^{\mu\nu})^2,\\
			\text{tr}(Y_1Y_2) &=  0,\\
			\text{tr}(Y_1Y_3) &=   4R_{\mu\nu}R^{\mu\nu}-2R_{\mu\nu\rho\sigma}\bar{F}^{\mu\nu}\bar{F}^{\rho\sigma},\\
			\text{tr}(Y_1Y_4) &= 4R_{\mu\nu}R^{\mu\nu}-2R_{\mu\nu\rho\sigma}\bar{F}^{\mu\nu}\bar{F}^{\rho\sigma},
		\end{split}
		\hspace{0.15in}
		\begin{split}
			\text{tr}(Y_2Y_3) &=  -R_{\mu\nu}R^{\mu\nu},\\
			\text{tr}(Y_2Y_4) &=  -R_{\mu\nu}R^{\mu\nu},\\
			\text{tr}(Y_3Y_4) &=  -16(\bar{F}_{\mu\nu}\bar{F}^{\mu\nu})^2,\\
			\text{tr}(Y_5Y_6) &=  R_{\mu\nu}R^{\mu\nu},\\
			\text{tr}(Y_5Y_7) &=  R_{\mu\nu}R^{\mu\nu},\\
			\text{tr}(Y_6Y_7) &=  R_{\mu\nu}R^{\mu\nu}-10(\bar{F}_{\mu\nu}\bar{F}^{\mu\nu})^2,\\
			\text{tr}(Y_8Y_{10}) &=  8R_{\mu\nu\rho\sigma}\bar{F}^{\mu\nu}\bar{F}^{\rho\sigma}-8R_{\mu\nu}R^{\mu\nu},\\
			\text{tr}(Y_8Y_{11}) &=  3R_{\mu\nu}R^{\mu\nu}-3R_{\mu\nu\rho\sigma}\bar{F}^{\mu\nu}\bar{F}^{\rho\sigma},\\
			\text{tr}(Y_9Y_{10}) &=  3R_{\mu\nu}R^{\mu\nu}-3R_{\mu\nu\rho\sigma}\bar{F}^{\mu\nu}\bar{F}^{\rho\sigma},\\
			\text{tr}(Y_9Y_{11}) &=  8R_{\mu\nu\rho\sigma}\bar{F}^{\mu\nu}\bar{F}^{\rho\sigma}-8R_{\mu\nu}R^{\mu\nu}.
		\end{split}
	\end{equation}
}
Therefore,
{
	\allowdisplaybreaks
	\begin{equation}\label{a19}
		\begin{split}
			\text{tr}\Big({(\Omega_{\rho\sigma})^{\psi_\mu}}_{\psi_\nu}{(\Omega^{\rho\sigma})^{\psi_\nu}}_{\psi_{\mu^\prime}}\Big) &= \sum_{i,j=1}^4 \text{tr}(Y_i Y_j) = -6R_{\mu\nu\rho\sigma}R^{\mu\nu\rho\sigma}+18 R_{\mu\nu}R^{\mu\nu}\\
			&\hspace{1.3in} -36(\bar{F}_{\mu\nu}\bar{F}^{\mu\nu})^2-8R_{\mu\nu\rho\sigma}\bar{F}^{\mu\nu}\bar{F}^{\rho\sigma},\\
			\text{tr}\Big({(\Omega_{\rho\sigma})^{\lambda}}_{\lambda}{(\Omega^{\rho\sigma})^{\lambda}}_{\lambda}\Big) &= \sum_{i,j=5}^7 \text{tr}(Y_i Y_j) \\
			&= - \frac{1}{2}R_{\mu\nu\rho\sigma}R^{\mu\nu\rho\sigma}+12 R_{\mu\nu}R^{\mu\nu}-24(\bar{F}_{\mu\nu}\bar{F}^{\mu\nu})^2,\\
			\text{tr}\Big({(\Omega_{\rho\sigma})^{\psi_\mu}}_{\lambda}{(\Omega^{\rho\sigma})^{\lambda}}_{\psi_\nu}\Big)&=\text{tr}\Big({(\Omega_{\rho\sigma})^{\lambda}}_{\psi_\nu}{(\Omega^{\rho\sigma})^{\psi_\nu}}_{\lambda}\Big)\\
			& = \sum_{i=8}^9\sum_{j=10}^{11} \text{tr}(Y_i Y_j)= 10R_{\mu\nu\rho\sigma}\bar{F}^{\mu\nu}\bar{F}^{\rho\sigma}-10 R_{\mu\nu}R^{\mu\nu}.
		\end{split}
	\end{equation}
}
Employing the data \eqref{a16} and \eqref{a19} in the definitions \eqref{a11}, one can obtain the trace results as provided in \cref{G11}.

%---------------------------------------------------------------------------------------
\subsection*{$\mathcal{N}=2,d=4$ EMSGT: vector multiplet}\label{b}

For the bosonic-sector fluctuations $\tilde{\xi}_m = \lbrace a_\mu,\phi_R,\phi_I \rbrace$ of the $\mathcal{N}=2$ vector multiplet (\cref{Vector}), we define the following traces
{
	\allowdisplaybreaks
	\begin{subequations}\label{b11}
		\begin{align}
			&\text{tr}(E) = \text{tr}\Big({E^{a_\mu}}_{a_\nu}+{E^{\phi_R}}_{\phi_R}+ {E^{\phi_I}}_{\phi_I}+ {E^{a_\mu}}_{\phi_R}+{E^{a_\mu}}_{\phi_I} \nonumber\\
			&\qquad\qquad\qquad + {E^{\phi_R}}_{a_\nu} +  {E^{\phi_R}}_{\phi_I}+ {E^{\phi_I}}_{a_\nu}+ {E^{\phi_I}}_{\phi_R}\Big), \\
			&\text{tr}(E^2) = \text{tr}\Big({E^{a_\mu}}_{a_\nu}{E^{a_\nu}}_{a_{\mu^\prime}}+ {E^{\phi_R}}_{\phi_R} {E^{\phi_R}}_{\phi_R}+ {E^{\phi_I}}_{\phi_I}{E^{\phi_I}}_{\phi_I}\nonumber\\
			&\qquad\qquad\quad + {E^{a_\mu}}_{\phi_R}{E^{\phi_R}}_{a_\nu}+ {E^{a_\mu}}_{\phi_I}{E^{\phi_I}}_{a_\nu}+{E^{\phi_R}}_{a_\nu}{E^{a_\nu}}_{\phi_R} \nonumber\\
			&\qquad\qquad\quad + {E^{\phi_R}}_{\phi_I}{E^{\phi_I}}_{\phi_R}+{E^{\phi_I}}_{a_\nu}{E^{a_\nu}}_{\phi_I}+ {E^{\phi_I}}_{\phi_R}{E^{\phi_R}}_{\phi_I}\Big), \\
			&\text{tr}(\Omega_{\rho\sigma}\Omega^{\rho\sigma}) = \text{tr}\Big({(\Omega_{\rho\sigma})^{a_\mu}}_{a_\nu}{(\Omega^{\rho\sigma})^{a_\nu}}_{a_{\mu^\prime}}+ {(\Omega_{\rho\sigma})^{\phi_R}}_{\phi_R} {(\Omega^{\rho\sigma})^{\phi_R}}_{\phi_R}+ {(\Omega_{\rho\sigma})^{\phi_I}}_{\phi_I}{(\Omega^{\rho\sigma})^{\phi_I}}_{\phi_I}\nonumber\\
			&\qquad\qquad\quad + {(\Omega_{\rho\sigma})^{a_\mu}}_{\phi_R}{(\Omega^{\rho\sigma})^{\phi_R}}_{a_\nu}+ {(\Omega_{\rho\sigma})^{a_\mu}}_{\phi_I}{(\Omega^{\rho\sigma})^{\phi_I}}_{a_\nu}+{(\Omega_{\rho\sigma})^{\phi_R}}_{a_\nu}{(\Omega^{\rho\sigma})^{a_\nu}}_{\phi_R} \nonumber\\
			&\qquad\qquad\quad + {(\Omega_{\rho\sigma})^{\phi_R}}_{\phi_I}{(\Omega^{\rho\sigma})^{\phi_I}}_{\phi_R}+{(\Omega_{\rho\sigma})^{\phi_I}}_{a_\nu}{(\Omega^{\rho\sigma})^{a_\nu}}_{\phi_I}+ {(\Omega_{\rho\sigma})^{\phi_I}}_{\phi_R}{(\Omega^{\rho\sigma})^{\phi_R}}_{\phi_I}\Big). 
		\end{align}
\end{subequations}}
With the aid of \cref{V6}, all the valid non-zero components of $E$ and $\Omega_{\rho\sigma}$ are expressed as
{
	\allowdisplaybreaks
	\begin{subequations}\label{b12}
		\begin{align}
			&{E^{a_\mu}}_{a_\nu} = \underbrace{-{R^{\mu}}_\nu}_{Q_1}\underbrace{+\bar{F}^{\mu\rho}\bar{F}_{\nu\rho}-\frac{1}{4}\bar{H}^{\mu\rho}\bar{H}_{\nu\rho}}_{Q_2},\\
			& {(\Omega_{\rho\sigma})^{a_\mu}}_{a_\nu} = \underbrace{{R^{\mu}}_{\nu\rho\sigma}}_{Q_3}\underbrace{+ \left({\bar{F^\mu}}_{\rho}{\bar{F_\sigma}}_{\nu}-{\bar{F^\mu}}_{\sigma}{\bar{F_\rho}}_{\nu}\right)}_{Q_4}\underbrace{-\frac{1}{4}\left({\bar{H^\mu}}_{\rho}{\bar{H_\sigma}}_{\nu}-{\bar{H^\mu}}_{\sigma}{\bar{H_\rho}}_{\nu}\right)}_{Q_5},\\
			& {(\Omega_{\rho\sigma})^{a_\mu}}_{\phi_R} =  \underbrace{\frac{i}{2}D^\mu \bar{H}_{\rho\sigma}}_{Q_6}, \thinspace {(\Omega_{\rho\sigma})^{\phi_R}}_{a_\nu} =  \underbrace{-\frac{i}{2}D_\nu \bar{H}_{\rho\sigma}}_{Q_7},\\
			& {(\Omega_{\rho\sigma})^{a_\mu}}_{\phi_I} =\underbrace{ D^\mu \bar{F}_{\rho\sigma}}_{Q_8}, \thinspace {(\Omega_{\rho\sigma})^{\phi_I}}_{a_\nu} = \underbrace{-D_\nu \bar{F}_{\rho\sigma}}_{Q_{9}},
		\end{align}
	\end{subequations}
}
%All other components vanish and hence their trace contributions will be zero. 
Executing similar kind of calculations as for the gravitino multiplet, we calculate the following traces
{
	\allowdisplaybreaks
	\begin{equation}\label{b13}
		\begin{split}
			\text{tr}(Q_1) &= 0,\\
			\text{tr}(Q_2) &= 0,\\
			\text{tr}(Q_1Q_1) &= R_{\mu\nu}R^{\mu\nu},\\
			\text{tr}(Q_2Q_2) &= R_{\mu\nu}R^{\mu\nu},\\
			\text{tr}(Q_1Q_2) &= -R_{\mu\nu}R^{\mu\nu},\\
			\text{tr}(Q_3Q_3) &= -R_{\mu\nu\rho\sigma}R^{\mu\nu\rho\sigma},\\
			\text{tr}(Q_4Q_4) &= \frac{1}{2}R_{\mu\nu}R^{\mu\nu}-\frac{3}{2}(\bar{F}_{\mu\nu}\bar{F}^{\mu\nu})^2,
		\end{split}
		\hspace{0.15in}
		\begin{split}
			\text{tr}(Q_5Q_5) &= \frac{1}{2}R_{\mu\nu}R^{\mu\nu}-\frac{3}{2}(\bar{F}_{\mu\nu}\bar{F}^{\mu\nu})^2,\\
			\text{tr}(Q_3Q_4) &= R_{\mu\nu\rho\sigma}\bar{F}^{\mu\nu}\bar{F}^{\rho\sigma},\\
			\text{tr}(Q_3Q_5) &= 2R_{\mu\nu}R^{\mu\nu}-R_{\mu\nu\rho\sigma}\bar{F}^{\mu\nu}\bar{F}^{\rho\sigma},\\
			\text{tr}(Q_4Q_5) &= \frac{3}{2}(\bar{F}_{\mu\nu}\bar{F}^{\mu\nu})^2-\frac{3}{2}R_{\mu\nu}R^{\mu\nu},\\
			\text{tr}(Q_6Q_7) &= R_{\mu\nu\rho\sigma}\bar{F}^{\mu\nu}\bar{F}^{\rho\sigma}-R_{\mu\nu}R^{\mu\nu},\\
			\text{tr}(Q_8Q_{9}) &= R_{\mu\nu}R^{\mu\nu}-R_{\mu\nu\rho\sigma}\bar{F}^{\mu\nu}\bar{F}^{\rho\sigma}.
		\end{split}
\end{equation}}
%Hence,
%{
%	\allowdisplaybreaks
%	\begin{equation}
%	\text{tr}({E^{a_\mu}}_{a_\nu})=\sum_{i=1}^3 \text{tr}(Q_i) = 0,\thinspace \text{tr}({E^{a_\mu}}_{a_\nu}{E^{a_\nu}}_{a_{\mu^\prime}}) = \sum_{i,j=1}^3 \text{tr}(Q_i Q_j) = 0.
%	\end{equation}}
Hence,%\footnote{All other trace components vanish for this particular case of the vector multiplet bosonic-sector.} 
{
	\allowdisplaybreaks
	\begin{equation}\label{b14}
		\begin{split}
			\text{tr}({E^{a_\mu}}_{a_\nu}) &=\sum_{i=1}^2 \text{tr}(Q_i) = 0,\\ \text{tr}({E^{a_\mu}}_{a_\nu}{E^{a_\nu}}_{a_{\mu^\prime}}) &= \sum_{i,j=1}^2 \text{tr}(Q_i Q_j) = 0,\\
			\text{tr}\Big({(\Omega_{\rho\sigma})^{a_\mu}}_{a_\nu}{(\Omega^{\rho\sigma})^{a_\nu}}_{a_{\mu^\prime}}\Big) &= \sum_{i,j=3}^5 \text{tr}(Q_i Q_j) = -R_{\mu\nu\rho\sigma}R^{\mu\nu\rho\sigma}+2 R_{\mu\nu}R^{\mu\nu}\\
			\text{tr}\Big({(\Omega_{\rho\sigma})^{a_\mu}}_{\phi_R}{(\Omega^{\rho\sigma})^{\phi_R}}_{a_\nu}\Big)&=\text{tr}\Big({(\Omega_{\rho\sigma})^{\phi_R}}_{a_\nu}{(\Omega^{\rho\sigma})^{a_\nu}}_{\phi_R}\Big) =  \text{tr}(Q_6 Q_7) \\
			&= R_{\mu\nu\rho\sigma}\bar{F}^{\mu\nu}\bar{F}^{\rho\sigma}-R_{\mu\nu}R^{\mu\nu},\\
			\text{tr}\Big({(\Omega_{\rho\sigma})^{a_\mu}}_{\phi_I}{(\Omega^{\rho\sigma})^{\phi_I}}_{a_\nu}\Big)&=\text{tr}\Big({(\Omega_{\rho\sigma})^{\phi_I}}_{a_\nu}{(\Omega^{\rho\sigma})^{a_\nu}}_{\phi_I}\Big) =  \text{tr}(Q_8 Q_{9})\\
			&= R_{\mu\nu}R^{\mu\nu}-R_{\mu\nu\rho\sigma}\bar{F}^{\mu\nu}\bar{F}^{\rho\sigma}.
		\end{split}
	\end{equation}
}
Employing the above data in the definitions \eqref{b11}, one can obtain the trace results as provided in \cref{V7}.
%------------------------------------------------------------------------------------
\subsection*{$\mathcal{N}=2,d=4$ EMSGT: hyper multiplet}\label{c}
For the fermionic-sector fluctuations $\tilde{\xi}_m = \lbrace \lambda_1,\lambda_2 \rbrace $ of the $\mathcal{N}=2$ hyper multiplet (\cref{hyper}), we define the following traces
{
	\allowdisplaybreaks
	\begin{subequations}\label{c11}
		\begin{align}
			&\text{tr}(E) = \text{tr}\Big({E^{\lambda_1}}_{\lambda_1}+ {E^{\lambda_2}}_{\lambda_2}+ {E^{\lambda_1}}_{\lambda_2}+ {E^{\lambda_2}}_{\lambda_1}\Big),\\
			&\text{tr}(E^2) = \text{tr}\Big({E^{\lambda_1}}_{\lambda_1}{E^{\lambda_1}}_{\lambda_1}+{E^{\lambda_2}}_{\lambda_2}{E^{\lambda_2}}_{\lambda_2}+{E^{\lambda_1}}_{\lambda_2}{E^{\lambda_2}}_{\lambda_1}+{E^{\lambda_2}}_{\lambda_1}{E^{\lambda_1}}_{\lambda_2}\Big),\\
			&\text{tr}(\Omega_{\rho\sigma}\Omega^{\rho\sigma}) = \text{tr}\Big({(\Omega_{\rho\sigma})^{\lambda_1}}_{\lambda_1}{(\Omega^{\rho\sigma})^{\lambda_1}}_{{\lambda_1}}+{(\Omega_{\rho\sigma})^{\lambda_2}}_{\lambda_2}{(\Omega^{\rho\sigma})^{\lambda_2}}_{{\lambda_2}}\nonumber\\
			&\qquad\qquad\qquad\quad + {(\Omega_{\rho\sigma})^{\lambda_1}}_{\lambda_2}{(\Omega^{\rho\sigma})^{\lambda_2}}_{{\lambda_1}}+{(\Omega_{\rho\sigma})^{\lambda_2}}_{\lambda_1}{(\Omega^{\rho\sigma})^{\lambda_1}}_{{\lambda_2}}\Big).
		\end{align}
\end{subequations}}
With the aid of \cref{H7}, all the components of $E$ and $\Omega_{\rho\sigma}$ are expressed as
{
	\allowdisplaybreaks
	\begin{subequations}\label{c12}
		\begin{align}
			&{E^{\lambda_1}}_{ \lambda_1}={E^{\lambda_2}}_{ \lambda_2} = \underbrace{\frac{1}{16} \gamma^\rho\gamma^\alpha\gamma^\beta\gamma^\theta\gamma^\phi\gamma_\rho \bar{F}_{\alpha\beta}\bar{F}_{\theta\phi}}_{P_1},\\
			&{E^{\lambda_1}}_{ \lambda_2}={E^{\lambda_2}}_{ \lambda_1} = 0,\\
			&{(\Omega_{\rho\sigma})^{\lambda_1}}_{ \lambda_1}={(\Omega_{\rho\sigma})^{\lambda_2}}_{ \lambda_2} = \underbrace{\frac{1}{4}\gamma^\alpha\gamma^\beta R_{\rho\sigma\alpha\beta}}_{P_2} \nonumber\\
			&\hspace{1.65in} \underbrace{-\frac{1}{16}(\gamma_\rho\gamma^\alpha\gamma^\beta+\gamma^\alpha\gamma^\beta\gamma_\rho) (\gamma_\sigma\gamma^\theta\gamma^\phi+\gamma^\theta\gamma^\phi\gamma_\sigma)\bar{F}_{\alpha\beta}\bar{F}_{\theta\phi}}_{P_3}\nonumber\\
			&\hspace{1.65in} \underbrace{+\frac{1}{16}(\gamma_\sigma\gamma^\alpha\gamma^\beta+\gamma^\alpha\gamma^\beta\gamma_\sigma)(\gamma_\rho\gamma^\theta\gamma^\phi+\gamma^\theta\gamma^\phi\gamma_\rho)\bar{F}_{\alpha\beta}\bar{F}_{\theta\phi}}_{P_4},
			\\
			&{(\Omega_{\rho\sigma})^{\lambda_1}}_{ \lambda_2} =\underbrace{-\frac{1}{4}(\gamma_\sigma\gamma^\alpha\gamma^\beta + \gamma^\alpha\gamma^\beta\gamma_\sigma)D_\rho\bar{F}_{\alpha\beta}}_{P_5} \underbrace{+\frac{1}{4}(\gamma_\rho\gamma^\alpha\gamma^\beta + \gamma^\alpha\gamma^\beta\gamma_\rho)D_\sigma\bar{F}_{\alpha\beta}}_{P_6},\\
			&{(\Omega_{\rho\sigma})^{\lambda_2}}_{ \lambda_1} =\underbrace{\frac{1}{4}(\gamma_\sigma\gamma^\alpha\gamma^\beta + \gamma^\alpha\gamma^\beta\gamma_\sigma)D_\rho\bar{F}_{\alpha\beta}}_{P_7} \underbrace{-\frac{1}{4}(\gamma_\rho\gamma^\alpha\gamma^\beta + \gamma^\alpha\gamma^\beta\gamma_\rho)D_\sigma\bar{F}_{\alpha\beta}}_{P_8}.
		\end{align}
	\end{subequations}
}
We then calculate the following trace values
{
	\allowdisplaybreaks
	\begin{equation}\label{c13}
		\begin{split}
			\text{tr}(P_1) &= -2\bar{F}_{\mu\nu}\bar{F}^{\mu\nu},\\
			\text{tr}(P_1P_1) &= 2(\bar{F}_{\mu\nu}\bar{F}^{\mu\nu})^2-R_{\mu\nu}R^{\mu\nu},\\
			\text{tr}(P_2P_2) &= -\frac{1}{2}R_{\mu\nu\rho\sigma}R^{\mu\nu\rho\sigma},\\
			\text{tr}(P_3P_3) &= 2R_{\mu\nu}R^{\mu\nu}-2(\bar{F}_{\mu\nu}\bar{F}^{\mu\nu})^2,\\
			\text{tr}(P_4P_4) &= 2R_{\mu\nu}R^{\mu\nu}-2(\bar{F}_{\mu\nu}\bar{F}^{\mu\nu})^2,\\
			\text{tr}(P_2P_3) &= 2R_{\mu\nu}R^{\mu\nu}-R_{\mu\nu\rho\sigma}\bar{F}^{\mu\nu}\bar{F}^{\rho\sigma},
		\end{split}
		\hspace{0.15in}
		\begin{split}
			\text{tr}(P_2P_4) &= 2R_{\mu\nu}R^{\mu\nu}-R_{\mu\nu\rho\sigma}\bar{F}^{\mu\nu}\bar{F}^{\rho\sigma},\\
			\text{tr}(P_3P_4) &= -4(\bar{F}_{\mu\nu}\bar{F}^{\mu\nu})^2,\\
			\text{tr}(P_5P_7) &= 4R_{\mu\nu\rho\sigma}\bar{F}^{\mu\nu}\bar{F}^{\rho\sigma}-4R_{\mu\nu}R^{\mu\nu},\\
			\text{tr}(P_5P_8) &= 2R_{\mu\nu}R^{\mu\nu}-2R_{\mu\nu\rho\sigma}\bar{F}^{\mu\nu}\bar{F}^{\rho\sigma},\\
			\text{tr}(P_6P_7) &= 2R_{\mu\nu}R^{\mu\nu}-2R_{\mu\nu\rho\sigma}\bar{F}^{\mu\nu}\bar{F}^{\rho\sigma},\\
			\text{tr}(P_6P_8) &= 4R_{\mu\nu\rho\sigma}\bar{F}^{\mu\nu}\bar{F}^{\rho\sigma}-4R_{\mu\nu}R^{\mu\nu}.
		\end{split}
\end{equation}}
Hence,
{
	\allowdisplaybreaks
	\begin{equation}\label{c14}
		\begin{split}
			&\text{tr}({E^{\lambda_1}}_{\lambda_1})= \text{tr}({E^{\lambda_2}}_{\lambda_2})=\text{tr}(P_1) = -2\bar{F}_{\mu\nu}\bar{F}^{\mu\nu},\\
			&\text{tr}({E^{\lambda_1}}_{\lambda_2})= \text{tr}({E^{\lambda_2}}_{\lambda_1}) = 0,\\
			&\text{tr}({E^{\lambda_1}}_{\lambda_1}{E^{\lambda_1}}_{\lambda_1})=\text{tr}({E^{\lambda_2}}_{\lambda_2}{E^{\lambda_2}}_{\lambda_2})=\text{tr}(P_1P_1) = 2(\bar{F}_{\mu\nu}\bar{F}^{\mu\nu})^2-R_{\mu\nu}R^{\mu\nu},\\
			&\text{tr}({E^{\lambda_1}}_{\lambda_2}{E^{\lambda_2}}_{\lambda_1})=\text{tr}({E^{\lambda_2}}_{\lambda_1}{E^{\lambda_1}}_{\lambda_2}) = 0.
		\end{split}
\end{equation}}
Also,
\begin{equation}\label{c15}
	\begin{split}
		\text{tr}\Big({(\Omega_{\rho\sigma})^{\lambda_1}}_{\lambda_1}{(\Omega^{\rho\sigma})^{\lambda_1}}_{\lambda_1}\Big)&=\text{tr}\Big({(\Omega_{\rho\sigma})^{\lambda_2}}_{\lambda_2}{(\Omega^{\rho\sigma})^{\lambda_2}}_{\lambda_2}\Big)\\
		&=\sum_{i,j=2}^4 \text{tr}(P_i P_j)=-\frac{1}{2}R_{\mu\nu\rho\sigma}R^{\mu\nu\rho\sigma}+12 R_{\mu\nu}R^{\mu\nu} \\
		&\qquad\qquad\qquad\qquad-4R_{\mu\nu\rho\sigma}\bar{F}^{\mu\nu}\bar{F}^{\rho\sigma}-12(\bar{F}_{\mu\nu}\bar{F}^{\mu\nu})^2,\allowdisplaybreaks[2]\\
		\text{tr}\Big({(\Omega_{\rho\sigma})^{\lambda_1}}_{\lambda_2}{(\Omega^{\rho\sigma})^{\lambda_2}}_{\lambda_1}\Big)&=\text{tr}\Big({(\Omega_{\rho\sigma})^{\lambda_2}}_{\lambda_1}{(\Omega^{\rho\sigma})^{\lambda_1}}_{\lambda_2}\Big) \\
		&=\sum_{i=5}^6\sum_{j=7}^{8} \text{tr}(P_i P_j)= 4R_{\mu\nu\rho\sigma}\bar{F}^{\mu\nu}\bar{F}^{\rho\sigma}-4 R_{\mu\nu}R^{\mu\nu}.
	\end{split}
\end{equation}
Employing the above data in the definitions \eqref{c11}, one can obtain the trace results as provided in \cref{H8}.



\begin{thebibliography}{99}

% Please avoid comments such as "For a review'', "For some examples",
% "and references therein" or move them in the text. In general,
% please leave only references in the bibliography and move all
% accessory text in footnotes.

% Also, please have only one work for each \bibitem.

\bibitem{Bekenstein:1973jb} J.D. Bekenstein, \emph{Black holes and entropy}, \href{https://journals.aps.org/prd/abstract/10.1103/PhysRevD.7.2333}{\emph{Phys. Rev.} \textbf{D7} (1973) 2333-2346} [\href{https://inspirehep.net/search?ln=en&as=1&m1=e&p1=Black+Holes+and+Entropy&f1=title&op1=a&m2=a&p2=Bekenstein&f2=author&op2=a&m3=o&p3=Phys.Rev.&f3=journal&action_search=Search&sf=&so=d&rm=&rg=25&sc=0&of=hb}{\scshape{in}SPIRE}].

\bibitem{Hawking:1975sh} S.W. Hawking, \emph{Particle creation by black holes}, \href{https://link.springer.com/article/10.1007%2FBF01608497}{\emph{Commun. Math. Phys.} \textbf{43} (1975) 199-220} [\href{https://inspirehep.net/search?ln=en&as=1&m1=e&p1=Particle+Creation+by+Black+Holes&f1=title&op1=a&m2=a&p2=Hawking&f2=author&op2=a&m3=o&p3=Commun.+Math.+Phys.&f3=journal&action_search=Search&sf=&so=d&rm=&rg=25&sc=0&of=hb}{\scshape{in}SPIRE}].	
	
\bibitem{Wald:1993rw} R.M. Wald, \emph{Black hole entropy is the Noether charge}, \href{https://journals.aps.org/prd/abstract/10.1103/PhysRevD.48.R3427}{\emph{Phys. Rev.} \textbf{D48} (1993) 3427-3431} [\href{https://arxiv.org/abs/gr-qc/9307038}{arXiv:gr-qc/9307038}] [\href{https://inspirehep.net/search?p=find+eprint+gr-qc/9307038}{\scshape{in}SPIRE}].
	
	
%------------------------------------------------------------------------------------------------------
\bibitem{Banerjee:2011oo} S. Banerjee, R.K. Gupta and A. Sen, \emph{Logarithmic corrections to extremal black hole entropy from quantum entropy function},  \href{https://link.springer.com/article/10.1007%2FJHEP03%282011%29147}
{\emph{JHEP} \textbf{03} (2011) 147} [\href{https://arxiv.org/abs/1005.3044}{arXiv:1005.3044}] [\href{https://inspirehep.net/search?p=find+EPRINT+arXiv:1005.3044}{\scshape{in}SPIRE}].
		
\bibitem{Banerjee:2011pp} S. Banerjee, R. K. Gupta, I. Mandal and A. Sen, \emph{Logarithmic corrections to N=4 and N=8 black hole entropy: a one loop test of quantum gravity}, \href{https://link.springer.com/article/10.1007%2FJHEP11%282011%29143}
{\emph{JHEP} \textbf{11} (2011) 143} [\href{https://arxiv.org/abs/1106.0080}{arXiv:1106.0080}] [\href{https://inspirehep.net/search?p=find+EPRINT+arXiv:1106.0080}{\scshape{in}SPIRE}].
			
\bibitem{Sen:2012qq} A. Sen, \emph{Logarithmic corrections to N=2 black hole entropy: an infrared window into the microstates}, \href{https://link.springer.com/article/10.1007%2Fs10714-012-1336-5}
{\emph{Gen. Rel. Grav.} \textbf{44} (2012) 1207} [\href{https://arxiv.org/abs/1108.3842}{arXiv:1108.3842}] [\href{https://inspirehep.net/search?p=find+EPRINT+arXiv:1108.3842}{\scshape{in}SPIRE}].
				
\bibitem{Gupta:2014ns} R.K. Gupta, S. Lal and S. Thakur, \emph{Logarithmic corrections to extremal black hole entropy in N = 2, 4 and 8 supergravity}, \href{https://link.springer.com/article/10.1007%2FJHEP11%282014%29072}
{\emph{JHEP} \textbf{11} (2014) 072} [\href{https://arxiv.org/abs/1402.2441}{arXiv:1402.2441}] [\href{https://inspirehep.net/search?p=find+EPRINT+arXiv:1402.2441}{\scshape{in}SPIRE}].
					
\bibitem{Ferrara:2012bp} S. Ferrara and A. Marrani, \emph{Generalized mirror symmetry and quantum black hole entropy}, \href{https://www.sciencedirect.com/science/article/pii/S0370269311014481?via%3Dihub}{\emph{Phys. Lett.} \textbf{B707} (2012) 173-177} [\href{https://arxiv.org/abs/1109.0444}{arXiv:1109.0444}] [\href{https://inspirehep.net/search?p=find+eprint+1109.0444}{\scshape{in}SPIRE}].
						
\bibitem{Keeler:2014nn} C. Keeler, F. Larsen and P. Lisbao, \emph{Logarithmic corrections to $\mathcal{N}\geq 2$ black hole entropy}, \href{https://journals.aps.org/prd/abstract/10.1103/PhysRevD.90.043011}{\emph{Phys. Rev.} \textbf{D90} (2014) 043011} [\href{https://arxiv.org/abs/1404.1379}{arXiv:1404.1379}] [\href{https://inspirehep.net/search?p=find+EPRINT+arXiv:1404.1379}{\scshape{in}SPIRE}].
						
\bibitem{Larsen:2015nx} F. Larsen and P. Lisbao, \emph{Quantum corrections to supergravity on $AdS_2 \times S^2$}, \href{https://journals.aps.org/prd/abstract/10.1103/PhysRevD.91.084056}{\emph{Phys. Rev.} \textbf{D91} (2015) 084056} [\href{https://arxiv.org/abs/1411.7423}{arXiv:1411.7423}] [\href{https://inspirehep.net/search?p=find+EPRINT+arXiv:1411.7423}{\scshape{in}SPIRE}].
						
\bibitem{Karan:2019sk} S. Karan, G. Banerjee, and B. Panda, \emph{Seeley-DeWitt coefficients in $\mathcal{N} = 2$ Einstein-Maxwell supergravity theory and logarithmic corrections to $\mathcal{N} = 2$ extremal black hole entropy},  \href{https://link.springer.com/article/10.1007%2FJHEP08%282019%29056}{\emph{JHEP} \textbf{08}  (2019) 056} [\href{https://arxiv.org/abs/1905.13058}{arXiv:1905.13058}] [\href{https://inspirehep.net/record/1737533}{\scshape{in}SPIRE}].
							
\bibitem{Banerjee:2020wbr} G. Banerjee, S. Karan and B. Panda, \emph{Logarithmic correction to the entropy of extremal black holes in $\mathcal{N}=1$ Einstein-Maxwell supergravity},  \href{https://link.springer.com/article/10.1007%2FJHEP01%282021%29090}{\emph{JHEP} \textbf{01} (2021) 090}  [\href{https://arxiv.org/abs/2007.11497}{arXiv:2007.11497}] [\href{https://inspirehep.net/literature?sort=mostrecent&size=25&page=1&q=find%20eprint%202007.11497}{\scshape{in}SPIRE}].	
								
\bibitem{Sen:2012rr} A. Sen, \emph{Logarithmic corrections to rotating extremal black hole entropy in four and five dimensions}, \href{https://link.springer.com/article/10.1007%2Fs10714-012-1373-0}
{\emph{Gen. Rel. Grav.} \textbf{44} (2012) 1947} [\href{https://arxiv.org/abs/1109.3706}{arXiv:1109.3706}] [\href{https://inspirehep.net/search?p=find+EPRINT+arXiv:1109.3706}{\scshape{in}SPIRE}].
									
\bibitem{Bhattacharyya:2012ss} S. Bhattacharyya, B. Panda and A. Sen, \emph{Heat kernel expansion and extremal Kerr-Newmann black hole entropy in Einstein-Maxwell theory}, \href{https://link.springer.com/article/10.1007%2FJHEP08%282012%29084} {\emph{JHEP} \textbf{08}  (2012) 084} [\href{https://arxiv.org/abs/1204.4061}{arXiv:1204.4061}] [\href{https://inspirehep.net/search?p=find+EPRINT+arXiv:1204.4061}{\scshape{in}SPIRE}].				
										
\bibitem{Chowdhury:2014np} A. Chowdhury, R.K. Gupta, S. Lal, M. Shyani and S. Thakur, \emph{Logarithmic corrections to twisted indices from the quantum entropy function}, \href{https://link.springer.com/article/10.1007%2FJHEP11%282014%29002}{\emph{JHEP} \textbf{11} (2014) 002} [\href{https://arxiv.org/abs/1404.6363}{arXiv:1404.6363}] [\href{https://inspirehep.net/search?p=find+EPRINT+arXiv:1404.6363}{\scshape{in}SPIRE}].
											
\bibitem{Jeon:2017ij} I. Jeon and S. Lal, \emph{Logarithmic corrections to entropy of magnetically charged AdS4 black holes}, \href{https://www.sciencedirect.com/science/article/pii/S0370269317307244?via%3Dihub}{\emph{Phys. Lett.} \textbf{B774} (2017) 41-45} [\href{https://arxiv.org/abs/1707.04208}{arXiv:1707.04208}] [\href{https://inspirehep.net/search?p=find+eprint+1707.04208}{\scshape{in}SPIRE}].								
												
\bibitem{Sen:2013ns} A. Sen, \emph{Logarithmic corrections to Schwarzschild and other non-extremal black hole entropy in different dimensions}, \href{https://link.springer.com/article/10.1007%2FJHEP04%282013%29156}{\emph{JHEP} \textbf{04} (2013) 156} [\href{https://arxiv.org/abs/1205.0971v2}{arXiv:1205.0971}] [\href{https://inspirehep.net/record/1113590}{\scshape{in}SPIRE}].	
													
\bibitem{Charles:2015nn} A.M. Charles and F. Larsen, \emph{Universal corrections to non-extremal black hole entropy in $\mathcal{N}\geq 2$ supergravity}, \href{https://link.springer.com/article/10.1007%2FJHEP06%282015%29200}{\emph{JHEP} \textbf{06} (2015) 200} [\href{https://arxiv.org/abs/1505.01156}{arXiv:1505.01156}] [\href{https://inspirehep.net/search?p=find+EPRINT+arXiv:1505.01156}{\scshape{in}SPIRE}].
														
\bibitem{Castro:2018tg} A. Castro, V. Godet, F. Larsen and Y. Zeng, \emph{Logarithmic corrections to black hole entropy: the non-BPS branch}, \href{https://link.springer.com/article/10.1007%2FJHEP05%282018%29079}{\emph{JHEP} \textbf{05} (2018) 079} [\href{https://arxiv.org/abs/1801.01926}{arXiv:1801.01926}] [\href{https://inspirehep.net/search?p=find+EPRINT+arXiv:1801.01926}{\scshape{in}SPIRE}].
%---------------------------------------------------------------------------------------------------------
\bibitem{Mohaupt:2000mj} T. Mohaupt, \emph{Black hole entropy, special geometry and strings}, \href{https://onlinelibrary.wiley.com/doi/abs/10.1002/1521-3978%28200102%2949%3A1/3%3C3%3A%3AAID-PROP3%3E3.0.CO%3B2-%23}{\emph{Fortsch. Phys.} \textbf{49} (2001) 3--161} [\href{https://arxiv.org/abs/hep-th/0007195v2}{arXiv:hep-th/0007195}] [\href{https://inspirehep.net/literature?sort=mostrecent&size=25&page=1&q=find%20eprint%20hep-th%2F0007195}{\scshape{in}SPIRE}].
%--------------------------------------------------------------------------------------------------------
\bibitem{Sen:2008wa} A. Sen, \emph{Entropy function and AdS(2) / CFT(1) correspondence}, \href{https://iopscience.iop.org/article/10.1088/1126-6708/2008/11/075}{\emph{JHEP} \textbf{11} (2008) 075} [\href{https://arxiv.org/abs/0805.0095}{arXiv:0805.0095}] [\href{https://inspirehep.net/search?p=find+EPRINT+arXiv:0805.0095}{\scshape{in}SPIRE}].
																
\bibitem{Sen:2009wb} A. Sen, \emph{Quantum entropy function from AdS(2)/CFT(1) correspondence}, \href{https://www.worldscientific.com/doi/abs/10.1142/S0217751X09045893}{\emph{Int. J. Mod. Phys.} \textbf{A24} (2009) 4225} [\href{https://arxiv.org/abs/0809.3304}{arXiv:0809.3304}] [\href{https://inspirehep.net/search?p=find+EPRINT+arXiv:0809.3304}{\scshape{in}SPIRE}].
																
\bibitem{Sen:2009wc} A. Sen, \emph{Arithmetic of quantum entropy function}, \href{https://iopscience.iop.org/article/10.1088/1126-6708/2009/08/068}{\emph{JHEP} \textbf{08} (2009) 068} [\href{https://arxiv.org/abs/0903.1477}{arXiv:0903.1477}] [\href{https://inspirehep.net/search?p=find+EPRINT+arXiv:0903.1477}{\scshape{in}SPIRE}].
%---------------------------------------------------------------------------------------------------------------------
%----------------------------------------------------------------------------------------------------------------	
\bibitem{Solodukhin:1995na} S.N. Solodukhin, \emph{The conical singularity and quantum corrections to entropy of black hole}, \href{https://journals.aps.org/prd/abstract/10.1103/PhysRevD.51.609}{\emph{Phys. Rev.} \textbf{D51} (1995) 609} [\href{https://arxiv.org/abs/hep-th/9407001}{arXiv:hep-th/9407001}] [\href{https://inspirehep.net/search?p=find+eprint+hep-th/9407001}{\scshape{in}SPIRE}].
																
\bibitem{Solodukhin:1995nb} S.N. Solodukhin, \emph{\enquote{Nongeometric} contribution to the entropy of a black hole due to quantum corrections}, \href{https://journals.aps.org/prd/abstract/10.1103/PhysRevD.51.618}{\emph{Phys. Rev.} \textbf{D51} (1995) 618} [\href{https://arxiv.org/abs/hep-th/9408068}{arXiv:hep-th/9408068}][\href{https://inspirehep.net/search?p=find+eprint+hep-th/9408068}{\scshape{in}SPIRE}].
																
\bibitem{Fursaev:1995df} D.V. Fursaev, \emph{Temperature and entropy of a quantum black hole and conformal anomaly}, \href{https://journals.aps.org/prd/abstract/10.1103/PhysRevD.51.R5352}{\emph{Phys. Rev.} \textbf{D51} (1995) 5352-5355} [\href{https://arxiv.org/abs/hep-th/9412161}{arXiv:hep-th/9412161}] [\href{https://inspirehep.net/search?p=find+eprint+hep-th/9412161}{\scshape{in}SPIRE}].
																
\bibitem{Mavromatos:1996kc} N.E. Mavromatos and E. Winstanley, \emph{Aspects of hairy black holes in spontaneously broken Einstein-Yang-Mills systems: Stability analysis and entropy considerations}, \href{https://journals.aps.org/prd/abstract/10.1103/PhysRevD.53.3190}{\emph{Phys. Rev.} \textbf{D53} (1996) 3190} [\href{https://inspirehep.net/record/400168}{arXiv:hep-th/9510007}] [\href{https://inspirehep.net/record/400168}{\scshape{in}SPIRE}].
																
\bibitem{Mann:1996bi} R.B. Mann and S.N. Solodukhin, \emph{Conical geometry and quantum entropy of a charged Kerr black hole}, \href{https://journals.aps.org/prd/abstract/10.1103/PhysRevD.54.3932}{\emph{Phys. Rev.} \textbf{D54} (1996) 3932} [\href{https://arxiv.org/abs/hep-th/9604118}{arXiv:hep-th/9604118}] [\href{https://inspirehep.net/search?p=find+eprint+hep-th/9604118}{\scshape{in}SPIRE}].
																
\bibitem{Mann:1998hm} R.B. Mann and S.N. Solodukhin, \emph{Universality of quantum entropy for extreme black holes}, \href{https://www.sciencedirect.com/science/article/pii/S0550321398000947?via%3Dihub}{\emph{Nucl. Phys.} \textbf{B523} (1998) 293} [\href{https://arxiv.org/abs/hep-th/9709064}{arXiv:hep-th/9709064}] [\href{https://inspirehep.net/search?p=find+eprint+hep-th/9709064}{\scshape{in}SPIRE}].
																	
\bibitem{Kaul:2000rk} R.K. Kaul and P. Majumdar, \emph{Logarithmic correction to the Bekenstein-Hawking entropy}, \href{https://journals.aps.org/prl/abstract/10.1103/PhysRevLett.84.5255}{\emph{Phys. Rev. Lett.} \textbf{84} (2000) 5255-5257} [\href{https://arxiv.org/abs/gr-qc/0002040}{arXiv:gr-qc/0002040}] [\href{https://inspirehep.net/search?p=find+eprint+gr-qc/0002040}{\scshape{in}SPIRE}].	
																	
\bibitem{Carlip:2000nv} S. Carlip, \emph{Logarithmic corrections to black hole entropy, from the Cardy formula}, \href{https://iopscience.iop.org/article/10.1088/0264-9381/17/20/302}{\emph{Class. Quant. Grav.} \textbf{17} (2000) 4175} [\href{https://arxiv.org/abs/gr-qc/0005017}{arXiv:gr-qc/0005017}] [\href{https://inspirehep.net/search?p=find+eprint+gr-qc/0005017}{\scshape{in}SPIRE}].
																	
\bibitem{Govindarajan:2001ee} T.R. Govindarajan, R.K. Kaul and V. Suneeta, \emph{Logarithmic correction to the Bekenstein-Hawking entropy of the BTZ black hole}, \href{https://iopscience.iop.org/article/10.1088/0264-9381/18/15/303}{\emph{Class. Quant. Grav.} \textbf{18} (2001) 2877} [\href{https://arxiv.org/abs/gr-qc/0104010}{arXiv:gr-qc/0104010}] [\href{https://inspirehep.net/search?p=find+eprint+gr-qc/0104010}{\scshape{in}SPIRE}].
																	
\bibitem{Gupta:2002bg} K.S. Gupta and S. Sen, \emph{Further evidence for the conformal structure of a Schwarzschild black hole in an algebraic approach}, \href{https://www.sciencedirect.com/science/article/pii/S0370269301015015?via%3Dihub}{\emph{Phys. Lett.} \textbf{B526} (2002) 121} [\href{https://arxiv.org/abs/hep-th/0112041}{arXiv:hep-th/0112041}] [\href{https://inspirehep.net/search?p=find+eprint+hep-th/0112041}{\scshape{in}SPIRE}].
																		
\bibitem{Medved:2004eh} A.J.M. Medved, \emph{A comment on black hole entropy or does Nature abhor a logarithm?}, \href{https://iopscience.iop.org/article/10.1088/0264-9381/22/1/009}{\emph{Class. Quant. Grav.} \textbf{22} (2005) 133} [\href{https://arxiv.org/abs/gr-qc/0406044}{arXiv:gr-qc/0406044}] [\href{https://inspirehep.net/search?p=find+eprint+gr-qc/0406044}{\scshape{in}SPIRE}].
																		
\bibitem{Page:2005xp} D.N. Page, \emph{Hawking radiation and black hole thermodynamics}, \href{https://iopscience.iop.org/article/10.1088/1367-2630/7/1/203}{\emph{New J. Phys.} \textbf{7} (2005) 203} [\href{https://arxiv.org/abs/hep-th/0409024}{arXiv:hep-th/0409024}] [\href{https://inspirehep.net/search?p=find+eprint+hep-th/0409024}{\scshape{in}SPIRE}].
																		
\bibitem{Banerjee:2008cf} R. Banerjee and B.R. Majhi, \emph{Quantum tunneling beyond semiclassical approximation}, \href{https://iopscience.iop.org/article/10.1088/1126-6708/2008/06/095}{\emph{JHEP} \textbf{06} (2008) 095} [\href{https://arxiv.org/abs/0805.2220}{arXiv:0805.2220}] [\href{https://inspirehep.net/search?p=find+eprint+0805.2220}{\scshape{in}SPIRE}].	
																		
\bibitem{Banerjee:2009fz} R. Banerjee and B.R. Majhi, \emph{Quantum tunneling and trace anomaly}, \href{https://www.sciencedirect.com/science/article/pii/S0370269309002834?via%3Dihub}{\emph{Phys. Lett.} \textbf{B674} (2009) 218} [\href{https://arxiv.org/abs/0808.3688}{arXiv:0808.3688}] [\href{https://inspirehep.net/search?p=find+eprint+0808.3688}{\scshape{in}SPIRE}].
																			
\bibitem{Majhi:2009gi} B.R. Majhi, \emph{Fermion tunneling beyond semiclassical approximation}, \href{https://journals.aps.org/prd/abstract/10.1103/PhysRevD.79.044005}{\emph{Phys. Rev.} \textbf{D79} (2009) 044005} [\href{https://arxiv.org/abs/0809.1508}{arXiv:0809.1508}] [\href{https://inspirehep.net/search?p=find+eprint+0809.1508}{\scshape{in}SPIRE}].
																			
\bibitem{Cai:2010ua} R.G. Cai, L.M. Cao and N. Ohta, \emph{Black holes in gravity with conformal anomaly and logarithmic term in black hole entropy}, \href{https://link.springer.com/article/10.1007%2FJHEP04%282010%29082}{\emph{JHEP} \textbf{04} (2010) 082} [\href{https://arxiv.org/abs/0911.4379}{arXiv:0911.4379}] [\href{https://inspirehep.net/search?p=find+eprint+0911.4379}{\scshape{in}SPIRE}].	
																				
\bibitem{Aros:2010jb} R. Aros, D.E. Diaz and A. Montecinos, \emph{Logarithmic correction to BH entropy as Noether charge}, \href{https://link.springer.com/article/10.1007%2FJHEP07%282010%29012}{\emph{JHEP} \textbf{07} (2010) 012} [\href{https://arxiv.org/abs/1003.1083}{arXiv:1003.1083}] [\href{https://inspirehep.net/search?p=find+eprint+1003.1083}{\scshape{in}SPIRE}].	
																					
\bibitem{Solodukhin:2010pk} S.N. Solodukhin, \emph{Entanglement entropy of round spheres}, \href{https://www.sciencedirect.com/science/article/pii/S0370269310010932?via%3Dihub}{\emph{Phys. Lett.} \textbf{B693} (2010) 605} [\href{https://arxiv.org/abs/1008.4314}{arXiv:1008.4314}] [\href{https://inspirehep.net/search?p=find+eprint+1008.4314}{\scshape{in}SPIRE}].
%-------------------------------------------------------------------------------------------------
																						
\bibitem{Dabholkar:2011ad} A. Dabholkar, J. Gomes and S. Murthy, \emph{Quantum black holes, localization and the topological string}, \href{https://link.springer.com/article/10.1007%2FJHEP06%282011%29019}{\emph{JHEP} \textbf{06} (2011) 019} [\href{https://arxiv.org/abs/1012.0265}{arXiv:1012.0265}] [\href{https://inspirehep.net/search?p=find+eprint+1012.0265}{\scshape{in}SPIRE}].
																							
\bibitem{Dabholkar:2013ad} A. Dabholkar, J. Gomes and S. Murthy, \emph{Localization \& exact holography}, \href{https://link.springer.com/article/10.1007%2FJHEP04%282013%29062}{\emph{JHEP} \textbf{04} (2013) 062} [\href{https://arxiv.org/abs/1111.1161}{arXiv:1111.1161}] [\href{https://inspirehep.net/search?p=find+eprint+1111.1161}{\scshape{in}SPIRE}].	
																								
\bibitem{Gupta:2013rkg} R.K. Gupta and S. Murthy, \emph{All solutions of the localization equations for N=2 quantum black hole entropy}, \href{https://link.springer.com/article/10.1007%2FJHEP02%282013%29141}{\emph{JHEP} \textbf{02} (2013) 141} [\href{https://arxiv.org/abs/1208.6221}{arXiv:1208.6221}] [\href{https://inspirehep.net/search?p=find+eprint+1208.6221}{\scshape{in}SPIRE}].	
																									
\bibitem{Dabholkar:2015ad} A. Dabholkar, J. Gomes and S. Murthy, \emph{Nonperturbative black hole entropy and Kloosterman sums}, \href{https://link.springer.com/article/10.1007%2FJHEP03%282015%29074}{\emph{JHEP} \textbf{03} (2015) 074} [\href{https://arxiv.org/abs/1404.0033}{arXiv:1404.0033}] [\href{https://inspirehep.net/search?p=find+eprint+1404.0033}{\scshape{in}SPIRE}].
																										
\bibitem{Murthy:2015sm} S. Murthy and V. Reys, \emph{Functional determinants, index theorems, and exact quantum black hole entropy}, \href{https://link.springer.com/article/10.1007%2FJHEP12%282015%29028}{\emph{JHEP} \textbf{12} (2015) 028} [\href{https://arxiv.org/abs/1504.01400}{arXiv:1504.01400}] [\href{https://inspirehep.net/search?p=find+eprint+1504.01400}{\scshape{in}SPIRE}].
																											
\bibitem{Gupta:2015rkg} R.K. Gupta, Y. Ito and I. Jeon, \emph{Supersymmetric localization for BPS black hole entropy: 1-loop partition function from vector multiplets}, \href{https://link.springer.com/article/10.1007%2FJHEP11%282015%29197}{\emph{JHEP} \textbf{11} (2015) 197} [\href{https://arxiv.org/abs/1504.01700}{arXiv:1504.01700}] [\href{https://inspirehep.net/search?p=find+eprint+1504.01700}{\scshape{in}SPIRE}].
																												
\bibitem{Murthy:2016sm} S. Murthy and V. Reys, \emph{Single-centered black hole microstate degeneracies from instantons in supergravity}, \href{https://link.springer.com/article/10.1007%2FJHEP04%282016%29052}{\emph{JHEP} \textbf{04} (2016) 052} [\href{https://arxiv.org/abs/1512.01553}{arXiv:1512.01553}] [\href{https://inspirehep.net/search?p=find+eprint+1512.01553}{\scshape{in}SPIRE}].						
																													
																													
%------------------------------------------------------------------------------------------------------------------
\bibitem{DeWitt:1965ff} B.S. DeWitt, \emph{Dynamical theory of groups and fields}, Gordon and Breach, New York, NY, U.S.A. (1965).
																													
\bibitem{DeWitt:1967gg} B.S. DeWitt, \emph{Quantum theory of gravity. 1. The canonical theory}, \href{https://journals.aps.org/pr/abstract/10.1103/PhysRev.160.1113}{\emph{Phys. Rev.} \textbf{160} (1967) 1113} [\href{https://inspirehep.net/search?p=find+J+%22Phys.Rev.,160,1113%22}{\scshape{in}SPIRE}].
																														
\bibitem{DeWitt:1967hh} B.S. DeWitt, \emph{Quantum theory of gravity. 2. The manifestly covariant theory}, \href{https://journals.aps.org/pr/abstract/10.1103/PhysRev.162.1195}{\emph{Phys. Rev.} \textbf{162} (1967) 1195} [\href{https://inspirehep.net/search?p=find+J+%22Phys.Rev.,162,1195%22}{\scshape{in}SPIRE}].
																															
\bibitem{DeWitt:1967ii} B.S. DeWitt, \emph{Quantum theory of gravity. 3. Applications of the covariant theory}, \href{https://journals.aps.org/pr/abstract/10.1103/PhysRev.162.1239}{\emph{Phys. Rev.} \textbf{162} (1967) 1239} [\href{https://inspirehep.net/search?p=find+J+%22Phys.Rev.,162,1239%22}{\scshape{in}SPIRE}].
																																
\bibitem{Seeley:1966tt} R.T. Seeley, \emph{Singular integrals and boundary value problems}, \href{https://www.jstor.org/stable/2373078?origin=crossref&seq=1#page_scan_tab_contents}{\emph{Amer. J. Math.} \textbf{88} (1966) 781}.
																																
\bibitem{Seeley:1969uu} R. Seeley, \emph{The resolvent of an elliptic boundary value problem}, \href{https://www.jstor.org/stable/2373309?origin=crossref&seq=1#page_scan_tab_contents}{\emph{Amer. J. Math.} \textbf{91} (1969) 889}.
																																
\bibitem{Vassilevich:2003ll} D.V. Vassilevich, \emph{Heat kernel expansion: user’s manual}, \href{https://www.sciencedirect.com/science/article/pii/S0370157303003545?via%3Dihub}{\emph{Phys. Rept.} \textbf{388} (2003) 279} [\href{https://arxiv.org/abs/hep-th/0306138}{arXiv:hep-th/0306138}] [\href{https://inspirehep.net/search?p=find+EPRINT+hep-th/0306138}{\scshape{in}SPIRE}].
																																	
%-------------------------------------------------------------------------------------------------------------------------------
\bibitem{Denef:2010nc} F. Denef, S. A. Hartnoll and S. Sachdev, \emph{Black Hole Determinants and Quasinormal Modes}, \emph{Class. Quant. Grav.} \textbf{27} (2010) 125001 [\href{https://arxiv.org/abs/0908.2657}{arXiv:0908.2657}] [\href{https://inspirehep.net/literature?sort=mostrecent&size=25&page=1&q=find%20eprint%200908.2657}{\scshape{in}SPIRE}].
																																		
\bibitem{David:2010xn} J. R. David, M. R. Gaberdiel and R. Gopakumar, \emph{The Heat Kernel on $AdS_3$ and Its Applications}, \emph{JHEP} \textbf{04} (2010) 125 [\href{https://arxiv.org/abs/0911.5085}{arXiv:0911.5085}] [\href{https://inspirehep.net/literature?sort=mostrecent&size=25&page=1&q=find%20eprint%200911.5085}{\scshape{in}SPIRE}].
																																			
\bibitem{Gopakumar:2011xi} R. Gopakumar, R. K. Gupta and S. Lal, \emph{The Heat Kernel on $AdS$}, \emph{JHEP} \textbf{11} (2011) 010 [\href{https://arxiv.org/abs/1103.3627}{arXiv:1103.3627}] [\href{https://inspirehep.net/literature?sort=mostrecent&size=25&page=1&q=find%20eprint%201103.3627}{\scshape{in}SPIRE}].
																																				
\bibitem{Lovrekovic:2016ni} I. Lovrekovic, \emph{One Loop Partition Function of Six Dimensional Conformal Gravity Using Heat Kernel on $AdS$}, \emph{JHEP} \textbf{10} (2016) 064 [\href{https://arxiv.org/abs/1512.00858}{arXiv:1512.00858}] [\href{https://inspirehep.net/literature?sort=mostrecent&size=25&page=1&q=find%20eprint%201512.00858}{\scshape{in}SPIRE}].
																																					
%-------------------------------------------------------------------------------------------------
\bibitem{Mandal:2010im} I. Mandal and A. Sen, \emph{Black hole microstate counting and its macroscopic counterpart}, \href{https://iopscience.iop.org/article/10.1088/0264-9381/27/21/214003}{\emph{Class. Quant. Grav.} \textbf{27} (2010) 214003} [\href{https://arxiv.org/abs/1008.3801}{arXiv:1008.3801}] [\href{https://inspirehep.net/search?p=find+eprint+1008.3801}{\scshape{in}SPIRE}].
																																					
\bibitem{Sen:2014nvt} A. Sen, \emph{Microscopic and macroscopic entropy of extremal black holes in string theory}, \href{https://link.springer.com/article/10.1007/s10714-014-1711-5}{\emph{Gen. Rel. Grav.} \textbf{46} (2014) 1711} [\href{https://arxiv.org/abs/1402.0109v1}{arXiv:1402.0109}] [\href{https://inspirehep.net/literature?sort=mostrecent&size=25&page=1&q=find%20eprint%201402.0109}{\scshape{in}SPIRE}].	
%------------------------------------------------------------------------------------------------------------
																																						%\bibitem{David:2010xn} J. R. David, M. R. Gaberdiel and R. Gopakumar, \emph{The heat kernel on $AdS_3$ and its applications}, \emph{JHEP} \textbf{04} (2010) 125 [\href{https://arxiv.org/abs/0911.5085}{arXiv:0911.5085}] [\href{https://inspirehep.net/record/838200}{\scshape{in}SPIRE}].
																																						
																																						%\bibitem{Gopakumar:2011xi} R. Gopakumar, R. K. Gupta and S. Lal, \emph{The heat kernel on $AdS$}, \emph{JHEP} \textbf{11} (2011) 010 [\href{https://arxiv.org/abs/1103.3627}{arXiv:1103.3627}] [\href{https://inspirehep.net/record/893128}{\scshape{in}SPIRE}].
																																						
																																						%\bibitem{Lovrekovic:2016ni} I. Lovrekovic, \emph{One loop partition function of six dimensional conformal gravity using heat kernel on $AdS$}, \emph{JHEP} \textbf{10} (2016) 064 [\href{https://arxiv.org/abs/1512.00858}{arXiv:1512.00858}] [\href{https://inspirehep.net/record/1407978}{\scshape{in}SPIRE}].
																																						
																																						%\bibitem{Denef:2010nc} F. Denef, S. A. Hartnoll and S. Sachdev, \emph{Black hole determinants and quasinormal modes}, \emph{Class. Quant. Grav.} \textbf{27} (2010) 125001 [\href{https://arxiv.org/abs/0908.2657}{arXiv:0908.2657}] [\href{https://inspirehep.net/record/829077}{\scshape{in}SPIRE}].	
																																						
%---------------------------------------------------------------------------------
\bibitem{Grana:2006mg} M. Grana, \emph{Flux compactifications in string theory: a comprehensive review}, \emph{Phys. Rept.} \textbf{423} (2006) 91 [\href{https://arxiv.org/abs/hep-th/0509003}{arXiv:hep-th/0509003}] [\href{https://inspirehep.net/literature?sort=mostrecent&size=25&page=1&q=find%20eprint%20hep-th%2F0509003}{\scshape{in}SPIRE}].
																																							
\bibitem{Freedman:2012xp} D. Z. Freedman and A. Van Proeyen, \emph{Supergravity}, Cambridge University Press, Cambridge,
U.K. (2012) [\href{https://doi.org/10.1017/CBO9781139026833}{DOI}].
																																							
%---------------------------------------------------------------------------------
\bibitem{Adamo:2014lk} T. Adamo, and E.T. Newman, \emph{The Kerr-Newman metric: A review}, \href{http://www.scholarpedia.org/article/Kerr-Newman_metric}{\emph{Scholarpedia} \textbf{9} (2014) 31791} [\href{https://arxiv.org/abs/1410.6626}{arXiv:1410.6626 }] [\href{https://inspirehep.net/record/1323609}{\scshape{in}SPIRE}].
%--------------------------------------------------------------------------------------------
\bibitem{Gibbons:1977ta} G.W. Gibbons and S.W. Hawking, \emph{Action integrals and partition functions in quantum gravity}, \href{https://journals.aps.org/prd/abstract/10.1103/PhysRevD.15.2752}{\emph{Phys. Rev.} \textbf{D15} (1977) 2752} [\href{https://inspirehep.net/search?p=find+J+%22Phys.Rev.,D15,2752%22}{\scshape{in}SPIRE}].
																																								
\bibitem{Hawking:1978td} S.W. Hawking, \emph{Quantum gravity and path integrals}, \href{https://journals.aps.org/prd/abstract/10.1103/PhysRevD.18.1747}{\emph{Phys. Rev.} \textbf{D18 } (1978) 1747} [\href{https://inspirehep.net/search?p=find+J+%22Phys.Rev.,D18,1747%22}{\scshape{in}SPIRE}].
																																									
\bibitem{Hawking:1977te} S.W. Hawking, \emph{Zeta function regularization of path integrals in curved space-time}, \href{https://link.springer.com/article/10.1007%2FBF01626516}{\emph{Commun. Math. Phys.} \textbf{55} (1977) 133} [\href{https://inspirehep.net/search?p=find+J+%22Comm.Math.Phys.,55,133%22}{\scshape{in}SPIRE}].
																																										
\bibitem{Denardo:1982tb} G. Denardo and E. Spallucci, \emph{Induced quantum gravity from heat kernel expansion}, \href{https://link.springer.com/article/10.1007%2FBF02902652}{\emph{Nuovo Cim.} \textbf{A69} (1982) 151} [\href{https://inspirehep.net/search?p=find+J+%22NuovoCim.,A69,151%22}{\scshape{in}SPIRE}].
																																											
\bibitem{Avramidi:1994th} I.G. Avramidi, \emph{The heat kernel approach for calculating the effective action in quantum field theory and quantum gravity} [\href{https://arxiv.org/abs/hep-th/9509077}{arXiv:hep-th/9509077}] [\href{https://inspirehep.net/search?p=find+EPRINT+hep-th/9509077}{\scshape{in}SPIRE}].	
																																											
\bibitem{Peixoto:2001wx} G. De Berredo-Peixoto, \emph{A note on the heat kernel method applied to fermions}, \href{https://www.worldscientific.com/doi/abs/10.1142/S0217732301005965}{\emph{Mod. Phys. Lett.} \textbf{A16} (2001) 2463} [\href{https://arxiv.org/abs/hep-th/0108223}{arXiv:hep-th/0108223}] [\href{https://inspirehep.net/search?p=find+EPRINT+hep-th/0108223}{\scshape{in}SPIRE}].	
																																											%-------------------------------------------------------------------------------------------------------------
																																											
\bibitem{Schwinger:1951sp} J.S. Schwinger, \emph{On gauge invariance and vacuum polarization}, \href{https://journals.aps.org/pr/abstract/10.1103/PhysRev.82.664}{\emph{Phys. Rev.} \textbf{82} (1951) 664} [\href{https://inspirehep.net/search?p=find+J+%22Phys.Rev.,82,664%22}{\scshape{in}SPIRE}].
																																												
\bibitem{DeWitt:1975ps} B.S. DeWitt, \emph{Quantum field theory in curved space-time}, \href{https://www.sciencedirect.com/science/article/abs/pii/0370157375900514?via%3Dihub}{\emph{Phys. Rept.} \textbf{19} (1975) 295} [\href{https://inspirehep.net/search?p=find+J+%22Phys.Rept.,19,295%22}{\scshape{in}SPIRE}].
																																													%----------------------------------------------------------------------------------------------------------------------------				
\bibitem{Karan:2018ac} S. Karan, S. Kumar and B. Panda, \emph{General heat kernel coefficients for massless free spin-3/2 Rarita-Schwinger field}, \href{https://www.worldscientific.com/doi/abs/10.1142/S0217751X1850063X}{\emph{Int. J. Mod. Phys.} \textbf{A33} (2018) 1850063} [\href{https://arxiv.org/abs/1709.08063}{arXiv:1709.08063}] [\href{https://inspirehep.net/search?p=find+EPRINT+arXiv:1709.08063}{\scshape{in}SPIRE}].
																																													
\bibitem{Henry:2000wd} R.C. Henry, \emph{Kretschmann scalar for a kerr-newman black hole}, \href{https://iopscience.iop.org/article/10.1086/308819}{\emph{Astrophys. J.} \textbf{535} (2000) 350} [\href{https://arxiv.org/abs/astro-ph/9912320}{arXiv:astro-ph/9912320}] [\href{https://inspirehep.net/search?p=find+EPRINT+astro-ph/9912320}{\scshape{in}SPIRE}].
																																													
\bibitem{Cherubini:2002we} C. Cherubini, D. Bini, S. Capozziello and R. Ruffini, \emph{Second order scalar invariants of the Riemann tensor: Applications to black hole space-times}, \href{https://www.worldscientific.com/doi/abs/10.1142/S0218271802002037}{\emph{Int. J. Mod. Phys.} \textbf{D11} (2002) 827} [\href{https://arxiv.org/abs/gr-qc/0302095}{arXiv:gr-qc/0302095}] [\href{https://inspirehep.net/search?p=find+EPRINT+gr-qc/0302095}{\scshape{in}SPIRE}].
																																													%--------------------------------------------------------------------------------------------------------------------
																																													
\bibitem{Bekenstein:2008bk} J.D. Bekenstein, \emph{Bekenstein-Hawking entropy}, \href{http://www.scholarpedia.org/article/Bekenstein-Hawking_entropy}{\emph{Scholarpedia} \textbf{3} (2008) 7375}.
																																													
\bibitem{Andrianopoli:2007nx} L. Andrianopoli, R. D'Auria, S. Ferrara and M. Trigiante, \emph{Black-hole attractors in N=1 supergravity}, \href{https://iopscience.iop.org/article/10.1088/1126-6708/2007/07/019}{\emph{JHEP} \textbf{07} (2007) 019} [\href{https://arxiv.org/abs/hep-th/0703178}{arXiv:hep-th/0703178}] [\href{https://inspirehep.net/search?p=find+eprint+hep-th/0703178}{\scshape{in}SPIRE}].
																																													
\bibitem{Andrianopoli:2002gm} L. Andrianopoli, R. D'Auria and S. Ferrara, \emph{Consistent reduction of $N=2 \to N=1$ four dimensional supergravity coupled to matter}, \href{https://www.sciencedirect.com/science/article/abs/pii/S0550321302000901?via%3Dihub}{\emph{Nucl. Phys.} \textbf{B628} (2002) 387} [\href{https://arxiv.org/abs/hep-th/0112192}{arXiv:arXiv:hep-th/0112192}] [\href{https://inspirehep.net/literature?sort=mostrecent&size=25&page=1&q=find%20eprint%20hep-th%2F0112192}{\scshape{in}SPIRE}].

\end{thebibliography}
\end{document}